\documentclass[trackchanges,twocolumn]{aastex701}

\usepackage{amsmath}

\begin{document}

\title{New Constraints on \(r\)-process Nucleosynthesis in Neutron Star Mergers\\from GW170817 Late-Phase Spectra}

\author[orcid=0009-0002-1232-243X]{Salma Rahmouni}
\affiliation{Astronomical Institute, Tohoku University, Aoba, Sendai 980-8578, Japan}
\email[show]{rahmouni.salma@astr.tohoku.ac.jp}  

\author[orcid=0000-0001-8253-6850]{Masaomi Tanaka} 
\affiliation{Astronomical Institute, Tohoku University, Aoba, Sendai 980-8578, Japan}
\affiliation{Division for the Establishment of Frontier Sciences, Organization for Advanced Studies, Tohoku University, Sendai 980-8577, Japan}
\email{masaomi.tanaka@astr.tohoku.ac.jp}

\author[orcid=0000-0002-5302-073X]{Daiji Kato}
\affiliation{National Institute for Fusion Science, 322-6 Oroshi-cho, Toki 509-5292, Japan}
\affiliation{Interdisciplinary Graduate School of Engineering Sciences, Kyushu University, Kasuga, Fukuoka 816-8580, Japan}
\email{kato.daiji@nifs.ac.jp} 

\author[orcid=0000-0003-0039-1163]{Gediminas Gaigalas}
\affiliation{Institute of Theoretical Physics and Astronomy, Vilnius University, Saulètekio Ave. 3, LT-10257 Vilnius, Lithuania}
\email{gediminas.gaigalas@tfai.vu.lt} 

\author[orcid=0000-0002-2502-3730]{Kenta Hotokezaka}
\affiliation{Research Center for the Early Universe (RESCEU), Graduate School of Science,
The University of Tokyo, 7-3-1 Hongo, Bunkyo, Tokyo 113-0033, Japan}
\email{kentah@resceu.s.u-tokyo.ac.jp}

\begin{abstract}

The neutron star merger event GW170817 provided the first direct evidence of \(r\)-process nucleosynthesis. Observed spectra of its electromagnetic counterpart AT2017gfo exhibited several features that encode information on the nature and abundance of the synthesized \(r\)-process elements.
In this study, we investigate the late nebular-phase spectral features of AT2017gfo, which provide important probes of the elemental abundance of the ejecta.
We construct an analytic spectral model that computes emission features produced by the radiative decay of collisionally excited ions through allowed and forbidden transitions. By comparing our model to AT2017gfo, we identify La III and Ce III as the main contributors to the emission features at \(1.4\,\mu\)m and \(1.6\,\mu\)m, respectively. We also confirm Te III as the dominant contributor to the \(2.1\,\mu\)m feature proposed in previous works. We infer mass fractions of  \(X({\rm La})\approx 0.025-0.05\), \(X({\rm Ce})\approx 0.05-0.1\), and \(X({\rm Te})\approx 0.04-0.08\), although the La and Ce abundance estimates remain tentative due to uncertainties in the radiation field. From the non-detections of Kr and Sb lines, we derive upper limits of \(X({\rm Kr})\lesssim 0.03\) and \(X({\rm Sb})\lesssim 0.003\). These results suggest that nucleosynthesis in the inner ejecta of GW170817 produced a suppressed first \(r\)-process peak and an enhanced heavy-element abundance compared to the solar \(r\)-process pattern, with an estimated lanthanide fraction of \(X_{\rm LN}\approx (3-6) \times 10^{-2}\). Our conclusions are consistent with the apparent universality of heavy \(r\)-process elements and the lanthanide fraction inferred from observations of \(r\)-enhanced metal-poor stars.

\end{abstract}


\section{Introduction} 

The coalescence of compact objects involving a neutron star has long been considered one of the primary sites for the nucleosynthesis of heavy elements through the rapid neutron capture process (\(r\)-process, e.g., \citealt{lattimer1974, symbalisty1982, eichler1989, freiburghaus1999, goriely2011, wanajo2014}). The decay of the freshly synthesized elements in such events powers an electromagnetic emission that can be observed in the infrared, optical, and ultraviolet wavelengths, known as kilonova \citep{li1998, metzger2010, roberts2011}.

The first direct observational evidence for \(r\)-process nucleosynthesis came in 2017 with the detection of gravitational waves (GW170817, \citealt{abbott2017a}) from a binary neutron star merger (NSM), where follow-up observations detected an electromagnetic counterpart (AT2017gfo, \citealt{abbott2017b}). The luminosity and color evolution of this event were consistent with theoretical expectations of kilonovae (e.g., \citealt{arcavi2017, pian2017, smartt2017, utsumi2017}), which proved that heavy elements were synthesized in this event (e.g., \citealt{kasen2017, shibata2017, tanaka2017, kawaguchi2018, rosswog2018, perego2019}).

Constraining the elemental abundance synthesized in GW170817 is necessary to investigate whether NSMs can account for all \(r\)-process elements in the Universe. 
To achieve this goal, several studies have identified and investigated atomic features in the early-phase spectra of AT2017gfo.
\cite{watson2019} proposed Sr II as the main candidate for the \(1\,\mu\)m absorption feature, an identification that was subsequently confirmed by radiative transfer simulations \citep{domoto2021, gillanders2021}. Recent findings suggest that this feature may be blended with He I when non-local thermodynamical equilibrium (non-LTE) populations are considered \citep{perego2022, tarumi2023, sneppen2024he, arya2026, chiba2026}. The \(0.7\,\mu\)m absorption feature was attributed to Y II (\citealt{sneppen2023}, but see also \citealt{pognan2023}). Two additional prominent infrared absorption features, detected at \(1.2\,\mu\)m and \(1.4\,\mu\)m, were attributed to La III and Ce III, respectively \citep{domoto2022, domoto2023, tanaka2023}, although the \(1.2\,\mu\)m feature may also contain contributions from Gd III \citep{rahmouni2025}. While the aforementioned studies have provided valuable abundance constraints (e.g., \citealt{domoto2021, domoto2022, gillanders2021, sneppen2023}), they mainly reflect the composition of the outer layer of the ejecta due to the optically thick conditions at early-phase.

While the early-phase spectra of AT2017gfo have been extensively studied, spectral features observed at late-phase remain more challenging to interpret. This difficulty arises from the increasing importance of non-thermal effects caused by high-energy particles produced during the decay of freshly synthesized elements. Identifying the late-phase features of kilonovae is particularly important because they provide a direct probe of the elemental abundance distribution synthesized during the merger due to the optically thin conditions of the ejecta. Several studies have investigated the late-phase evolution of the ejecta while accounting for non-LTE effects (e.g., \citealt{hotokezaka2021, pognan2022steady, pognan2023, pognan2025, jerkstrand2025}).
Using physical conditions characteristic of the late-phase inferred from such simulations, \cite{hotokezaka2023} proposed Te III as a likely candidate for the emission feature observed at \(2.1\,\mu\)m. This identification was subsequently confirmed by \cite{jerkstrand2025} and \cite{pognan2025}, although the latter argued that additional elements may also contribute to the feature. Two other prominent late-phase features are observed near \(1.4\,\mu\)m and \(1.6\,\mu\)m. For these features, \cite{gillanders2024} proposed several possible candidates but did not reach definitive identifications. Recent non-LTE spectral simulations do not predict pronounced excess emission at these wavelengths \citep{jerkstrand2025, pognan2025}, leaving the origin of these features largely unexplored.

One of the major challenges in the spectral investigation of kilonovae is the lack of complete and accurate atomic data. Solving for the ejecta evolution and modeling the late-phase spectra require transition probabilities and collisional rates for individual lines, as well as recombination and ionization rates for individual \(r\)-process elements. Although significant effort has focused on improving the atomic data required for non-LTE modeling (e.g., \citealt{mulholland2024, mulholland2024te, mulholland2025, mulholland2025te, mulholland2026ce, banerjee2025, dougan2025, bromley2026}), considerable uncertainties remain. This limitation is further compounded by the computational cost of treating large atomic datasets and solving for the populations of complex multi-level systems in non-LTE calculations. As a result, current late-phase spectral models typically include only a limited number of heavy elements and often rely on approximations for adopted atomic data (e.g., \citealt{jerkstrand2025, pognan2025}). These challenges make it difficult to fully reproduce the observed spectra.

In this work, we construct an analytic spectral model capable of treating a much broader range of \(r\)-process elements than is currently feasible in non-LTE models, with the aim of reproducing and identifying the late-phase features observed in AT2017gfo.
By inferring the relevant physical conditions directly from the observed spectra, we evaluate the expected luminosities of individual transitions using updated atomic data. The details of the model are introduced in Section \ref{sec:model}. By comparing the model's predictions with the observed spectra, we identify La III, Ce III, and Te III as the main contributors to the emission features observed at \(1.4\,\mu\)m, \(1.6\,\mu\)m, and \(2.1\,\mu\)m, respectively. The relative importance of these species and the corresponding mass fraction constraints are discussed in Section \ref{sec:cand}. We examine the broader implications of our results in Section \ref{sec:disc}, and summarize our conclusions in Section \ref{sec:concl}.

\section{Spectral Model}\label{sec:model}
Our spectral model computes the luminosity emitted through the radiative decay of collisionally excited elements in an optically thin ejecta. In this section, we describe the basic equations underlying the model and the atomic data adopted to compute the emission lines of all species considered in this work.

\subsection{Luminosity Evaluation}
We assume that each transition can be treated independently by considering that it originates from a two-level energy system with energy separation \(\Delta E\). Under this approximation, transitions to and from other atomic levels are neglected. The balance between collisional excitation, collisional de-excitation, and radiative decay is therefore given by
\begin{equation}
n_{\rm e}n_1q_{12} = n_{\rm e}n_2q_{21} + n_2\beta A_{21},
\end{equation}
where \(n_{\rm e}\) is the electron density, \(n_k\) is the population of level \(k\), and \(q_{12}\) is the collisional excitation rate coefficient, such that
\begin{equation}
q_{12} = \frac{8.629\times10^{-6}}{T_{\rm e}^{1/2}}\,\frac{\Upsilon_{12}}{g_1}e^{-\Delta E/kT_{\rm e}}.\label{eq:col_rate}
\end{equation}
Here \(T_{\rm e}\) is the electron temperature, \(g_1\) is the statistical weight of the lower level, and \(\Upsilon_{12}\) is the velocity-averaged collision strength,
\begin{equation}
\Upsilon_{12} = \int_0^{\infty}\Omega_{12}(E) e^{-E/kT_{\rm e}}\,d\left(\frac{E}{kT_{\rm e}}\right).\label{eq:upsilon}
\end{equation}
\(E\) is the kinetic energy of the electron after the interaction, and \(\Omega_{12}\) is the energy-specific collision strength, an atomic property related to the collisional excitation cross section of the transition in question. The collisional excitation and de-excitation rate coefficients satisfy the detailed-balance relation
\begin{equation}
q_{12} = \frac{g_2}{g_1}q_{21}e^{-\Delta E/kT_{\rm e}}.
\end{equation} 

The radiative rate from the upper level is given by \(n_2 \beta A_{21}\), where \(A_{21}\) is the Einstein coefficient for spontaneous emission, while \(\beta\) accounts for the photon self-absorption and can be expressed as
\begin{equation}
\beta = \frac{1-e^{-\tau}}{\tau},
\end{equation}
where \(\tau\) is taken as the Sobolev optical depth in a homologously expanding ejecta
\begin{equation}
	\tau = \frac{\pi e^2}{m_{\rm e} c}n_1 t \lambda f.\label{eq:sobolev}
\end{equation}
Here \(f\) and \(\lambda\) denote the oscillator strength and the wavelength of the transition, respectively.

The luminosity emitted per unit volume by a given transition is then obtained from the radiative decay rate,
\begin{equation}
L = \Delta E \beta n_2 A_{21}.
\end{equation}
It is useful to define the critical density of a transition as
\begin{equation}
n_{\rm cr} = \frac{\beta A_{21}}{q_{21}},\label{eq:ncr}
\end{equation}
which corresponds to the density at which radiative decay and collisional de-excitation occur at comparable rates.

If \(n_{\rm cr} < n_{\rm e}\), collisional de-excitation becomes important, and the upper level population can be approximated by a Boltzmann distribution \(n_2\approx n_i \frac{g_2}{Z_{\rm p}} e^{-E_2/kT_{\rm e}}\). The luminosity emitted per volume is therefore
\begin{equation}
L=\Delta E n_i \frac{g_2}{Z_{\rm p}(T_e)} e^{-E_2/kT_{\rm e}}\beta A_{21},\label{eq:lum1}
\end{equation}
where \(n_i\) is the number density of the ion, \(Z_{\rm p} (T_e)\) is its partition function, and \(E_k\) is the energy of level \(k\).

Conversely, if \(n_{\rm cr}> n_{\rm e}\), radiative decay dominates over collisional de-excitation, and we may assume \(n_2\beta A_{21}\approx n_2n_1q_{12}\). In our model, we consider only excitations occurring from the ground or from meta-stable levels. The population of the lower level can therefore be approximated by a Boltzmann distribution, and the luminosity emitted per volume is
\begin{equation}
L=\Delta E n_{\rm e} n_i \frac{g_1}{Z_{\rm p}(T_e)}e^{-E_1/kT_{\rm e}}q_{12}.\label{eq:lum2}
\end{equation}

\subsection{Atomic data}\label{sec:model_atom}
\subsubsection{Radiative Transitions}

\begin{figure*}[t]
    \centering
    \begin{tabular}{cc}
    \includegraphics[width=0.45\linewidth]{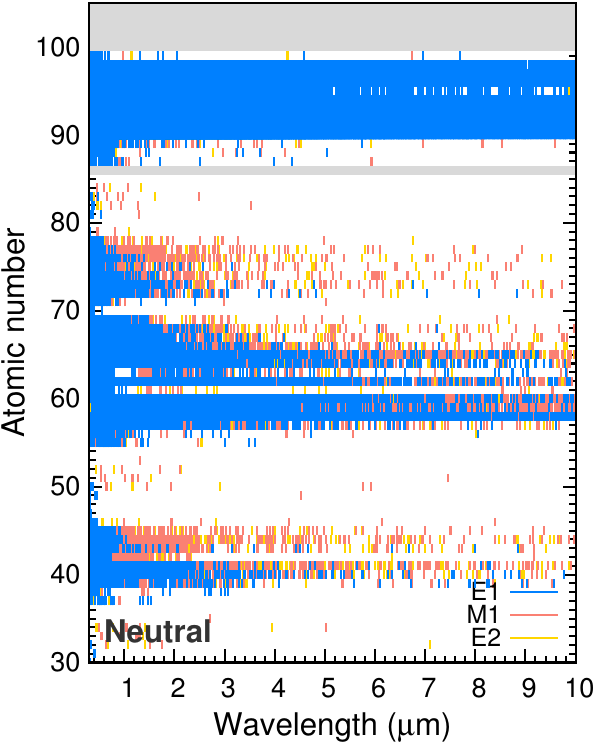} &
    \includegraphics[width=0.45\linewidth]{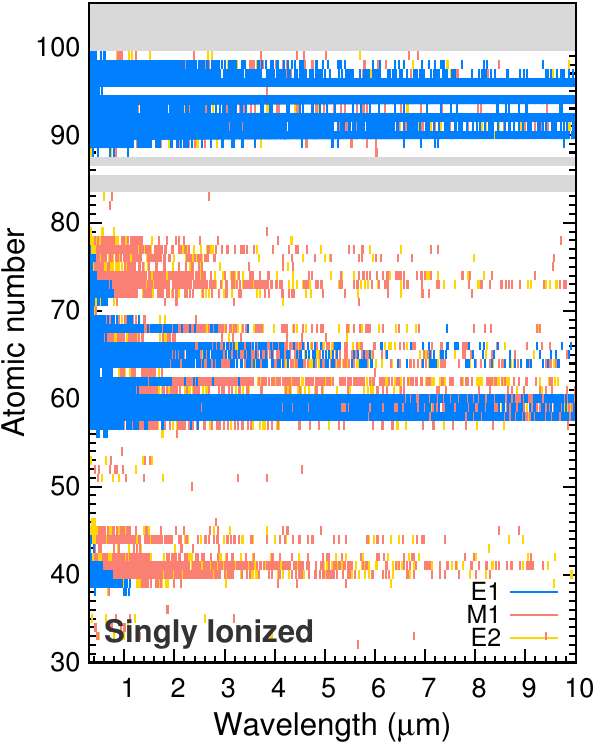} \\
     \includegraphics[width=0.45\linewidth]{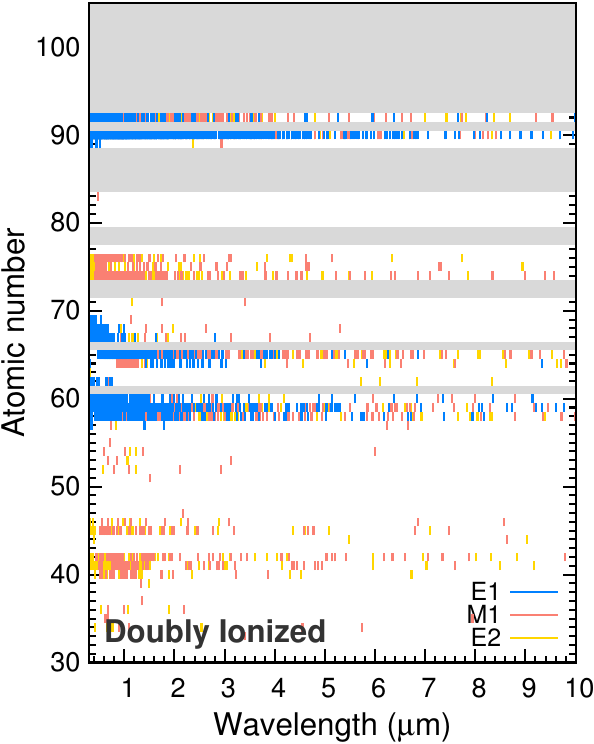} &
    \includegraphics[width=0.45\linewidth]{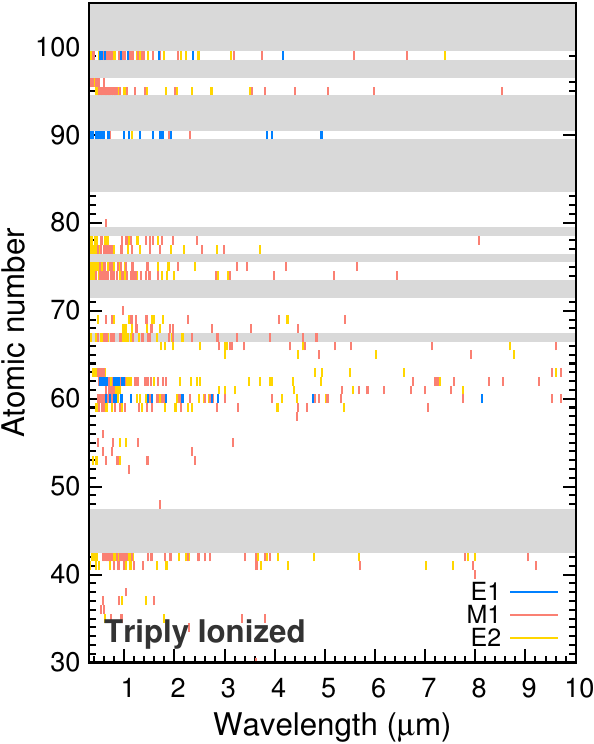} \\
    \end{tabular}
\caption{
  \label{fig:transitions}
	Selected transitions occurring below an energy threshold of 2 eV for neutral, singly, doubly, and triply ionized elements in the wavelength range \(\lambda = 0.3\,\mu{\rm m} - 10\,\mu{\rm m}\). E1, M1, and E2 transitions are highlighted in blue, red, and yellow, respectively. Elements excluded from our selection due to a lack of available energy levels are shown as gray-shaded areas.
}
\end{figure*}

Due to their complex atomic structure, collisionally excited energy levels of heavy elements subsequently decay through both allowed and forbidden channels. Therefore, we include in our model allowed electric dipole (E1), and forbidden magnetic-dipole (M1) and electric quadrupole (E2) transitions, selected using the method introduced by \cite{rahmouni2025}.

Energy-level data of each ion with atomic number \(30\leq Z \leq 90\) in our model are extracted from the National Institute of Standards and Technology Atomic Spectra Database (NIST ASD; \citealt{NIST_ASD}). For elements with  \(91\leq Z \leq 99\), we use the Selected Constants Energy Levels and Atomic Spectra of Actinides (SCASA; \citealt{scasa}). For Nd III (\(Z = 60\)), we adopt the energy levels from \cite{ding2024} due to the completeness of their data. Extracted energy levels are constructed from radiative transitions measured using laboratory spectroscopic experiments and are sufficiently accurate for performing spectral identifications. For completeness, we include in our model all neutral, singly ionized, doubly ionized, and triply ionized species available in the aforementioned atomic databases. 

Allowed and forbidden transitions of each element are selected based on the rigorous selection rules governing the change in parity and total angular momentum
\footnote{The rigorous selection rules are:\\ E1: (1) parity change (2) \(\Delta J=0,\pm 1\) (except (\(0\leftrightarrow 0\)),\\ M1: (1) no parity change (2) \(\Delta J=0,\pm1\) (except \(0\leftrightarrow0\)),\\ E2: (1) no parity change (2) \(\Delta J=0,\pm1,\pm2\) (except \(0\leftrightarrow0, 1/2\leftrightarrow1/2, 0\leftrightarrow1 \)).}, and the wavelength of each transition is calculated using the energy differences between the extracted energy levels. We restrict our analysis to transitions occurring below an energy threshold of 2 eV \citep{rahmouni2025}, as higher energy levels are unlikely to be populated at the typical plasma conditions at late-phase even when population by non-LTE effects is taken into account \citep{pognan2022opacity}.

The wavelengths of the selected E1, M1, and E2 transitions for all elements are shown in Figure \ref{fig:transitions}, while elements excluded due to the lack of atomic data are indicated by gray-shaded regions. Selected E1 transitions are mainly due to lanthanides (\(Z=57-71\)) and actinides (\(Z=89-103\)). This is a consequence of the complex atomic structure of such elements, which shows a strong overlap of the low-lying configuration manifolds involving \(f\), \(d\), and \(s\)-orbitals. These overlapping configurations provide many low-lying levels of different parity within the adopted excitation-energy range, allowing many E1 transitions to be selected.  
In contrast, selected forbidden transitions may arise from a wide range of elements, with a particularly large number of M1 and E2 transitions for lanthanides and actinides, as well as for transition-metal regions with \(Z=40-45\) and \(Z=72-78\). This result is due to the many low-lying energy levels associated with open \(d\) and \(s\)-shell configurations for such species. Since these levels have essentially the same parity, E1 transitions between them are forbidden, and radiative decay mainly proceeds through M1 and E2 transitions.

It is also noteworthy that the number of selected transitions decreases with increasing ionization state. Electrons in higher charged ions experience a larger effective nuclear charge, which results in more tightly bound electrons and an overall increase in the atomic energy scale. Consequently, neutral and weakly ionized species tend to possess a larger number of low-lying excited states and therefore more transitions below our 2 eV threshold, whereas triply ionized species generally have fewer transitions satisfying this criterion.

\subsubsection{Collisional and Radiative Rates}
Computing the luminosity of individual transitions requires both collision strengths and Einstein \(A\)-coefficients. However, comprehensive experimental measurements of these quantities are not available for all elements and ionization states considered in this work. We therefore adopt a two-step approach: (1) First, we employ a set of physically motivated approximations to estimate the collisional and radiative rates of all transitions in order to perform a comprehensive search for candidate spectral features. (2) Second, for the transitions identified as potentially important, we compute more accurate collision strengths and Einstein coefficients using the atomic structure code HULLAC (\citealt{hullac}; see Appendix \ref{app:hullac}). In the following, we describe the approximations adopted in the first step of our analysis.

For E1 transitions, the collision strengths are estimated using the van Regemorter semi-empirical formula \citep{regemorter}
\begin{equation}
	\Upsilon_{12;E1} = 2.39P(x)\left(\frac{\lambda}{1\,\mu\text{m}}\right)^3\left(\frac{g_1A_{21}}{10^6\,\text{s}^{-1}}\right),\label{eq:regemorter}
\end{equation}
where \(x=\Delta E/kT\). This approximation assumes the incident electron can be described as a plane wave and treats the electron-ion interaction as a perturbation. The Gaunt factor \(P(x)\) provides a semi-empirical correction that accounts for additional interaction channels between the electron and ion. \(P(x)\) is integrated over a range of electron velocities and has values typically within \(0.2-1.04\), tabulated in \cite{regemorter}. For the radiative rates, we adopt the Einstein \(A\) coefficient from the calibrated theoretical calculations of \cite{flors2025} for singly and doubly ionized lanthanides. For all other ions, we assume \(g_1A_{21}=10^6\,\text{s}^{-1}\), a value chosen to be broadly representative of strong transitions while lying toward the higher end of the distribution found in atomic databases \citep{vald1, vald2, vald3}
For Th III (\(Z=90\)), transition probabilities are available for a subset of transitions in Table 2 of \cite{domoto2025}. We adopt these tabulated values when available and use the representative value above for all other Th III transitions.

For M1 transitions, \cite{mulholland2025} found that the collision strengths depend on the dominant open shell of the ion. Their calculations yield \(\Upsilon_{12;M1} = 0.001-0.01\) for open \(d\)-shell systems and \(\Upsilon_{12;M1}=0.1-1\) for open \(s\)-shell systems.
Since the first step in our analysis is to identify all potentially relevant contributors to the observed spectra, we adopt a rather optimistic value of \(\Upsilon_{12;M1}=1\) for all M1 transitions. The Einstein \(A\) coefficient for fine-structure M1 transitions is calculated as follows \citep{pasternack1940, shortley1940, bahcall1968}
\begin{equation}
	A_{21;M1} = 82.3 \left(\frac{\lambda}{1\,\mu\text{m}}\right)^{-3}f(J,L,S)\;\text{s}^{-1}
\end{equation}
where \(f(J,L,S)\) is an algebraic factor depending on the orbital angular momentum \(L\), spin \(S\), and the total angular momentum \(J\) of the upper level, such as
\begin{equation}
\small
f(J, L, S) = \frac{(J^2-(L-S)^2)((L+S+1)^2-J^2)}{12J(2J+1)}
\end{equation}
if \(J_{2} = J_{1}+1\) and
\begin{equation}
\small
f(J,L,S) = \frac{((J+1)^2-(L-S)^2)((L+S+1)^2-(J+1)^2)}{12(J+1)(2J+1)}
\end{equation}
 if \(J_2 = J_{1}-1\), where \(J_2\) and \(J_1\) are the total angular momentum of the upper and lower levels, respectively. For inter-configuration M1 transitions for which the above expressions are not applicable, we adopt a representative value of \(A_{21;M1}= 1\,\text{s}^{-1}\), following previous works \citep{hotokezaka2021, ricigliano2025}.

For E2 transitions, previous calculations indicate collision strengths are of similar magnitude to M1 transitions, but generally have smaller Einstein coefficients (e.g., \citealt{mulholland2024te, mulholland2025te}; see also Table \ref{tab:hullac}). We therefore adopt \(\Upsilon_{12;E2} = 1\) and \(A_{21;E2} = 0.1\,{\rm s}^{-1}\) for all E2 transitions.

Since our objective in this first step is to identify candidate contributors, the adopted approximations are intended to provide order-of-magnitude estimates. More accurate atomic data are employed for the key transitions discussed in Section \ref{sec:cand}.

\subsection{Resulting Spectrum}\label{sec:model_result}

\begin{figure}[t!]
    \centering
    \includegraphics[width=\columnwidth]{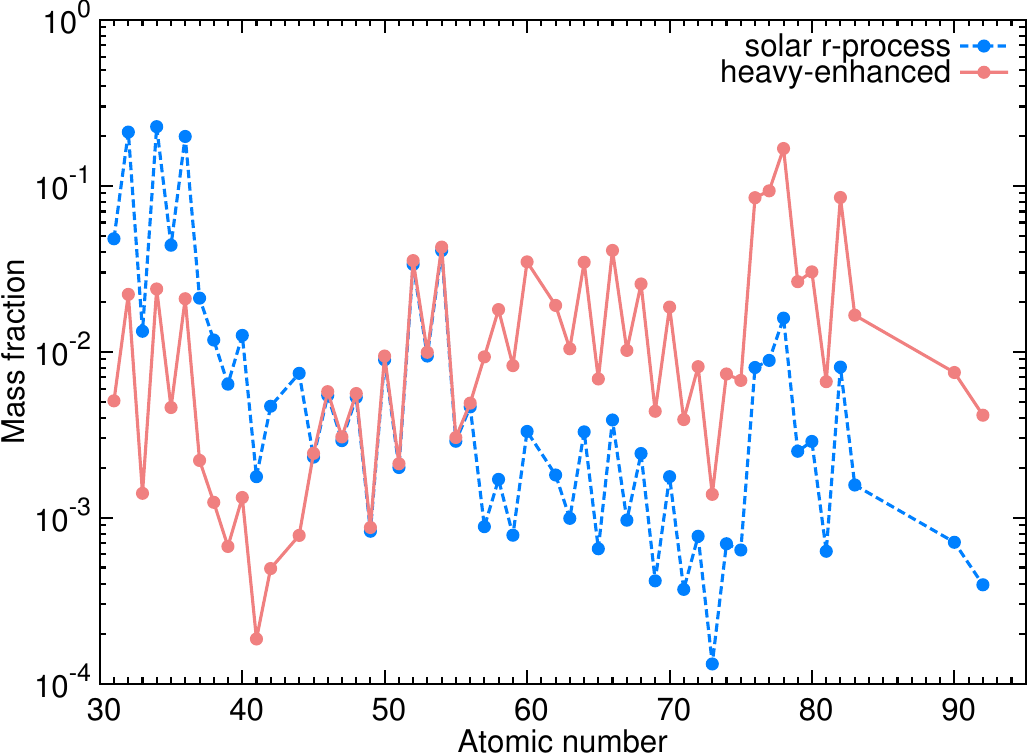}
    \caption{Abundance distributions used in this work. The blue line shows the solar \(r\)-process abundance pattern \citep{prantzos2020}, while the red line shows the {\it heavy-enhanced} pattern constructed by suppressing elements \(Z<45\) and enhancing elements \(Z>56\) each by a factor of ten, then renormalizing the mass fractions.
    \label{fig:abund_enhanced}}
\end{figure}

\begin{figure*}[t]
    \centering
    \begin{tabular}{cc}
    \includegraphics[width=0.45\linewidth]{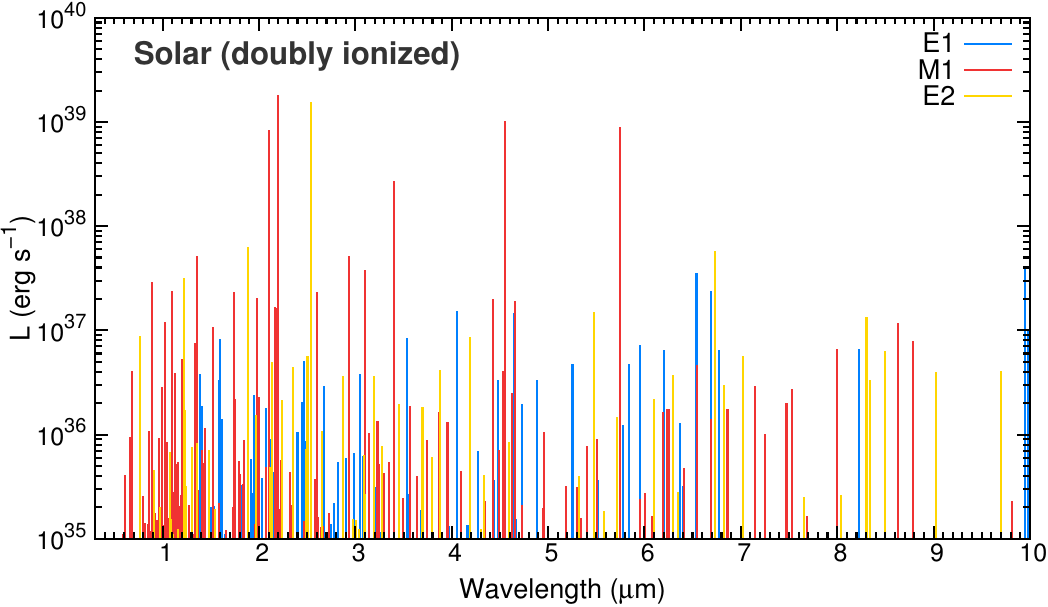} &
    \includegraphics[width=0.45\linewidth]{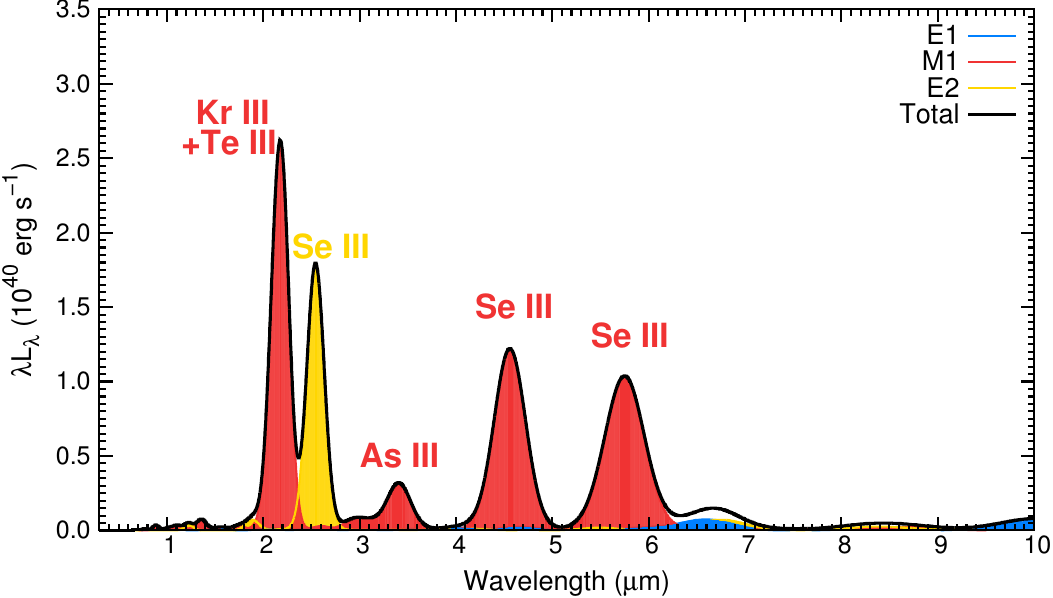} \\    
    \includegraphics[width=0.45\linewidth]{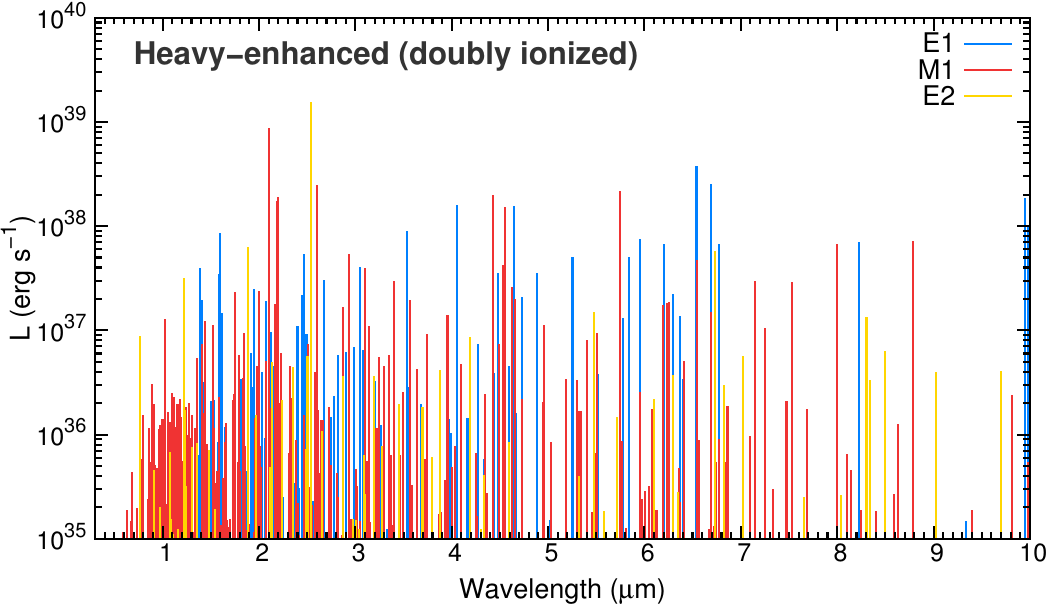} &
    \includegraphics[width=0.45\linewidth]{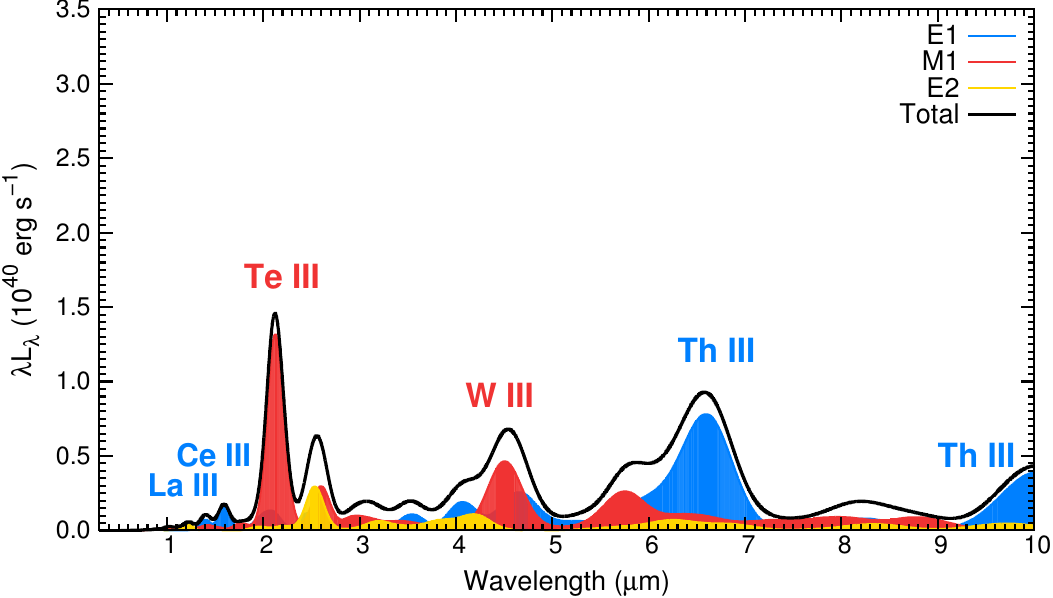} \\    
    \end{tabular}
\caption{
  \label{fig:spec_23_full}
  	{\it Left}: Calculated luminosity of individual transitions across the wavelength range \(\lambda = 0.3\,\mu{\rm m} - 10\,\mu{\rm m}\) assuming an electron temperature \(T_{\rm e}=2000\,{\rm K}\) and \(n_{\rm e} = 10^7\,{\rm cm}^{-3}\) as fiducial model. The contributions from E1, M1, and E2 transitions are shown in red, blue, and yellow, respectively. A doubly ionized composition is assumed, and the solar \(r\)-process and heavy-enhanced abundance patterns are assumed for the upper and lower panels, respectively. {\it Right}: The corresponding model spectra computed by assuming a Gaussian broadening of each line of \(v=0.07\,c\). The red, blue, and yellow regions show the contributions of E1, M1, and E2 transitions, and the total spectrum is shown in black. The strongest contributors for each produced feature are highlighted.
}
\end{figure*}

The model spectrum is obtained by summing the contributions from all selected E1, M1, and E2 transitions. As fiducial case, we adopt an electron temperature of \(T_{\rm e}=2000\,\text{K}\) and an electron density of \(n_{\rm e} =10^{7}\,\text{cm}^{-3}\) \citep{hotokezaka2023, gillanders2024}. Each transition is broadened using a Gaussian profile corresponding to an expansion velocity of \(v=0.07\,c\).

The abundance distributions adopted in our calculations are shown in Figure \ref{fig:abund_enhanced}. The blue line shows the solar \(r\)-process abundance distribution \citep{prantzos2020}, while the red line shows an abundance pattern with a suppressed first \(r\)-process peak and enhanced lanthanide fraction and third-peak abundances, which we refer to hereafter as the {\it heavy-enhanced} abundance distribution. The heavy-enhanced distribution was constructed by suppressing elements \(Z<45\) by a factor of ten, and enhancing those with \(Z>56\) by a factor of ten, then renormalizing the mass fractions.
These two distributions were selected as representative examples to explore how qualitatively different abundance patterns may affect the resulting spectra.

Previous non-LTE studies have shown that singly and doubly ionized species dominate at the epochs of interest \citep{hotokezaka2021, pognan2022steady, pognan2025, jerkstrand2025}. Nevertheless, to avoid introducing assumptions about the ionization state when performing line identifications, we compute separate spectra for neutral, singly ionized, doubly ionized, and triply ionized species. As an illustrative example, Figure \ref{fig:spec_23_full} shows the spectrum obtained for a doubly ionized composition. The left and right panels of the figure display the individual line contributions together with the total spectrum, where contributions from E1, M1, and E2 transitions are highlighted in blue, red, and yellow, respectively.

Figure \ref{fig:spec_23_full} shows that both allowed and forbidden transitions can produce prominent emission features throughout the near- and mid-infrared wavelength ranges. The upper panels show the spectra assuming the solar \(r\)-process abundance pattern, where we find that the strongest features originate from fine-structure transitions of first-peak \(r\)-process elements such as Kr, Se, and As.
In contrast, the spectra computed with the heavy-enhanced abundance distribution (lower panels) shows a few forbidden features, the strongest of which are produced by Te and W, both discussed in previous works \citep{hotokezaka2022, hotokezaka2023}. The model also predicts a few E1 features, most notably those of La III and Ce III in the near-infrared, together with stronger Th III features at mid-infrared wavelengths.

A notable characteristic of all our model spectra is the suppression of emissions at optical wavelengths (\(\lambda<1\,\mu\)m). This behavior arises from their low collisional excitation rates at the typical temperature range in the kilonova ejecta at late-phase. Transitions at shorter wavelengths generally involve large excitation energies, and their collisional rates are exponentially suppressed as shown in Equation \ref{eq:col_rate}. 
Our results therefore indicate that the strongest collisionally excited emission features in the late-phase kilonovae are expected to emerge mainly at infrared wavelengths.

Overall, Figure \ref{fig:spec_23_full} illustrates the potential of mid-infrared (\(\lambda>3\,\mu\)m) observations for constraining the abundance distribution of the kilonova ejecta. An example is the kilonova candidate associated with GRB230307A, which was observed with JWST out to \(\sim5\,\mu\)m at approximately 29 days after the GRB detection \citep{gillanders2023grb, levan2023}.
An emission feature consistent with both [Se III] \(4.555\,\mu\)m and [W III] \(4.432\,\mu\)m was reported (\citealt{levan2023, gillanders2024grb}, see also \citealt{mulholland2026ce}). However, the emergence of a continuum component, argued to be associated with dust \citep{domoto2026}, complicated the identification of individual spectral features. These observations suggest that infrared spectra at earlier epochs may provide stronger constraints on the abundance distribution in future events.

\section{AT2017gfo Spectral Analysis}\label{sec:cand}

\subsection{Line Identification}\label{sec:line_id}

\begin{figure*}[t]
    \centering
    \begin{tabular}{cc}
    \includegraphics[width=0.45\linewidth]{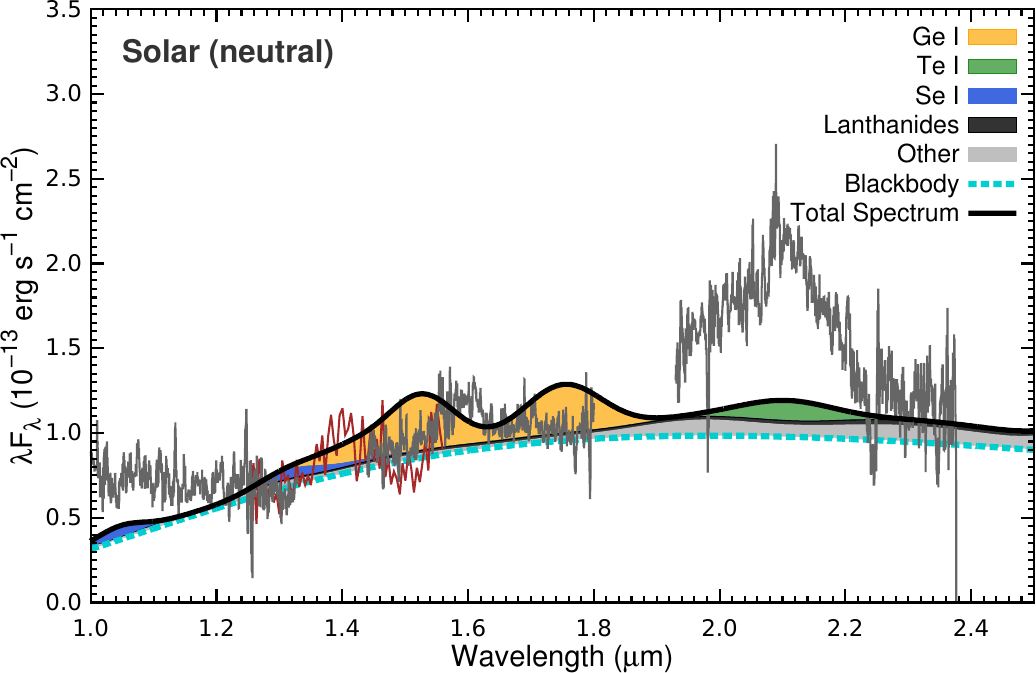} &
    \includegraphics[width=0.45\linewidth]{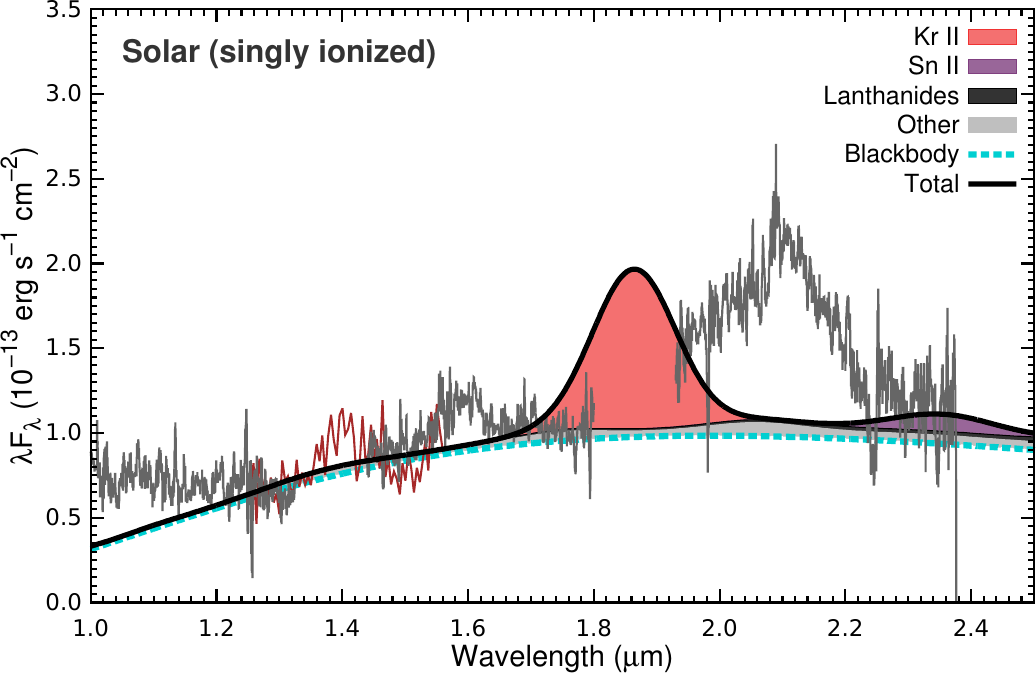} \\
     \includegraphics[width=0.45\linewidth]{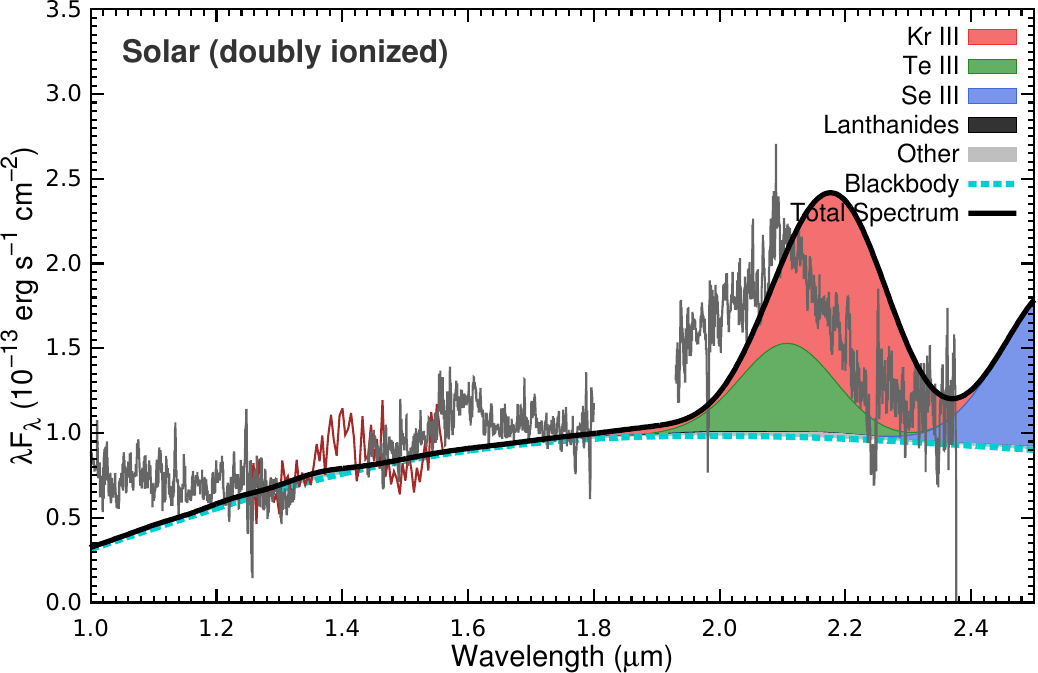} &
    \includegraphics[width=0.45\linewidth]{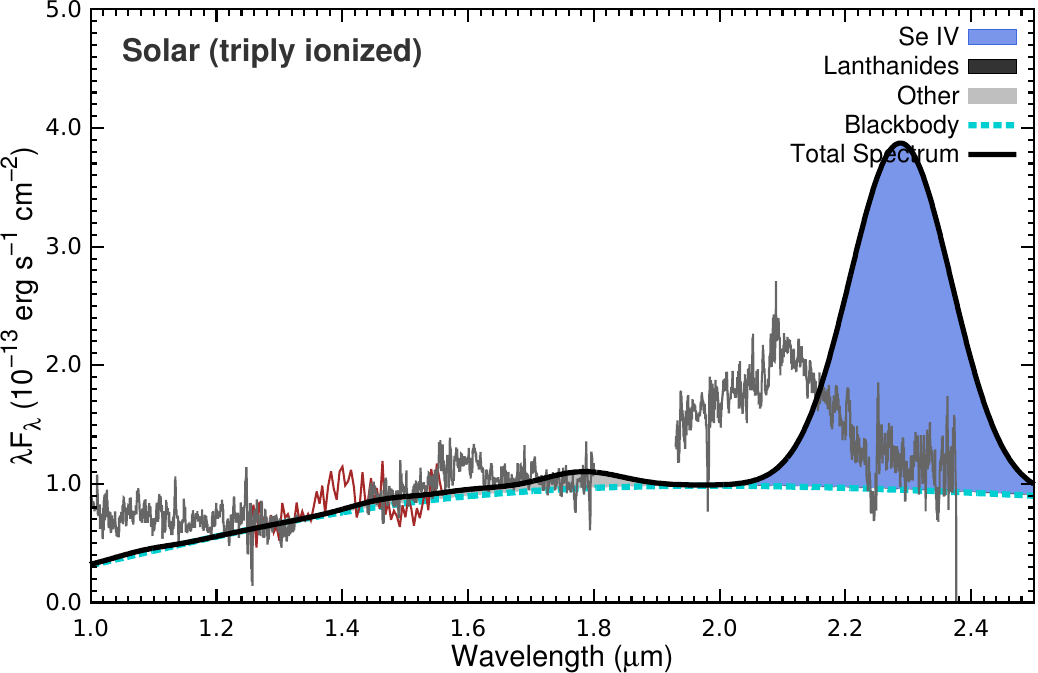} \\
    \end{tabular}
\caption{
  \label{fig:spec_solar}
	Comparison of our synthetic models calculated using the solar \(r\)-process abundance pattern to the 9.4 days spectrum of AT2017gfo observed with the VLT (black; \citealt{pian2017, smartt2017}) and the HST (brown; \citealt{tanvir2017}). The continuum is reproduced assuming a blackbody of \(T=1900\,\)K. Each panel shows the results assuming a different ionization pattern, and the contribution of individual elements with prominent transitions is highlighted in different colors. The total spectrum in each panel is shown with a black line. We note that the synthetic spectra shown here are for the sake of comparison and are not intended to reproduce the observations. 
}
\end{figure*}

\begin{figure*}[t]
    \centering
    \begin{tabular}{cc}
    \includegraphics[width=0.45\linewidth]{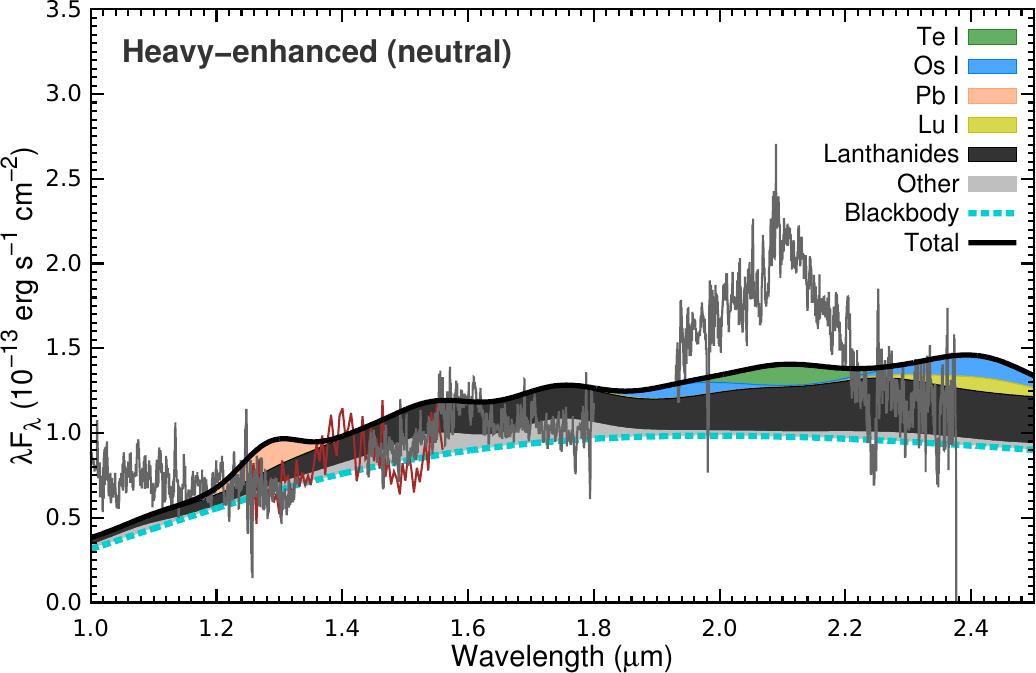} &
    \includegraphics[width=0.45\linewidth]{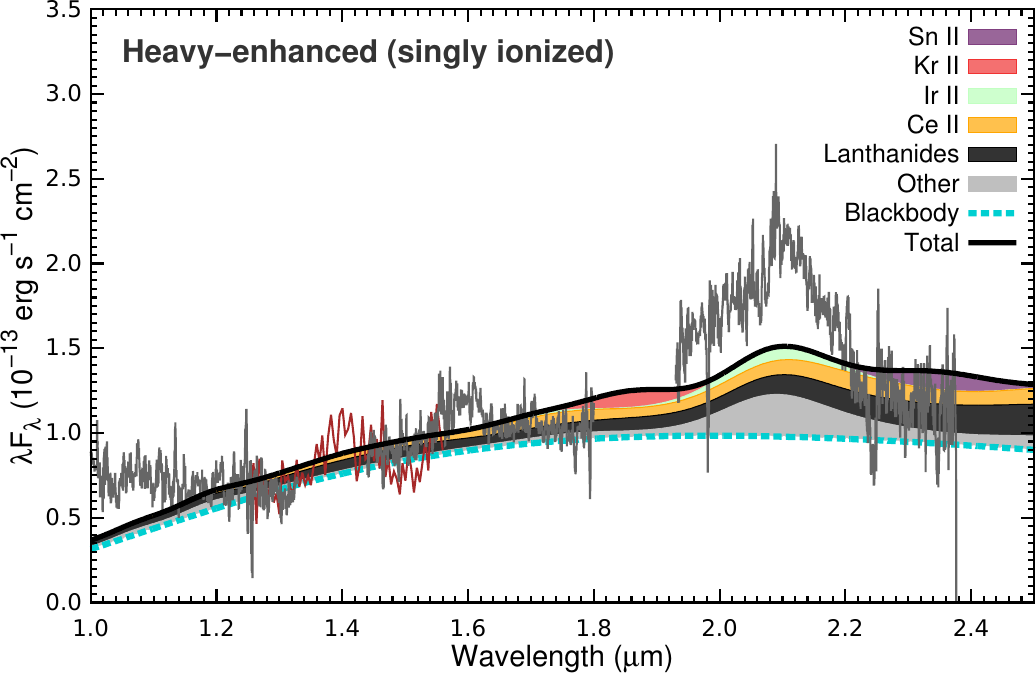} \\ 
     \includegraphics[width=0.45\linewidth]{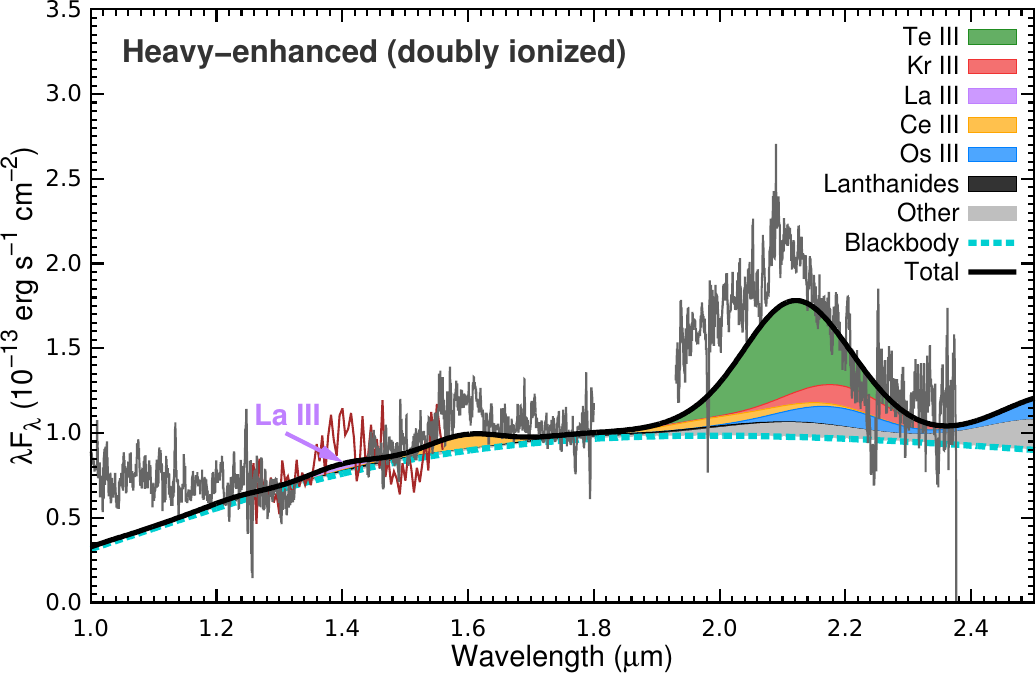} &
    \includegraphics[width=0.45\linewidth]{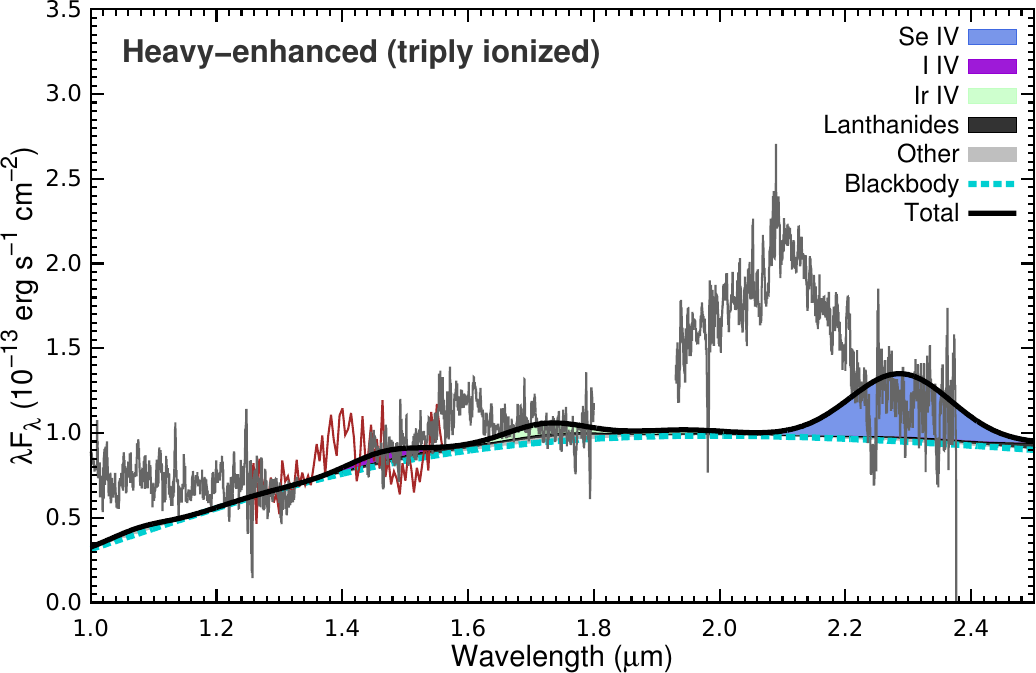} \\
    \end{tabular}
\caption{
  \label{fig:spec_heavy}
	Same as Figure \ref{fig:spec_solar} but assuming the heavy-enhanced abundance distribution.
}
\end{figure*}

The late-phase spectra of AT2017gfo shows three main emission features centered near \(1.4\,\mu\)m, \(1.6\,\mu\)m, and \(2.1\,\mu\)m. The \(1.6\,\mu\)m and \(2.1\,\mu\)m features appear consistently and evolve throughout the late-phase spectrum observed with the Very Large Telescope (VLT; \citealt{pian2017, smartt2017}), while the \(1.4\,\mu\)m feature can only be observed in the spectrum taken with the Hubble Space Telescope (HST; \citealt{tanvir2017}) due to the strong telluric absorption in the ground-based observations. We compare our model spectrum to the observed AT2017gfo to investigate whether we can reproduce the observed features. We use the observed spectrum at 9.4 days due to the available HST spectrum at the same epoch.

Figures \ref{fig:spec_solar} and \ref{fig:spec_heavy} show the synthetic spectra assuming the solar \(r\)-process abundance and the heavy-enhanced abundance patterns, respectively, with individual contributions of the most prominent transitions shown in colors. Each panel corresponds to a different ionization composition. The continuum is reproduced using a blackbody of temperature \(T=1900\,K\). We must note that a blackbody assumption may not hold at late-phase since the ejecta is expected to be at least partially optically thin at these epochs. In fact, the fit does not reproduce the observed continuum equally well at all wavelengths, and an emission excess can particularly be seen in the optical, as shown in Figures \ref{fig:spec_solar} and \ref{fig:spec_heavy}. As discussed in Section \ref{sec:model}, we do not expect emission lines from collisional excitation in the optical wavelength range at the adopted temperature range, suggesting that this excess is unlikely to originate from blended line emission. Nevertheless, we consider that the blackbody model provides a reasonable fit to the continuum in the infrared where the features of interest are located, and we leave the discussion of its origin to future work.

As discussed in Section \ref{sec:model_result}, the first \(r\)-process peak elements exhibit strong emission features when assuming the solar \(r\)-process abundance pattern. However, Figure \ref{fig:spec_solar} shows that none of the synthesized features provide a good match to the observed emission features of AT2017gfo. The results obtained with the heavy-enhanced abundance distribution in Figure \ref{fig:spec_heavy} show several features from lanthanides, along with contributions from species belonging to the first (Kr, Se), second (Te, Sn), and third (Ir, Os) \(r\)-process peaks. We find that our synthetic model assuming doubly ionized species gives a good match to the observed emission features, with the \(1.4\,\mu\)m, \(1.6\,\mu\)m, and \(2.1\,\mu\)m features attributed to La III, Ce III and Te III, respectively. The mass fractions of each ion required to fully reproduce the features, as well as the effect of blending with other elements, are discussed in Section \ref{sec:detection}.  We refer the reader to Appendix \ref{app:cand} for an in-depth discussion on other possible candidates for each feature and their reasons for exclusion.

The importance of La and Ce among lanthanides stems from their relatively simple valence-electron structures. As a result, these elements exhibit several strong electric dipole transitions that can be observed in near-infrared \citep{domoto2022}. Indeed, these two elements have been proposed as the main candidates for two absorption features observed at \(1.2-1.4\,{\mu}\)m during the photospheric phase of AT2017gfo \citep{domoto2022}. As the ejecta expands and becomes optically thin, the same transitions evolve from absorption to emission, making La III and Ce III strong candidates for the late-phase features identified here. This was also noted by \cite{gillanders2024}, who suggested La III and Ce III, among other candidates, as possible candidates for the \(1.4\,\mu\)m and \(1.6\,\mu\)m features.

The identification of Te III is motivated by a different mechanism. Te III possesses a strong fine-structure transition originating from its ground state, which was previously highlighted by \cite{hotokezaka2023}. In addition, Te belongs to the second \(r\)-process peak and is therefore expected to be synthesized in large quantities. The combination of a strong fine-structure transition and a potentially high abundance makes Te III a compelling explanation for the \(2.1\,\mu{\rm m}\) feature.

\subsection{Constraints on the Ionization Degree}\label{sec:ion}

Our search for candidate elements for each feature considers transitions from multiple ionization states (see also Appendix \ref{app:cand}). Nevertheless, the preferred identifications for all three observed features correspond to doubly ionized species. Previous works solving for the non-LTE evolution of kilonova ejecta generally predict that the relevant heavy elements are predominantly singly or doubly ionized at the epochs and velocities of interest \citep{pognan2022steady, pognan2025, jerkstrand2025}. Although a rigorous determination of the ionization composition requires a full non-LTE treatment with accurate atomic data, several observational arguments suggest that doubly ionized species dominate the formation of the observed emission features.

Ce II exhibits a dense forest of transitions extending from \(1.4\,\mu\)m up to and beyond \(2.4\,\mu\)m (upper right panel of Figure \ref{fig:spec_heavy}). If a large fraction of cerium was singly ionized, these Ce II lines would blend and obscure the isolated Ce III feature at \(1.6\,\mu\)m. The presence of a distinct and relatively narrow emission feature therefore suggests that Ce is predominantly doubly ionized during the late phase of AT2017gfo. Other lanthanides share similar ionization potentials as Ce, and they are all expected to occupy similar regions in the ejecta since they are synthesized under similar nucleosynthesis conditions. Therefore, other lanthanides, including La, most likely share the same dominant ionization state and are likewise mainly doubly ionized.

A similar argument can be given for Te III. Although we do not identify particularly strong transitions from other ionization states of Te, we find that Sn (\(Z=50\)), an element neighboring Te (\(Z=52\)) in the periodic table, has a strong feature at 2.3521\(\,\mu{\rm m}\), produced in our model (see upper right panel of Figure \ref{fig:spec_heavy}). Moreover, our HULLAC calculations predict a collision strength more than three times higher than our original assumptions (Table \ref{tab:hullac}), meaning that this transition would produce a prominent feature if a large fraction of Sn were singly ionized. Because both Sn and Te are part of the second \(r\)-process peak, they are expected to be produced in broadly similar nucleosynthesis conditions. The absence of [Sn II] 2.3521\(\,\mu{\rm m}\) in the observed spectra suggests that the amount of Sn\(^{+}\) is relatively small. This in turn favors a doubly ionized state of Te and other second \(r\)-process peak elements, since the two elements have comparable ionization potentials.

Motivated by these arguments, we hereafter adopt a doubly ionized composition of elements in our calculations. Nevertheless, we emphasize that robust constraints on the ionization degree of the ejecta ultimately require full non-LTE calculations that include the impact of all heavy elements.

\subsection{Abundance Estimations from Detected Lines}\label{sec:detection}
\subsubsection{La III \((Z=57)\)}

La III is our best candidate for the \(1.4\,\mu\)m emission feature as it has two strong transitions at \(1.3898\,\mu{\rm m}\) and \(1.4100\,\mu{\rm m}\). Both lines are E1 transitions with high transition probabilities, and critical densities satisfying \(n_{\rm cr}>n_{\rm e}\). Therefore, their total luminosities are computed in the low-density limit described by Equation \ref{eq:lum2}, yielding
{
\small
\begin{align*}
	L_{\text{La III}} = \Delta E_{1.38} & \,n_{\rm e} \frac{N_{\text{La III}}}{Z_{p;\text{La III}}} \frac{8.6\times10^{-4}}{\sqrt{T_{\rm e}}}\Upsilon_{1.38}\, e^{-E_{2;1.38}/k_BT_{\rm e}} \\
	+ \Delta E_{1.41} & \,n_{\rm e} \frac{N_{\text{La III}}}{Z_{p;\text{La III}}} \frac{8.6\times10^{-4}}{\sqrt{T_{\rm e}}}\Upsilon_{1.41}\,e^{-E_{2;1.41}/k_BT_{\rm e}},\label{eq:lum_la3}
\end{align*}
}
where the subscripts 1.38 and 1.41 refer to La III \(1.3898\,\mu{\rm m}\) and \(1.4100\,\mu{\rm m}\) respectively, and \(E_{2;1.38}\) and \(E_{2;1.41}\) indicate the upper levels of each of the two transitions. \(N_{\text{La III}}\) is the number of \({\rm La}^{2+}\) ions and \(Z_{p;\text{La III}}\) is the partition function. For the collision strengths \(\Upsilon_{1.38}\) and \(\Upsilon_{1.41}\), we used the results of our calculations using the atomic structure code HULLAC shown in Table \ref{tab:hullac}.

To reproduce the observed emission feature at each observed epoch, the electron density is parametrized as \(n_{\rm e}=10^7(t/9.5\,{\rm days})^{-3}\,{\rm cm}^{-3}\), following previous works \citep{hotokezaka2023}. The electron temperature, on the other hand, is assumed to be the best-fit blackbody temperature of the observed continuum, namely \(T=2400, 2100, 1900, \text{and } 1700\,{\rm K}\) for the spectra observed at 7.4, 8.4, 9.4, and 10.4 days, respectively.  

\begin{figure*}[t]
    \centering
    \begin{tabular}{ccc}
    \includegraphics[width=0.3\linewidth]{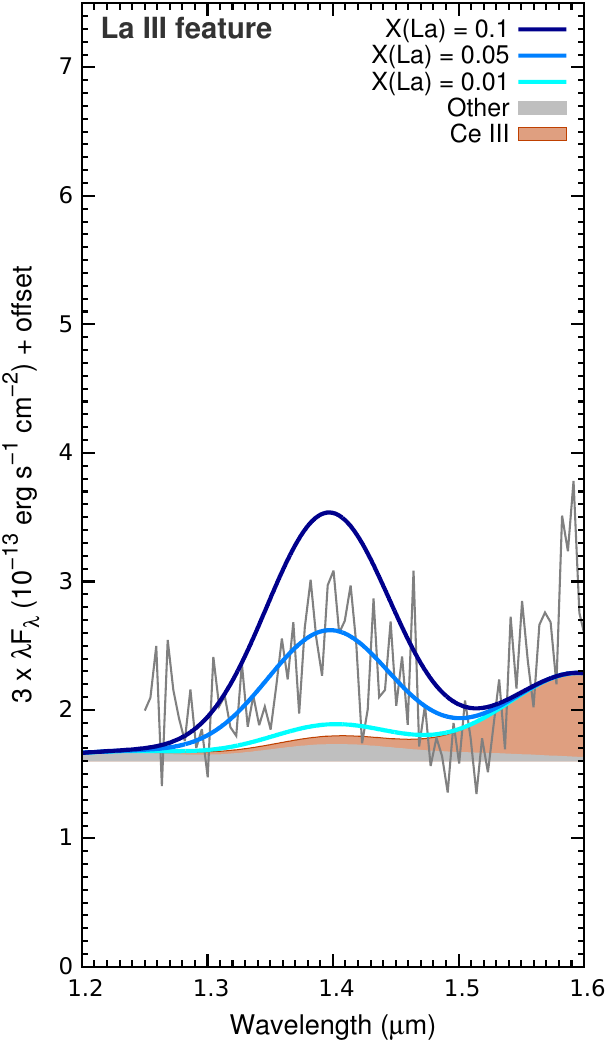} &
    \includegraphics[width=0.3\linewidth]{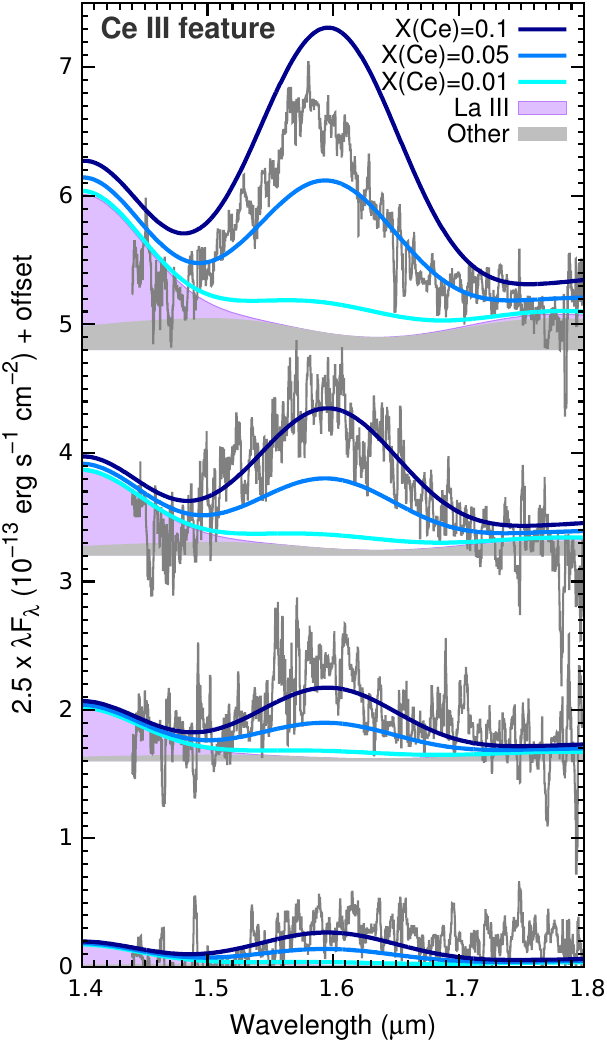} &
    \includegraphics[width=0.3\linewidth]{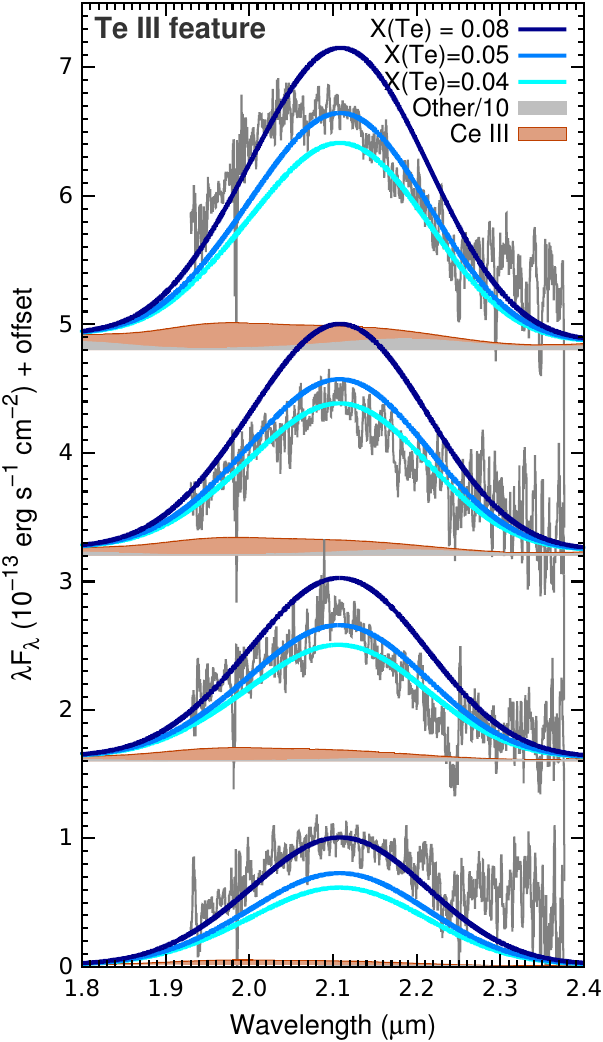} \\
    \end{tabular}
\caption{Comparison of our synthetic model to the observed features of AT2017gfo. The continua have been removed by assuming a blackbody of temperatures \(T=2400, 2100, 1900, \text{and } 1700\,{\rm K}\) for the observations at 7.4, 8.4, 9.4, and 10.4 days, respectively. The synthetic spectra are computed assuming electron temperatures equal to the blackbody temperature and an electron density of \(n_{\rm e}=10^7(t/9.5\,{\rm days})^{-3}\,{\rm cm}^{-3}\).
{\it Left}: Comparison with the \(1.4\,\mu\)m feature in the HST spectrum at 9.4 days after the merger. Models with different La mass fractions are shown in different shades of blue with line broadening of \(v=0.05\,c\). The contribution from Ce III is shown in orange, while the combined contribution from all other elements is shown in gray. The flux is boosted by a factor of 3 for better visualization.
{\it Center:} Comparison with the \(1.6\,\mu\)m feature in with the VLT spectra at 7.4, 8.4, 9.4, and 10.4 days after the merger (top to bottom). Models with different Ce mass fractions are shown in shades of blue with line broadening of \(v=0.05\,c\). The contribution from La III is shown in purple, and that from all other elements is shown in gray. The flux is boosted by a factor of 2.5 for better visualization
{\it Right:} Comparison with the \(2.1\,\mu\)m feature in the VLT spectra. Models with different Te mass fractions are shown in shades of blue with line broadening of \(v=0.07\,c\). The contribution from Ce III is highlighted in orange. The combined contribution of other elements is shown in gray and has been suppressed by a factor of ten (see text).
  \label{fig:abund_la_ce_te}
}
\end{figure*}

The left panel of Figure \ref{fig:abund_la_ce_te} compares the synthetic La III with the observations for different La mass fractions. We adopt a line broadening of \(v=0.05\,c\), which gives the best match to the observed line profile. Assuming that La is predominantly doubly ionized \(N_{\rm La}\), the mass fraction is computed as \(X_{\rm La} = N_{\rm La}A_{\rm La}m_{\rm u}/M_{\rm{ej}}\), where \(A_{\rm La}\) is La mass number, \(m_{\rm u}\) is the atomic mass unit, and the ejecta mass is taken as \(M_{\rm ej}=0.05\,M_\odot\) \citep{hotokezaka2020}.

The \(1.4\mu\)m feature is observed at a single epoch at 9.4 days after the merger with HST \citep{tanvir2017}. To isolate the emission feature, we subtract a continuum obtained from a blackbody fit with \(T=1900\,{\rm K}\).
Comparing the synthetic spectra with observations, we find that a mass fraction of \(X({\rm La})\approx X(\text{La III})\approx0.05\) provides the best match to the observed \(1.4\,\mu\)m feature. This value exceeds the lanthanum mass fraction expected from a solar \(r\)-process abundance distribution by more than two orders of magnitude (Figure \ref{fig:abund_enhanced}).

\subsubsection{Ce III \((Z=58)\)}

Ce III is our best candidate for the \(1.6\,\mu\)m feature as it has three lines at wavelengths \(1.5961\,\mu{\rm m}\), \(1.5851\,\mu{\rm m}\), and \(1.6133\,\mu{\rm m}\). All three transitions have critical densities satisfying \(n_{\rm cr}>n_{\rm e}\), and Ce III contribution can therefore be computed as
{
\small
\begin{align*}
	L_{\text{Ce III}} = \Delta E_{1.59} &  \,n_{\rm e} \frac{N_{\text{Ce III}}}{Z_{p;\text{Ce III}}} \frac{8.6\times10^{-4}}{\sqrt{T_{\rm e}}}\Upsilon_{1.59}\,e^{- E_{2;1.59}/k_BT_{\rm e}} \\
	+ \Delta E_{1.58} & \,n_{\rm e} \frac{N_{\text{Ce III}}}{Z_{p;\text{Ce III}}} \frac{8.6\times10^{-4}}{\sqrt{T_{\rm e}}}\Upsilon_{1.58}\,e^{-E_{2;1.58}/k_BT_{\rm e}} \\
	+ \Delta E_{1.61} & \,n_{\rm e} \frac{N_{\text{Ce III}}}{Z_{p;\text{Ce III}}}  \frac{8.6\times10^{-4}}{\sqrt{T_{\rm e}}}\Upsilon_{1.61}\,e^{-E_{2;1.61}/k_BT_{\rm e}},
\end{align*}
}
where the subscripts 1.59, 1.58, and 1.61 refer to Ce III \(1.5961\,\mu{\rm m}\), \(1.5851\,\mu{\rm m}\), and \(1.6133\,\mu{\rm m}\), respectively. The energy-average collision strengths are taken from our HULLAC calculations (Table \ref{tab:hullac}).

The middle panel of Figure \ref{fig:abund_la_ce_te} shows the comparison of Ce III contribution to the observed 1.6\(\,\mu\)m feature for different Ce mass fractions. The emission is isolated by subtracting a blackbody continuum of temperatures \(T=2400, 2100, 1900, \text{and } 1700\,{\rm K}\) for the observations at 7.4, 8.4, 9.4, and 10.4 days, respectively. Similarly to the \(1.4\,\mu\)m feature, a velocity broadening of \(v=0.05\,c\) provides the best match to the observed line profile. This suggests that elements responsible for both \(1.4\,\mu\)m and \(1.6\,\mu\)m features originate from ejecta regions with comparable velocity distributions, further supporting our identifications. 

We find that a cerium mass fraction in the range \(X({\rm Ce})\approx X(\text{Ce III})\approx0.05-0.1\) best reproduces the observed 1.6\(\,\mu\)m feature. The variation in the preferred value between epochs may reflect changes in the ionization degree, departures from the assumed electron density, or uncertainties in the continuum subtraction. Nevertheless, the temporal evolution of the feature is broadly consistent with the gradual weakening predicted by the model as the ejecta cools and expands. At 9.4 days, the observed feature is best reproduced with \(X({\rm Ce}) =  0.1\). Combined with the previously derived La abundance, we find a mass ratio between La and Ce of \(X({\rm La})/X({\rm Ce}) \approx 0.5\), consistent with the value expected from a solar \(r\)-process abundance pattern, \(\left(X({\rm La})/X({\rm Ce})\right)_\odot = 0.52\). We also find that Ce III feature is slightly affected by an emission line on its blue side, highlighted in gray in the middle panel of Figure \ref{fig:abund_la_ce_te}. This feature is attributed to Sb III, and is further discussed in Section \ref{sec:upper_limits}.

\subsubsection{Te III \((Z=52)\)}

We identify [Te III] \(2.1048\,\mu{\rm m}\) as the main contributor to the \(2.1\,\mu\)m feature, consistent with previous studies \citep{hotokezaka2023}. For our estimates, we adopt the atomic data computed for this line by \cite{mulholland2024te}. Their atomic structure calculations are performed with the R-matrix method, which captures resonance interactions in the cross sections. This gives a collision strength of \(\Upsilon_{\text{Te III}}\approx 3.5\) for \(T_e=2000\,{\rm K}\), a factor of a few higher compared to our HULLAC calculations (Table \ref{tab:hullac}).

The [Te III] \(2.1048\,\mu{\rm m}\) transition satisfies \(n_{\rm cr}<n_{\rm e}\) under the conditions relevant to AT2017gfo, its luminosity is therefore evaluated as
\begin{equation*}
	L_{\text{Te III}} = \Delta E_{2.10} N_{\text{Te III}} \frac{g_{2;2.10}}{Z_{p;\text{Te III}}} e^{-\Delta E_{2;2.10}/k_BT_{\rm e}}\beta A_{2.1}
\end{equation*}
where  \(g_{2;2.1}\) and \(E_{2;2.1}\) are the statistical weight and energy of the upper level of [Te III] \(\,2.1048\,\mu{\rm m}\). The Einstein coefficient \(A_{2.1}\) is also taken from \cite{mulholland2024te}.

The right-hand panel of Figure \ref{fig:abund_la_ce_te} compares the synthetic spectra with the observed 2.1\(\,\mu\)m feature for different tellurium abundances. We find that a velocity broadening of \(0.07\,c\) provides the best match to the observed line profile. The Te III feature in our model is blended with other lines, notably those of Kr III, Os III, and Ce III (see bottom left panel of Figure \ref{fig:spec_heavy}). [Kr III] \(2.1985\,\mu{\rm m}\) and [Os III] \(2.1841\,\mu{\rm m}\) are both redward of the observed feature. The effect of Kr III line is further discussed in Section \ref{sec:upper_limits}. The [Os III] \(2.1841\,\mu{\rm m}\) line is an inter-configuration M1 line that does not follow \(LS\)-coupling rules, and its contribution may be expected to be weaker than predicted by our fiducial model. Since we cannot provide updated atomic data for Os III due to its complex atomic structure (Appendix \ref{app:hullac}), we leave further discussion of possible constraints on Os abundance to future work.

Although weak blending with other elements may be expected, Te III remains the main contributor to the \(2.1\,\mu\)m feature, as shown in the bottom left panel of Figure \ref{fig:spec_heavy}. Therefore, to minimize the dependence of our Te abundance estimate on assumptions regarding the abundances and atomic data of neighboring species, we reduce their contribution relative to the fiducial heavy-enhanced abundance distribution. 
We highlight that the presence of Kr III and Os III in the spectra causes a spectral profile that is redshifted compared to observations, and reducing their abundance does not largely affect our estimates of \(X(\rm Te)\) (see Section \ref{sec:upper_limits} for a discussion on the effect of Kr III). In contrast, Ce III is retained in the calculation because its abundance is independently constrained by the \(1.6\,\mu\)m feature. 
Also, collision strengths for the Ce III lines from our HULLAC calculations are available, providing a more physically motivated estimate than the approximations adopted initially. The strongest Ce III transitions contributing in this wavelength range occur at wavelengths 1.9146, 1.9503, 1.9975, 2.0691, and 2.0987\(\,\mu{\rm m}\).

We find that Te mass fractions in the range \(X({\rm Te}) \approx X(\text{Te III}) \approx 0.04-0.08\) reproduce the observed feature across the available epochs. The variation in the required mass fraction may be due to changes in the ionization state or uncertainties in the subtracted continuum level. Our inferred abundance is somewhat larger than, but consistent with, the value obtained by \cite{hotokezaka2023}, who found \(X(\text{Te}^{2+})\sim 0.02\). This difference is due to the different atomic data adopted in the two studies. Notably, the observed spectra show that the temporal evolution of the luminosity of the \(2.1\,\mu\)m feature is much weaker than that of the \(1.6\,\mu\)m feature. This behavior is consistent with our model, as the Te III emission does not depend on the electron density of the ejecta as long as \(n_{cr}<n_e\) is satisfied, while Ce III emission declines rapidly since \(L_{\text{Ce III}}\propto n_e\).

Overall, the synthetic spectra provide a good match to the observed \(2.1\,\mu\)m feature at most epochs. The main exception occurs at 7.4 days, where the observed feature is blueshifted by approximately \(0.03\,\mu{\rm m}\) compared to later observations. Similar behavior was previously noted by \cite{gillanders2024}, who suggested that this change in the line shape may result from the blending of Te III emission line with an additional emission feature at \(\sim 2.06\,\mu\)m. Although Ce III in our model contributes several transitions blueward to the Te III feature, its abundance is already constrained by the \(1.6\,\mu\)m emission and does not appear sufficient to explain the observed evolution. More recently, \cite{chiba2026} proposed that He I \(\;2.059\,\mu\)m line emission produced by non-LTE level population may contribute to the observed feature. Investigating such effects requires a self-consistent treatment of non-LTE populations and is therefore beyond the scope of this work.

\subsection{Upper Limits from Non-detected Lines}\label{sec:upper_limits}

\subsubsection{Kr II/Kr III \((Z=36)\)}

\begin{figure*}[t]
    \centering
    \begin{tabular}{ccc}
    \includegraphics[width=0.32\linewidth]{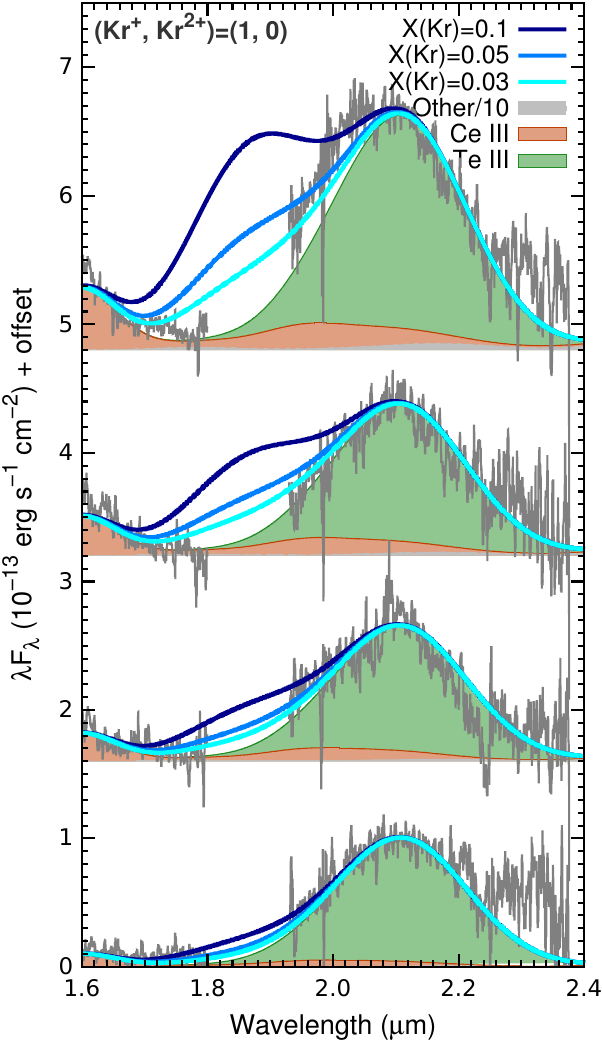} &
    \includegraphics[width=0.32\linewidth]{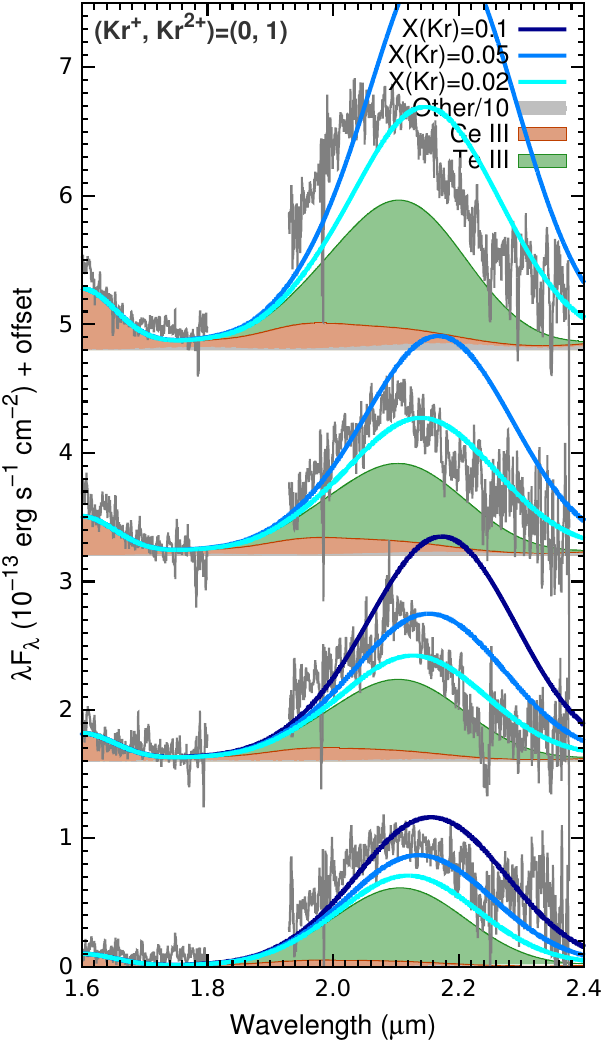} &
    \includegraphics[width=0.30\linewidth]{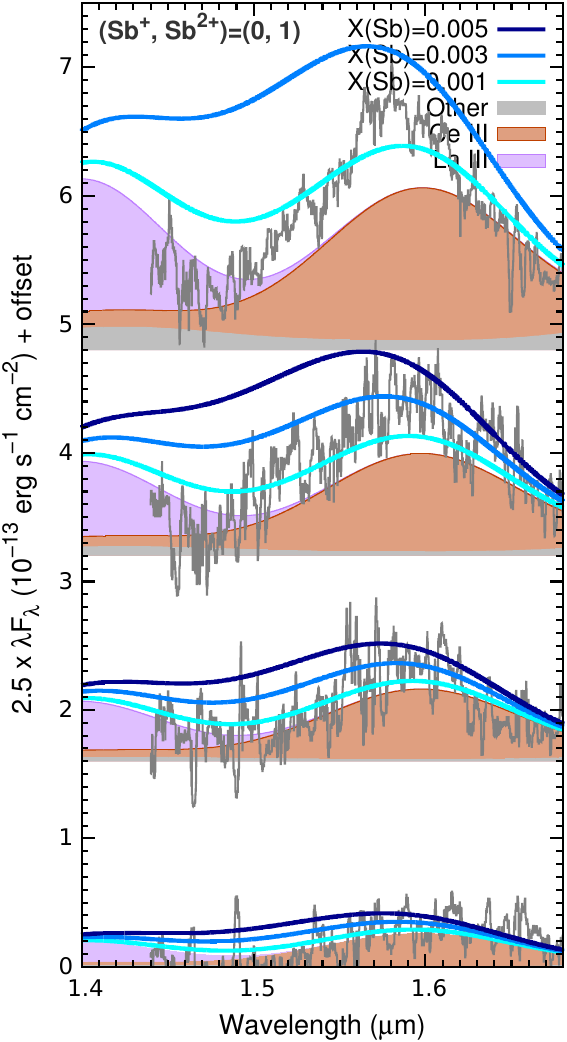} \\
    \end{tabular}
\caption{
  \label{fig:abund_kr_sb_os}
	Same as Figure \ref{fig:abund_la_ce_te} but showing the effect of varying the abundances of Kr and Sb.
	{\it Left}: Comparison of the synthetic spectra with the observed \(2.1\,\mu\)m feature of AT2017gfo, assuming different Kr mass fractions with \(({\rm Kr}^{+}, {\rm Kr}^{2+}) = (1,0)\) and line broadening of \(v=0.07\,c\). The contribution of Te is shown in green, while the contribution of other elements is suppressed by a factor of ten and is shown in gray.
	{\it Center}: Same as the left panel, but assuming \(({\rm Kr}^{+}, {\rm Kr}^{2+}) = (0,1)\). Here half of the best-fit Te abundance (\(X({\rm Te})/2\)) is assumed to investigate whether [Kr III] \(2.1986\,\mu{\rm m}\) could be considered an additional contributor to the \(2.1\,\mu\)m feature.
	{\it Right}: Comparison of the synthetic spectra with the observed \(1.6\,\mu\)m feature of AT2017gfo, assuming different Sb mass fractions with \(({\rm Sb}^{+}, {\rm Sb}^{2+}) = (0,1)\) and line broadening of \(v=0.07\,c\). The contribution from Ce is shown in orange, while the combined contribution of other elements is shown in gray.
}
\end{figure*}

Our model assuming the solar \(r\)-process abundance pattern predicts several prominent Kr emission lines that are not observed in the spectra of AT2017gfo (upper right and bottom left panels of Figure \ref{fig:spec_solar}). The absence of these features can therefore be used to place upper limits on the Kr abundance in the ejecta. The non-detection of Kr was also previously discussed by \cite{hotokezaka2023} and \cite{jerkstrand2025}.

Non-LTE calculations of \cite{jerkstrand2025} show that Kr is expected to be predominantly singly ionized at the velocity and timescale of interest. This behavior differs from that of other elements discussed in this work, mainly because the ionization potential of Kr\(^{+}\) is 24.3 eV, relatively high compared to 11.0 and 18.6 eV for Ce\(^{+}\) and Te\(^{+}\), respectively \citep{NIST_ASD}. Nevertheless, given the uncertainties in the ionization composition of the ejecta, we consider both singly and doubly ionized Kr when deriving abundance constraints.
The transition probabilities and collisional cross sections are taken from our HULLAC calculations (Table \ref{tab:hullac}), while we use the R-matrix calculation results for the collision strength of [Kr III] \(2.1986\,\mu{\rm m}\) (\(\Upsilon_{\text{Kr III}}\approx 3.8\); \citealt{schoning1997}).

The left panel of Figure \ref{fig:abund_kr_sb_os} shows the case of a purely singly ionized Kr (\(({\rm Kr}^{+}, {\rm Kr}^{2+}) = (1,0)\)). Although the peak of [Kr II] \(1.8622\,\mu{\rm m}\) falls within a region strongly affected by telluric absorption, its broad wings extend into neighboring wavelengths and become inconsistent with the observations for Kr mass fractions exceeding \(X({\rm Kr})\sim0.03\).

The middle panel of Figure \ref{fig:abund_kr_sb_os} shows the case of \(({\rm Kr}^{+}, {\rm Kr}^{2+}) = (0,1)\) .
Here we adopt half of the best-fit Te abundance, \(X({\rm Te})/2\), in order to investigate whether [Kr III] \(2.1986\,\mu{\rm m}\) could be considered an additional contributor to the observed \(2.1\,\mu{\rm m}\) feature. We find that increasing the Kr abundance produces a noticeable redward shift of the synthetic feature relative to the observations. This allows an upper limit of approximately \(X({\rm Kr})\lesssim0.02\).

Given the uncertainty in the Kr ionization state, we adopt the more conservative limit, \(X({\rm Kr})\lesssim0.03\), as our final abundance constraint.

\subsubsection{Sb III \((Z=51)\)}

Sb III possesses a strong fine-structure line with a wavelength of \(1.5207\,\mu{\rm m}\), located bluewards of the Ce III feature. Because no corresponding emission feature is observed in AT2017gfo, the non-detection of this line can be used to constrain the abundance of Sb.

To derive an upper limit, we compare our synthetic spectra with observations, as shown in the right panel of Figure \ref{fig:abund_kr_sb_os}, while varying the Sb abundance. Since Sb belongs to the second \(r\)-process peak and is expected to share a similar ionization degree with Te, we assume that Sb is predominantly doubly ionized. A velocity broadening of \(v=0.07\,c\) is adopted, and the Ce and La abundance is fixed to the best-fit value derived in Section \ref{sec:detection}.

Increasing the Sb abundance produces a broad emission feature that blends with Ce III and La III, shifting the synthetic spectrum away from the observed profile. We therefore find that the mass fraction of Sb should satisfy \(X({\rm Sb})\lesssim0.001\) at 7.4 days, and \(X({\rm Sb})\lesssim0.003\) at 8.4 days and later epochs. Using the Te abundance derived at 8.4 days in Section \ref{sec:detection}, \(X({\rm Sb})\lesssim 0.003\) gives \(X({\rm Sb})/X({\rm Te})\lesssim0.075\), consistent with the solar \(r\)-process mass ratio \(\left(X({\rm Sb})/X({\rm Te})\right)_\odot = 0.06\). It is noteworthy that our derived upper limit for Sb III at 7.4 days would give \(X({\rm Sb})/X({\rm Te})\lesssim0.025\) assuming \(X({\rm Te})=0.04\), which falls below the solar \(r\)-process ratio. However, as discussed in Section \ref{sec:detection}, the modeling of the \(2.1\,\mu\)m feature at 7.4 days is subject to significant uncertainties because the observed line profile is not well reproduced by our Te III model. We therefore regard the later-epoch constraint as more robust.

\begin{figure*}[t!]
    \centering
    \includegraphics[width=0.8\textwidth]{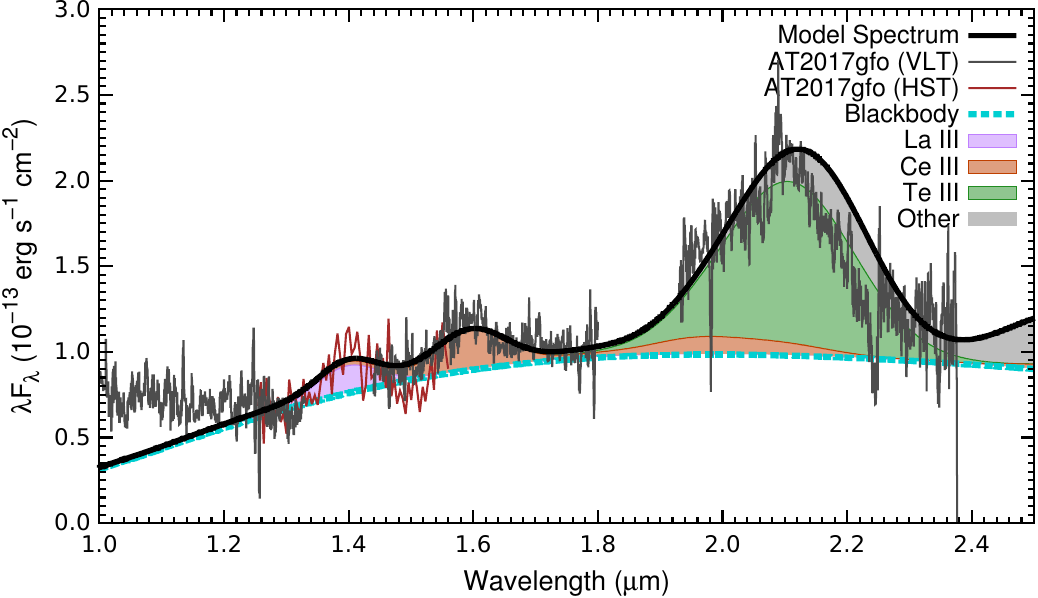}
    \caption{Comparison of our best-fit synthetic spectrum to the 9.4 days spectrum of AT2017gfo. The continuum is reproduced assuming a blackbody of \(T=1900\,\)K. The contributions of La III, Ce III, and Te III are computed with our best-fit mass fractions and are shown in purple, orange, and green, respectively. The combined contribution of all other elements with a doubly ionized state and heavy-enhanced abundance distribution is shown in gray.
    \label{fig:best_fit}}
\end{figure*}

Overall, the absence of detectable Sb III emission indicates that Sb is not significantly enhanced relative to Te and remains consistent with the abundance ratios expected for a solar-like second \(r\)-process peak, with a mass fraction upper limit of \(X({\rm Sb})\lesssim 0.003\).

\section{Discussion}\label{sec:disc}

\subsection{Overall Picture}\label{sec:overall}
Using our analytical spectral model, we were able to give robust identifications to all three emission features observed in the late-phase spectrum of AT2017gfo, where we find La III, Ce III, and Te III to be the main candidates for the \(1.4\,\mu\)m, \(1.6\,\mu\)m, and \(2.1\,\mu\)m features, respectively. The corresponding mass fractions are \(X({\rm Te})\approx 0.04-0.08\), \(X({\rm La})\approx 0.05\), and \(X({\rm Ce})\approx 0.05-0.1\). From the non-detection of Kr and Sb lines, we also infer mass fraction upper limits of \(X({\rm Kr})\lesssim 0.03\) and \(X({\rm Sb})\lesssim 0.003\). Figure \ref{fig:best_fit} shows the comparison of our model computed with the derived best-fit mass fractions, to the observed AT2017gfo spectrum. We use the observed spectra at 9.4 days due to the available HST spectrum at the same epoch.

\begin{figure*}[t!]
    \centering
    \includegraphics[width=0.75\textwidth]{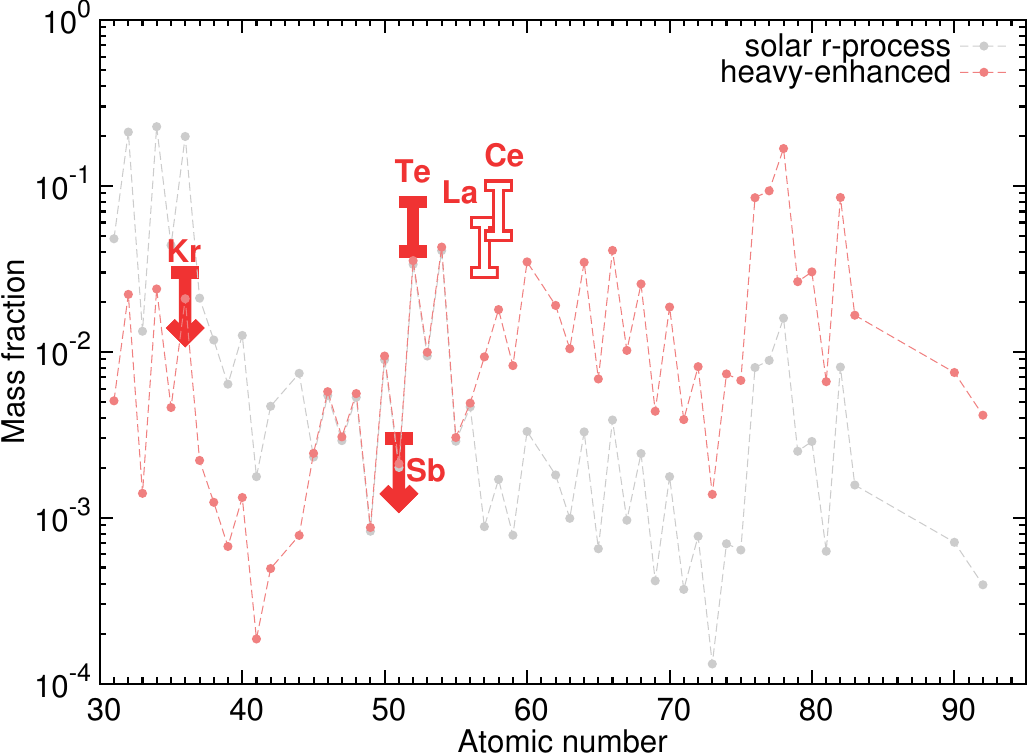}
    \caption{Summary of the abundance constraints inferred in this work. The constraints on La and Ce are tentative due to uncertainties of the radiation field and are shown as open symbols. The solar \(r\)-process and heavy-enhanced abundance patterns are shown in gray and red, respectively. We note that the error bars do not represent the uncertainties but rather the range of mass fractions inferred from our analysis (see Section \ref{sec:caveats}).
    \label{fig:abund_constraints}}
\end{figure*}

The mass fractions and upper limits derived in this work are summarized in Figure \ref{fig:abund_constraints}. 
Since the mass fraction of La is estimated by a single epoch (\(X({\rm La})\approx 0.05\)), we derive the range of \(X({\rm La})\) by assuming a constant mass ratio of La to Ce throughout all epochs. Therefore, based on the mass ratio \(X({\rm La})/X({\rm Ce})\approx 0.05\) inferred from the 9.4 days observations, we find an estimated mass fraction range of \(X({\rm La})\approx 0.025-0.05\).

Figure \ref{fig:abund_constraints} shows that our constraints favor an abundance pattern characterized by a suppressed first \(r\)-process peak and enhanced production of heavy \(r\)-process elements. The low Kr abundance provides direct evidence against a solar-like first peak. This was previously suggested by \cite{hotokezaka2023} and \cite{jerkstrand2025}. The detections of lanthanide species, on the other hand, require larger heavy-element abundances than expected from the solar pattern.

We highlight that the derived mass fractions of La and Ce may be uncertain, and we show them as open bars in Figure \ref{fig:abund_constraints}. Unlike Te III, which is observed through a forbidden transition, the La III and Ce III features arise from strong E1 transitions that may remain sensitive to the radiation field. If part of the late-time continuum originates from an optically thick inner region, radiative excitation may still affect these transitions. In that case, P-Cygni profiles may form even at late-phase.
The absence of obvious absorptions blueward of the emission peaks suggests that such effects are not dominant. However, this conclusion is difficult to establish because the relevant wavelength regions are affected by telluric absorption, and because the continuum level is uncertain. To illustrate the sensitivity of these transitions to optically thick conditions, we consider an outer region of density \(\rho=10^{-17}\,{\rm g\,cm}^{-3}\) and temperature \(T=2000\,\text{K}\) at \(t=10\,\)days. Using Equation \ref{eq:sobolev}, we find that the Sobolev optical depth of the strongest La III and Ce III lines reaches \(\tau\sim1\) for abundances as low as \(X=10^{-3}\), comparable to expectations from the solar \(r\)-process pattern. Nevertheless, such estimates are subject to uncertainties in the radiation field and the density of the possible line-forming region, and providing further abundance constraints is challenging. It is noteworthy, however, that the gaussian-like shape of the forbidden line [Te III] suggests that the inner part of the ejecta is optically thin enough to emit collisionally excited forbidden transitions. This in turn supports collisional excitation as the main origin for Ce III and La III features.

This is an interesting physical situation unique to lanthanides (and actinides) that stems from the low-lying energy levels of Ce III and La III. This causes their strong E1 lines to slowly shift through time from radiative excitation by the radiation field to collisional excitation by free electrons. Quantifying when this shift occurs requires full understanding of the continuum and ejecta properties of the merger. We leave such discussion of the continuum to future work, and we propose our La and Ce mass fraction estimates as tentative.

\subsection{Implications}

Our work finds that the inner ejecta of GW170817 had a nucleosynthesis pattern with a suppressed first \(r\)-process peak, and enhanced heavy-element production relative to the solar \(r\)-process abundance distribution.
Such an abundance pattern is consistent with expectations for the nucleosynthesis site responsible for heavy \(r\)-process element enrichments observed in \(r\)-enhanced metal-poor stars.
It has long been recognized that the abundance pattern of heavy elements (\(Z>56\)) in these stars closely follows the solar \(r\)-process pattern, suggesting a common nucleosynthesis origin, while lighter elements (\(31<Z<50\)) show a much larger star-to-star scatter, and do not show the same degree of universality (e.g., \citealt{sneden2003, roederer2018, roederer2022}). Our results therefore support a scenario in which NSMs are the dominant production site of the second \(r\)-process peak elements and beyond, while an additional astrophysical source may be required to explain the observed solar abundance of first-peak \(r\)-process elements.

A useful quantity for characterizing heavy-element production is the {\it lanthanide fraction}, defined as the total mass fraction of synthesized lanthanides. By analyzing the spectra of a sample of \(r\)-enhanced metal-poor stars, \cite{ji2019} measured the lanthanide fraction and found a typical \(X_{\rm LN}\approx3\times10^{-2}\). They argued that their estimates are higher compared to the lanthanide fraction in GW170817, inferred from kilonova light-curve modeling, which gives \(X_{\rm LN}=(2-20)\times10^{-3}\) (e.g., \citealt{chornock2017, kilpatrick2017, tanaka2017}).
However, \cite{kitamura2025} recently cautioned against using the red and blue components of kilonovae to infer ejecta parameters due to the reprocessing of light from different components. This would therefore introduce uncertainties in the lanthanide fraction derived from light curve modeling alone. \cite{gillanders2025} recently modeled the observed early-phase spectra of AT2017gfo using updated atomic data from \cite{flors2025}, and found that a lanthanide fraction of \(X_{{\rm LN}} \approx 2.5\times10^{-3}\) gives the best match to observations. Similarly, \cite{domoto2022} inferred cerium mass fractions of \(10^{-5}-10^{-3}\) from radiative-transfer simulations of the early photospheric spectra, from which they derived \(X_{{\rm LN}} \approx (2-200)\times10^{-4}\). However, due to the optically thick conditions at early-phase, such estimates are only representative of the outer ejecta layer.

Estimating the lanthanide fraction from our late-phase analysis is challenging because only a limited number of elements can currently be identified. Nevertheless, under the assumption that the relative abundances of heavy \(r\)-process elements follow the solar \(r\)-process abundance distribution, we extrapolate the abundance of lanthanides using the solar \(r\)-process pattern with our Te abundance estimate, where we find \(X_{\rm LN}=(3-6)\times10^{-2}\). We choose not to use the La and Ce abundances for this estimate because their inferred mass fractions may be affected by uncertainties associated with the radiation field, as discussed in Section \ref{sec:overall}. Interestingly, our inferred lanthanide fraction is consistent with the characteristic lanthanide fraction measured in \(r\)-enhanced metal-poor stars, \(X_{\rm LN}\approx3\times10^{-2}\) \citep{ji2019}. This agreement further supports the idea that NSMs are the primary source of the lanthanides observed in these stellar populations.

The apparent discrepancy between our inferred lanthanide fraction and previous estimates based on early-phase spectra \citep{domoto2022, gillanders2021, gillanders2025} may suggest different abundance patterns in different layers of the ejecta. During the photospheric phase, the ejecta is optically thick, and the observed absorption features mainly probe the outermost ejecta within our line of sight. In the case of GW170817, observations indicate that the event was viewed at an angle close to the polar region \citep{mooley2018, ghirlanda2019}. Numerical simulations suggest that the polar ejecta are characterized by relatively high electron fractions and consequently lower lanthanide abundances compared to the equatorial ejecta (e.g., \citealt{fujibayashi2023}). Abundance estimates derived from the photospheric phase may reflect the composition of the lanthanide-poor outer polar ejecta. In contrast, the late-time spectra analyzed in this work originate from the optically thin phase, where emission from a much larger fraction of the ejecta volume contributes to the observed spectrum. Consequently, our abundance estimates may better represent the overall nucleosynthesis yields of the event.

Finally, we revisit the question of whether NSMs can account for all of the \(r\)-process elements observed in the Milky Way. 
Assuming that the mean \(r\)-process in the Galaxy follows the solar \(r\)-process abundance pattern, with a total Galactic baryonic mass of \(\sim6\times10^{10}\,M_\odot\) \citep{mcmillan2011}, then the total mass of \(r\)-process elements in the Milky Way is \(M_{r,A>69}\sim 23000\,M_\odot\). Our main findings suggest that GW170817 mainly synthesized elements of the second \(r\)-process peak and beyond (\(A>115\); \(Z>49\)). Therefore, if the nucleosynthesis in GW170817 is representative of NSMs, then such events would be responsible for producing \(M_{r,A>115}\sim 4000\,M_\odot\) of heavy \(r\)-process elements over the history of the Milky Way. The rate of NSMs needed to explain this observed amount of Galactic heavy \(r\)-process can be estimated such \citep{hotokezaka2018, rosswog2018}

\begin{equation}
R \sim 8\,{\rm Myr}^{-1} \left(\frac{M_{r,A>115}}{4000\,M_\odot}\right) \left(\frac{m_r}{0.05M_\odot}\right)^{-1}\left(\frac{t_{\rm MW}}{10\,{\rm Gyr}}\right)^{-1},
\end{equation}
where \(m_r\) is the ejecta mass per merger event, and is taken as \(0.05\,M_\odot\) as was inferred for GW170817 \citep{hotokezaka2020}.

The inferred rate is consistent with estimates from gravitational-wave observations, including \(0.76-25\,{\rm Myr}^{-1}\) \citep{ligo2025} and the more conservative estimate of \(0.28-30\,{\rm Myr}^{-1}\) reported by \cite{fishbach2026}, inferred assuming a number density of Milky Way-like galaxies of \(\approx 0.01\,{\rm Mpc}^{-3}\). Our estimate assumes a constant star-formation history for the Milky Way, and may further decrease by a factor of a few if the star-formation burst expected during the early evolution of the Galaxy is taken into account \citep{fishbach2026}. Although uncertainties remain, our results indicate that, if GW170817 is representative of NSMs, then a low inferred rate of these events can still explain the Galactic inventory of heavy \(r\)-process elements.

\subsection{Caveats}\label{sec:caveats}

Our work provides robust identification of all observed late-phase features of AT2017gfo, from which we derive constraints on the ejecta abundance distribution of key elements. In this Section, we discuss the principal limitations of our analysis.

Our results depend strongly on the adopted atomic data. We use the HULLAC atomic structure code to provide the radiative and collisional rates necessary for our model. However, HULLAC calculations are restricted by the limited configuration space chosen to reduce the computational cost, as well as uncertainties regarding the choice of adequate configuration sets (Table \ref{tab:configs}). Therefore, the La III and Ce III collision strengths may be subject to uncertainties associated with the calculated inter-configuration energies and wavefunctions, which may affect the inferred Ce and La mass fraction. For fine-structure forbidden lines, the effective collision strength may instead be underestimated because our HULLAC calculations do not capture resonance contributions to the cross sections. To mitigate this effect, R-matrix results of previous works are used \citep{schoning1997, mulholland2024te} when available. Nevertheless, we emphasize that theoretical calculations are generally subject to systematic uncertainties when treating heavy, multi-electron elements, and experimental measurements remain necessary.

In addition to uncertainties in atomic data, our treatment of the emission features adopts a two-level approximation for each transition. In reality, the populations of energy levels are coupled through radiative and collisional processes involving several levels, and cascades from higher-lying states may contribute to the population of the emitting levels (e.g., \citealt{pognan2025}). The treatment of such effects is challenging, particularly for complex \(r\)-process elements. Further theoretical atomic calculations and laboratory measurements of atomic data, together with spectral models incorporating multi-level treatment of heavy elements, are necessary to improve the abundance estimates presented in this work.

Our abundance constraints are also affected by uncertainties in the physical conditions of the ejecta. In particular, we assume an electron density of \(n_{\rm e} = 10^7 (t/9,{\rm days})^{-3}\,{\rm cm}^{-3}\), following previous studies \citep{hotokezaka2023}. Lower electron densities result in lower luminosities, which suppress all emission features in the spectra. Higher electron densities, on the other hand, may arise if the ejecta are significantly clumped \citep{gillanders2024}. Increasing \(n_{\rm e}\) would strengthen the emission by La III and Ce III, while it would not impact the Te III line as its emission does not depend on \(n_{\rm e}\) as long as \(n_{\rm cr}<n_{\rm e}\) is satisfied.

The electron temperature and ionization composition adopted in our model are both constrained observationally. The temperature is derived from our blackbody fits, while the ionization state is constrained by the absence of specific emission features (see Sections \ref{sec:ion} and \ref{sec:detection}). However, the assumption that the electron temperature follows the blackbody temperature may not hold at late-phase. Similarly, our ionization constraints depend on the accuracy of the atomic data, since inaccurately modeled transitions could affect the interpretation of non-detections.

A self-consistent determination of the ejecta properties would require non-LTE calculations that solve for the temperature and ionization evolution of the ejecta while accounting for the contributions of all synthesized elements. Currently, such calculations are limited both by uncertainties in the ejecta's hydrodynamic evolution and by the lack of atomic data required for heavy elements, including recombination coefficients and collision strengths. Therefore, by constraining the model parameters as directly as possible from the observed AT2017gfo spectra, our simple analytic model establishes a benchmark for future studies on the interpretation of kilonova spectra.

\section{Summary}\label{sec:concl}

We have built an analytic spectral model that computes the luminosity emitted by transitions of heavy elements in an optically-thin regime, with the goal of interpreting late-phase kilonova spectra. The model includes selected allowed E1 and forbidden M1 and E2 transitions for all elements with available energy levels in the NIST \citep{NIST_ASD} and SCASA \citep{scasa} atomic databases. By individually computing the luminosity of each transition, our model predicts emission spectra extending to large wavelengths, with distinct features that depend sensitively on the assumed abundance distribution. Our results indicate that mid-infrared observations will be essential for placing robust constraints on the ejecta composition of future kilonovae.

We apply our model to the late-phase spectra of AT2017gfo, associated with the NSM event GW170817, which mainly exhibits three prominent emission features. We identify La III, Ce III, and Te III as the main contributors to the observed \(1.4\,\mu\)m, \(1.6\,\mu\)m, and \(2.1\,\mu\)m features, respectively. Using radiative and collisional rates newly computed with the atomic structure code HULLAC, and R-matrix results of previous works when available, we infer mass fractions of \(X({\rm Te})\approx 0.04-0.08\), \(X({\rm La})\approx 0.025-0.05\), and \(X({\rm Ce})\approx 0.05-0.1\). The constraints on La and Ce remain tentative, however, due to uncertainties in the origin of the radiation field. From the non-detection of Kr and Sb features, we further derive upper limits of \(X({\rm Kr})\lesssim 0.03\) and \(X({\rm Sb})\lesssim 0.003\). 
These abundance constraints suggest that the nucleosynthesis in the inner ejecta of GW170817 produced a suppressed first \(r\)-process peak and enhanced production of heavier elements relative to the solar \(r\)-process abundance pattern. This result is consistent with nucleosynthesis inferred from observations of metal-poor stars enhanced in \(r\)-process. Furthermore, under the assumption that the relative abundances of heavy \(r\)-process elements follow the solar pattern, our Te abundance implies a lanthanide fraction of \(X_{\rm LN}\approx (3-6)\times10^{-2}\), comparable to values inferred for \(r\)-enhanced metal-poor stars (\(X_{\rm LN}\approx3\times10^{-2}\); \citealt{ji2019}).

Our work provides robust identifications of all late-phase spectral features observed in AT2017gfo and new abundance constraints derived from those features. Further progress will require improved atomic data for heavy elements and fully self-consistent non-LTE calculations of the ejecta evolution. The results presented here provide a benchmark for such efforts and a framework for interpreting late-time spectra from future kilonova events.

\begin{acknowledgments}
We thank the anonymous referee for constructive feedback. S.R. acknowledges support from the Graduate Program on Physics for the Universe (GP-PU) at Tohoku University. This work is supported by the Grant-in-Aid for Scientific research from JSPS (grant Nos. 23H00127, 23H04894, 23H04891, 23H05432, 26K00732, 26K21726) and the JST FOREST Program (grant No. JPMJFR212Y).
\end{acknowledgments}


\appendix
\twocolumngrid

\section{Candidate Elements}\label{app:cand}

We discuss here the viable candidate species for each feature, all summarized in Table \ref{tab:candidates} with their relevant transitions and reason for exclusion. In this Section, we adopt the atomic data from our HULLAC atomic structure calculations when available (Appendix \ref{app:hullac}) to test for the relevance of selected elements. To avoid confusion, we refer hereafter to our analytic model that uses the original atomic data approximations introduced in Section \ref{sec:model_atom} as {\it approximated model}, while our model computed with the updated HULLAC atomic data is referred to as the {\it HULLAC model}.

\subsection{1.4 \(\mu\)m feature}

\begin{figure}[t]
    \centering
    \includegraphics[width=\columnwidth]{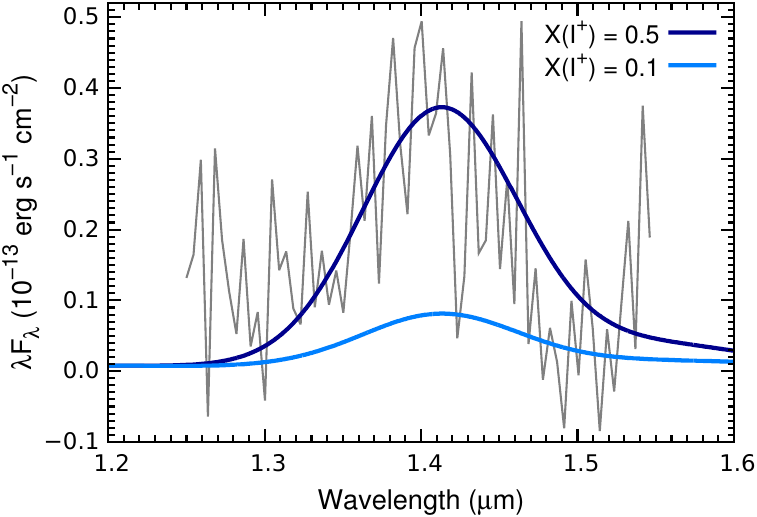} \\
    \includegraphics[width=\columnwidth]{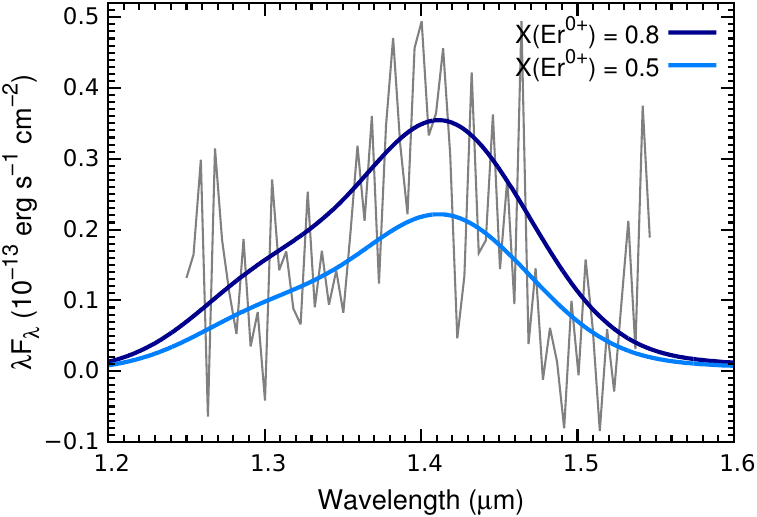}\\
    \includegraphics[width=\columnwidth]{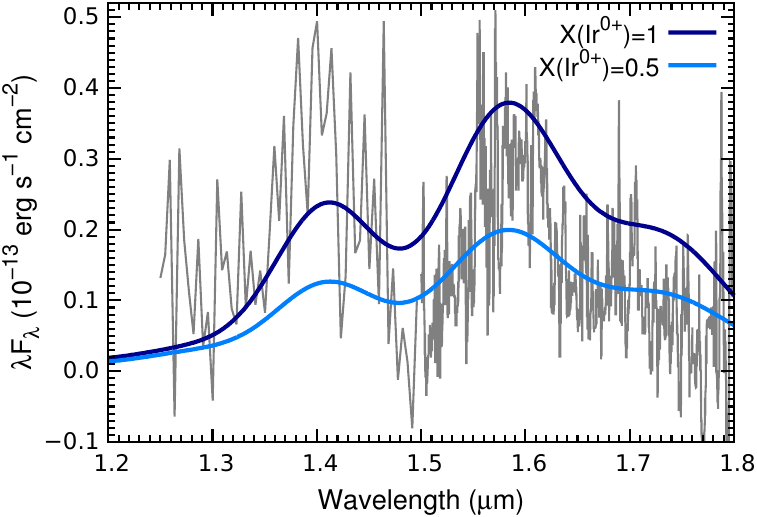} \\
\caption{
  \label{fig:abund_i_er_ir}
	{\it Top}: Comparison of the synthetic I II feature to the 1.4\(\,\mu\)m feature in the HST spectrum at 9.4 days post-merger, assuming different mass fractions of I\(^{+}\) and a broadening of \(v=0.05\,c\).
	{\it Middle}: same as upper but for Er I feature.
	{\it Bottom}: Comparison of the synthetic Ir I features to the 1.4\(\,\mu\)m and 1.6\(\,\mu\)m features in the HST and VLT spectra at 9.4 days, assuming different mass fractions of Ir\(^{+}\) and a broadening of \(v=0.05\,c\).
}
\end{figure}

{\bf Ge I (\(\mathbf{Z=32}\)):} Assuming the solar abundance distribution, we find that [Ge I] \(1.4034\,\mu{\rm m}\) may explain the \(1.4\,\mu\)m feature. However, with the collision strength and transition probabilities from HULLAC, we find that the approximated model overestimates the luminosity emitted by this element's transition by two orders of magnitude (see Table \ref{tab:hullac}). Furthermore, Figure \ref{fig:spec_solar} shows that Ge I has other stronger M1 transitions that do not match the observed features. This excludes Ge I as a candidate for the \(1.4\,\mu\)m feature.

{\bf I II (\(\mathbf{Z=53}\)):} Singly ionized iodine has one fine-structure M1 transition at \(1.4111\,\mu{\rm m}\), which is slightly shifted from the observed peak wavelength. I II atomic data computed with HULLAC is overall consistent with our original approximations, and does not exclude this element as a candidate. We therefore estimate the required iodine fraction necessary to reproduce the observation with the HULLAC model, and we show the result in the upper panel of Figure \ref{fig:abund_i_er_ir}. We find that an unphysical mass fraction of \(X(\text{I}^{+}) = 0.5\) is required to explain the observations. The strong Te III line may suggest that the fraction of singly ionized second \(r\)-process elements may be small, which would further increase the required iodine mass fraction. We conclude that [I II] \(1.4111\,\mu{\rm m}\) is most likely not the main candidate for the \(1.4\,\mu\)m feature.

{\bf La III (\(\mathbf{Z=57}\)):} This element is our best candidate for the observed 1.4 \(\mu\)m emission feature. It has two strong E1 transitions at \(1.3898\,\mu{\rm m}\) and \(1.4100\,\mu{\rm m}\), which average matches nicely with the emission line peak wavelength at \(1.4\,\mu{\rm m}\). These two lines were also discussed as important candidates for the early-phase absorption feature at \(\sim 1.2\,\mu{\rm m}\) in AT2017gfo \citep{domoto2022}.

{\bf Dy I (\(\mathbf{Z=66}\)): } Dysprosium has one E2 transition at \(1.4183\,\mu{\rm m}\). With HULLAC calculations, we find that this transition has an Einstein \(A\)-coefficient six orders of magnitude lower compared to our original approximations. Therefore, the approximated model largely overestimates the contribution of this transition, and Dy I is excluded as a candidate.

{\bf Er I (\(\mathbf{Z=68}\)):} This lanthanide element has one E1 transition at \(1.3934\,\mu{\rm m}\), matching the observed feature's peak wavelength. In the approximated model, this element's contribution was calculated using the Regemorter formula (Equation \ref{eq:regemorter}) with an Einstein coefficient of \(A=10^6\,{\rm s}^{-1}\). We could not compute the relevant atomic data of this transition as the upper level's \(LS\)-term was not available in the NIST database, which complicates the calibration of theoretical data. However, Er I has another E1 transition and fine-structure M1 transition from the ground level at wavelengths \(1.2992\,\mu{\rm m}\) and \(1.4371\,\mu{\rm m}\), respectively. A higher fraction of Er I would therefore cause an emission feature that does not match the observations, as shown in the middle panel of Figure \ref{fig:abund_i_er_ir}. Reproducing the observed flux would also require an unphysical mass fraction of \(X(\text{Er}^{0+})>0.5\) with our approximated model. We therefore exclude Er I as a candidate, but we encourage further efforts to compute the necessary atomic data to estimate its importance.

{\bf Ir I (\(\mathbf{Z=77}\)):} Iridium has one M1 transition at \(1.4071\,\mu{\rm m}\), matching the observations. Due to limitations in HULLAC atomic structure calculations in treating heavy elements, we could not produce better estimates for the collision strengths and radiative transition probabilities for this element. However, since this element has another transition  as a candidate for the \(1.6\,\mu\)m feature ([Ir I] \(1.5812\,\mu\)m; see below), we try either way to model the two features with our approximated model. The bottom panel of Figure \ref{fig:abund_i_er_ir} shows the comparison of our results to the 9.4 days observations, where we find that an unphysical \(X(\text{Ir}^{0+})=1\) is needed to explain the two features. Furthermore, Ir has other M1 and E2 transitions at \(1.7245\,\mu{\rm m}\) and \(1.7287\,\mu{\rm m}\), respectively, which do not match the observations as shown in Figure \ref{fig:abund_i_er_ir}. We conclude that Ir I is most likely not the main contributor for both \(1.4\,\mu\)m and \(1.6\,\mu\)m features, but more accurate atomic data for each of its transitions remain necessary.

\subsection{1.6 \(\mu\)m feature}

\begin{figure}[t]
    \centering
    \includegraphics[width=\columnwidth]{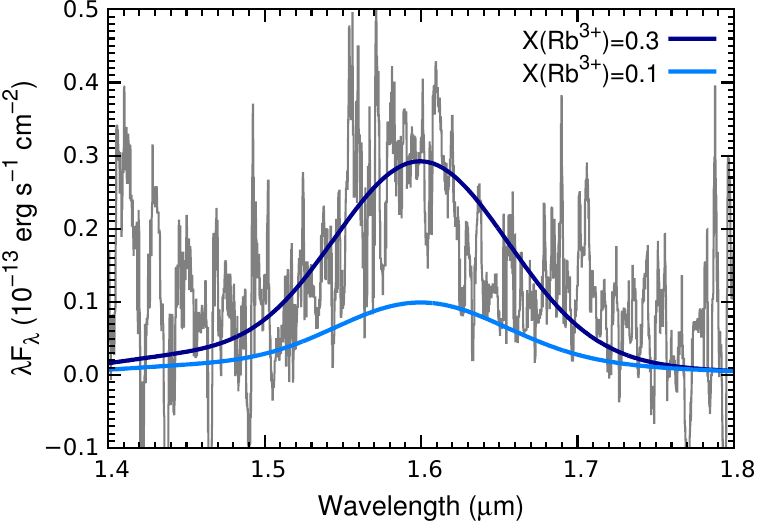} \\
\caption{
  \label{fig:abund_rb}
	Comparison of the Rb IV feature to the 1.6\(\,\mu\)m feature in the VLT spectrum at 9.4 days post-merger, assuming different mass fractions of Rb\(^{3+}\) and a line broadening of \(v=0.05\,c\).
}
\end{figure}

{\bf Rb IV (\(\mathbf{Z=37}\)):} Assuming the solar abundance distribution, [Rb IV] \(1.5973\,\mu{\rm m}\) may be considered a candidate for the \(1.6\,\mu\)m feature. Figure \ref{fig:abund_rb} shows the results of our HULLAC model to the \(1.6\,\mu\)m feature with Rb IV transition, where we find that a high mass fraction of \(X(\text{Rb}^{3+}) = 0.3\) is necessary to explain the observations. Considering our upper limits on the first \(r\)-process peak from the non-detection of Kr (\(Z=36\)), and the fact that we find no convincing candidate for the \(1.4\,\mu\)m feature assuming the solar abundance distribution pattern, we conclude that Rb is most likely not the main candidate for the feature.

{\bf Ce III (\(\mathbf{Z=58}\)):} This is our strongest candidate for the feature at \(1.6\,\mu\)m since it has multiple strong emission lines in the wavelength range of interest. It has three E1 lines at \(1.5961\,\mu{\rm m}\), \(1.5851\,\mu{\rm m}\), and \(1.6133\,\mu{\rm m}\) which average matches the emission line peak wavelength. These three lines were discussed previously as important candidates for the absorption feature at \(\sim 1.5\,\mu{\rm m}\) in the early-phase \citep{domoto2022}.

{\bf Ir I (\(\mathbf{Z=77}\)):} Iridium has one fine-structure line at \(1.5812\,\mu{\rm m}\), slightly blueshifted compared to the observed feature's peak. Although we could not provide more accurate atomic data for this element, we find that it is unlikely to be the main candidate for the observed features (see above and Figure \ref{fig:abund_i_er_ir}).

\subsection{2.1 \(\mu\)m feature}

{\bf Te III (\(\mathbf{Z=52}\)):} This element has a significantly strong M1 transition at \(2.1048\,\mu{\rm m}\), and is our best candidate for the observed feature. It has been previously discussed as the main contributor for the 2.1 \(\mu\)m feature \citep{hotokezaka2023}.

{\bf Te I (\(\mathbf{Z=52}\)):} Neutral tellurium has also a strong M1 transition at \(2.1049\,\mu{\rm m}\), similar wavelength to that of Te III line. As we do not solve for ionization, we cannot draw firm conclusions on which of the two lines is dominant. Previous works solving for non-LTE ionization structure find that the fraction of Te\(^{0+}\) is less than 0.1 at around 10 days post-merger \citep{pognan2022steady, jerkstrand2025}, meaning that the contribution of Te I line towards this feature may be ignored. We note, however, that the abundance inferred in our analysis for Te would remain roughly consistent even if this feature entirely originated from Te I, since both Te III and Te I lines have a similar wavelength and comparable collision strengths and transition probabilities (Table \ref{tab:hullac}).

\begin{figure}[t]
    \centering
    \includegraphics[width=\columnwidth]{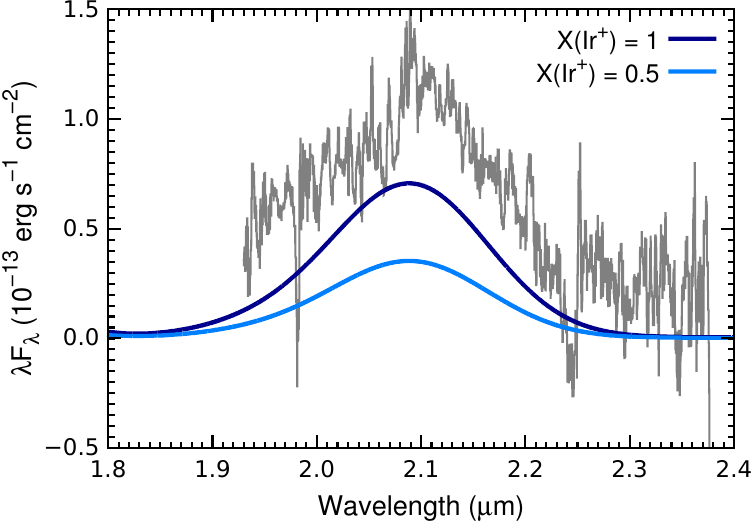} \\
\caption{
  \label{fig:abund_th_ir}
	Comparison of the Ir II feature to the 2.1\(\,\mu\)m feature in the VLT spectrum at 9.4 days post-merger, assuming different mass fractions of Ir\(^{+}\) and a line broadening of \(v=0.07\,c\).
}
\end{figure}

{\bf Ir II (\(\mathbf{Z=77}\)):} singly ionized Iridium has a fine-structure transition at \(2.0885\,\mu{\rm m}\). This element appears as a candidate for several observed features due to its high abundance as an element belonging to the third \(r\)-process peak. By assuming that the emission feature is entirely due to Ir II, our approximated model predicts that the required mass fraction is \(X(\text{Ir}^+)>1\) as shown in Figure \ref{fig:abund_th_ir}. Such predictions are unphysical, excluding Ir II as a candidate.

\begin{deluxetable*}{ccccccc}
\tablecaption{List of all candidates for each feature discussed in this work, with their reason for exclusion. Our best candidates for each feature are shown in bold.\label{tab:candidates}}
\tablewidth{0pt}
\tablehead{
\colhead{Feature} & \colhead{Abundance} & \colhead{Candidate} & \colhead{\(\lambda(\mu{\rm m})\)} & \colhead{Transition} & \colhead{Type} & \colhead{Reason for exclusion} 
}
\startdata
\(1.4\,\mu\)m & Solar & [Ge I] & 1.4034 & \(4s^24p^2\;(^1D_2) \rightarrow 4s^24p^2\;(^3P_0)\) & E2 & Too weak + non-detection of\\
& & & & & & other transitions \\
 	&      & [I II]\(^a\) & 1.4111 & \(5s^25p^4\;(^3P_1)\rightarrow5s^25p^4\;(^3P_2)\) & M1 & Too weak\\
 	& Heavy & {\bf La III} & {\bf1.3898} & \(\mathbf{4f\;(^2F^\circ_{5/2})\rightarrow 5d\;(^2D_{3/2})}\) & {\bf E1} & - \\
 	&	 & 	& {\bf 1.4100} &  \(\mathbf{4f\;(^2F^\circ_{7/2})\rightarrow5d\;(^2D_{5/2})}\) &  {\bf E1} &  - \\
	&	 & [Dy I] & 1.4183 & \(4f^{10}6s^2\;(^5I_{6})\rightarrow4f^{10}6s^2\;(^5I_{8})\) & E2 & Too weak\\
	&	 & Er I & 1.3934 & \(4f^{11}5d6s^2\rightarrow4f^{12}6s^2\;(^3H_{6})\) & E1 & Too weak + non-detection of\\
& & & & & & other transitions \\
 	&	 & [Ir I] & 1.4071 & \(5d^86s^1\;(^4F_{7/2})\rightarrow 5d^76s^2\;(^4F_{9/2})\) & M1 & Too weak + non-detection of\\
& & & & & &  other transitions \\
 \hline
\(1.6\,\mu\)m & Solar  & [Rb IV] & 1.5973 & \(4s^24p^4\;(^3P_1)\rightarrow 4s^24p^4\;(^3P_2)\) & M1 & Too weak \\
		    & Heavy & {\bf Ce III} & {\bf 1.5961} &  \(\mathbf{4f5d\;(^3G^\circ_3)\rightarrow4f^2\;(^3H_4)}\) & {\bf E1} &  - \\
 		    & 		  & 	      & {\bf 1.5852} &  \(\mathbf{4f5d\;(^3G^\circ_4)\rightarrow4f^2\;(^3H_5)}\) & {\bf E1} & -   \\
 		    & 		  &	      & {\bf 1.6133} & \(\mathbf{4f5d\;(^3G^\circ_5)\rightarrow4f^2\;(^3H_6)}\) & {\bf E1} & - \\
 		    & 		  & [Ir I] & 1.5813 &  \(5d^76s^2\;(^4F_{7/2})\rightarrow 5d^76s^2\;(^4F_{9/2})\) & M1 &  Too weak + non-detection of \\
& & & & & & other transitions \\
 \hline
 \(2.1\,\mu\)m & Solar & {\bf [Te III]\(^a\)} & {\bf2.1048} &  \(\mathbf{5s^25p^2\;(^3P_1)\rightarrow5s^25p^2\;(^3P_0)}\) & {\bf M1} & - \\
	& 	& [Te I]\(^a\) & 2.1049& \(5s^25p^4\;(^3P_1)\rightarrow5s^25p^4\;(^3P_2)\) & M1& Different ionization \\
	& Heavy & [Ir II] & 2.0885 & \(5d^76s\;(^5F_4)\rightarrow 5d^76s\;(^5F_{5})\) & M1 & Too weak \\
\enddata
\tablecomments{\(^a\) [I II], [Te III], and [Te I] appear as candidates for both the solar \(r\)-process and the heavy-enhanced abundance distributions, but we only show them here for solar.}
\end{deluxetable*}

\section{HULLAC calculations}\label{app:hullac}

\begin{deluxetable*}{ccl}
\tablecaption{Configurations used to compute the atomic data for featured species \label{tab:configs}}
\tablewidth{0pt}
\tablehead{
\colhead{Ion} & \colhead{\(N_{\text{level}}^a\)} & \colhead{Configurations} 
}
\startdata
Ge I & 156 & \(4s^24p^2,\;4s^2 4p^1 5s^1,\; 4s^2 4p^1 5p^1,\; 4s^2 4p^1 4d^1,\; 4s^2 4p^1 6s^1,\; 4s^2 4p^1 6p^1,\;  4s^2 4p^1 5d^1,\) \\
        & 	& \(4s^2 4p^1 4f^1,\;  4s^2 4p^1 7s^1,\; 4s^2 4p^1 7p^1,\, 4s^2 4p^1 8s^1,\; 4s^2 4p^1 6d^1,\; 4p^4,\; 4s^1 4p^2 5s^1,\)\\
        & 	& \(4s^1 4d^2 5s^1,\, 4s^1 4p^1 5s^2\)\\
Kr II &  354  & \(4s^2 4p^5,\; 4s^1 4p^6,\;  4s^2 4p^4 5s^1,\; 4s^2 4p^4 4d^1,\; 4s^2 4p^4 5p^1,\; 4s^2 4p^4 5d^1,\; 4s^2 4p^4 6s^1,\)\\
	&	& \(4s^2 4p^4 6p^1,\; 4s^2 4p^4 4f^1,\; 4s^2 4p^4 7s^1,\; 4s^2 4p^4 6d^1,\; 4s^2 4p^4 5f^1,\; 4s^2 4p^4 8s^1,\; 4s^2 4p^4 6f^1,\)\\
	& 	& \(4s^2 4p^4 7f^1,\; 4s^2 4p^4 8f^1,\; 4s^2 4p^3 5s^2,\; 4s^1 4p^5 5s^1,\; 4s^1 4p^5 4d^1,\; 4s^1 4p^4 5s^2\)\\
Kr III & 1669 & \(4s^2 4p^4,\; 4s^1 4p^5,\; 4s^2 4p^3 4d^1,\; 4s^2 4p^3 5s^1,\; 4s^2 4p^3 5p^1,\; 4s^2 4p^3 6s^1,\; 4s^2 4p^3 6d^1,\) \\
        &	& \(4p^6,\; 4s^2 4p^1 4d^3,\; 4s^2 4d^4,\; 4s^1 4p^3 4d^2,\; 4s^1 4p^2 4d^3,\; 4s^1 4p^1 4d^4,\; 4p^4 4d^2,\) \\
	&	&  \(4s^2 4p^2 4d^2,\; 4s^2 4p^2 5s^2\)\\
Rb IV & 406 & \(4s^2 4p^4,\;4s^1 4p^5,\;4s^2 4p^3 4d^1,\; 4s^2 4p^3 5s^1,\;4s^2 4p^3 5p^1,\; 4s^2 4p^3 5d^1,\; 4s^2 4p^3 6s^1\)\\
Sn II & 61 & \(5s^2 5p^1,\;5s^1 5p^2,\;5s^2 6s^1,\;5s^2 5d^1,\;5s^2 6p^1,\;5s^2 7s^1,\;5s^2 4f^1,\;5s^2 6d^1,\; 5s^2 7p^1,\;5s^2 8s^1,\;\)\\
	&	& \(5s^2 5f^1,\;5s^2 7d^1,\;5s^2 8p^1,\;5s^2 9p^1,\;5s^1 5p^1 6s^1,\;5s^1 5p^1 5d^1\)\\
Sb III & 88 & \(5s^2 5p^1,\; 5s^1 5p^2,\; 5s^2 6s^1,\; 5s^2 5d^1,\; 5s^2 6p^1,\; 5s^2 7s^1,\; 5s^2 6d^1,\;5s^2 4f^1,\; 5s^2 8s^1,\; 5p^3,\)\\
	&	& \(5s^1 5p^1 6s^1,\; 5s^1 5p^1 5d^1,\; 5s^1 5p^1 7s^1,\; 5s^1 6s^2,\;5s^1 6d^2,\; 5p^2 6s^1\)\\
Te I & 602 & \(5p^4,\; 5p^3 6s^1,\; 5p^2 6s^2,\; 5p^3 6p^1,\; 5p^3 5d^1,\; 5p^3 7s^1,\; 5p^3 7p^1,\;5p^3 6d^1,\; 5p^3 8s^1,\; 5p^3 8p^1,\)\\
	&	& \(5p^3 5f^1,\; 5p^2 5d^2,\; 5p^1 5d^3,\; 5p^2 6s^1 6p^1,\;5p^1 6s^2 6p^1,\; 5p^2 6s^1 5d^1,\; 5p^2 6s^1 7s^1,\; 5p^1 6s^2 5d^1,\)\\
	&	& \(6s^2 6p^2\)\\
Te III & 273 & \(5s^2 5p^2,\; 5s^1 5p^3,\; 5s^2 5p^1 5d^1,\; 5s^2 5p^1 6s^1,\; 5s^2 5p^1 7s^1,\; 5s^2 5p^1 6d^1,\; 5s^2 5p^1 4f^1,\)\\
	&	& \(5s^2 5p^1 6p^1,\;  5p^4,\; 5s^1 5p^2 5d^1,\; 5s^1 5p^2 6s^1,\; 5s^1 5p^1 6s^2,\; 5s^1 5p^2 7s^1,\; 5s^2 5d^2,\)\\
	&	& \(5d^4,\; 5s^2 6s^2,\; 5s^1 5p^2 5d^1,\; 5p^3 5d^1,\; 5p^3 6s^1,\; 5p^3 7s^1,\; 5p^2 6s^2\)\\
I II & 406 &  \(5s^2 5p^4,\; 5s^2 5p^3 6s^1,\; 5s^1 5p^5,\; 5s^2 5p^3 5d^1,\; 5s^2 5p^3 6p^1,\; 5s^2 5p^3 7s^1,\; 5s^2 5p^3 6d^1,\)\\
     &		& \(5s^2 5p^3 4f^1,\; 5s^2 5p^3 7p^1,\; 5s^2 5p^3 8s^1,\; 5s^2 5p^3 6d^1,\; 5s^2 5p^3 7d^1,\; 5s^2 5p^3 5f^1,\; 5s^2 5p^3 8s^1,\)\\
     &		& \(5p^6,\; 5s^2 5p^2 6s^2,\; 5s^2 5p^2 5d^2\)\\
La III & 137 & \(5p^6 5d^1,\; 5p^6 4f^1,\; 5p^6 6s^1,\; 5p^6 6p^1,\; 5p^6 7p^1,\; 5p^6 6p^1,\; 5p^6 7s^1,\;5p^6 6d^1,\; 5p^6 5f^1,\)\\
	&	& \(5p^6 8s^1,\; 5p^6 7d^1,\; 5p^6 5g^1,\; 5p^6 6f^1,\; 5p^6 8p^1,\;5p^6 9s^1,\; 5p^6 8d^1,\; 5p^5 5d^2,\; 5p^5 5d^1 6s^1,\)\\
	&	&  \(5p^5 6s^2,\; 5p^5 6s^1 6p^1,\; 5p^5 5d^1 7s^1\)\\
Ce III & 230 & \(4f^2,\; 4f^1 5d^1,\; 4f^1 6s^1,\; 5d^2,\; 4f^1 6p^1,\; 5d^1 6s^1,\;4f^1 6d^1,\;4f^1 7s^1,\; 5d^1 6p^1,\; 4f^1 5f^1\)\\
	&	& \(4f^1 7p^1,\; 4f^1 8s^1,\;4f^1 6f^1,\; 4f^1 5g^1,\;6p^2,\; 5d^1 6d^1,\; 6s^2,\; 6s^1 6p^1,\; 5d^1 7s^1,\)\\
	&	&  \(6s^1 6d^1\)\\
Dy I & 5161 & \(4f^{10} 6s^2,\; 4f^9 5d^1 6s^2,\; 4f^{10} 6s^1 6p^1,\; 4f^{10} 5d^1 6s^1\)\\
\enddata
\tablecomments{\(^a\)  Number of levels computed with HULLAC for each ion}
\end{deluxetable*}

We summarize here the atomic data used in our abundance estimates, computed with the atomic structure code HULLAC \citep{hullac}.  HULLAC is an integrated code for calculating the energy levels, cross sections, and radiative transition probabilities of atoms. Orbital functions are derived by solving the single-electron Dirac equation with a central-field potential that includes both a nuclear field and a spherically-averaged potential due to electron-electron interactions. Using these orbital functions, N-electron configuration state functions are then built as antisymmetric combinations of the orbitals. Relativistic configuration interaction (RCI) calculations are carried out based on these configuration state functions, where the accuracy of the RCI calculations is improved by increasing the number of configurations. This in turn increases the computational cost, especially for complex elements such as lanthanides. In this study, the number of configurations is chosen so that a good balance between accuracy and computational time can be achieved. The accuracy is further improved by optimizing the effective potential following \cite{kato2024}. The elements for which we perform HULLAC calculations, as well as the configurations used for the RCI calculations for each atom, are summarized in Table \ref{tab:configs}. We note here that HULLAC cannot achieve good accuracy for the heaviest elements with \(Z\gtrsim72\) due to their complex atomic structure. Therefore, we choose not to include such calculations in this study.

The transition probabilities and cross sections are calculated with HULLAC for the relevant elements. The energy levels in our analysis are taken from the NIST database and are identified by the \(LS\)-coupling scheme, while the calculations with HULLAC are performed in the \(jj\)-coupling scheme. We therefore use the JJ2LSJ extension of the HULLAC code recently developed by \cite{gaigalas2026jj2lsj} to transform HULLAC results into the \(LS\)-coupling scheme. The energy levels of each transition of interest are then matched to their corresponding energy levels in HULLAC, and the computed transition probabilities are corrected for the change in the transition energy after calibration, such that:

\begin{equation}
\begin{split}
A_{E1} & = \left(\frac{\lambda}{\lambda_{\text{HULLAC}}}\right)^2\,A_{E1;\text{HULLAC}} \\
A_{M1} & = \left(\frac{\lambda}{\lambda_{\text{HULLAC}}}\right)^3\,A_{M1;\text{HULLAC}} \\
A_{E2} & = \left(\frac{\lambda}{\lambda_{\text{HULLAC}}}\right)^5\,A_{E2;\text{HULLAC}} 
\end{split}
\end{equation}
where \(A_{E1}\), \(A_{M1}\), and \(A_{E2}\) are the Einstein \(A\) coefficients for electric dipole, magnetic dipole, and electric quadrupole transitions, respectively. The transition energies \(\Delta E\) calculated with HULLAC and the corresponding values from NIST \citep{NIST_ASD}, along with the corrected transition probabilities of key transitions are summarized in Table \ref{tab:hullac}. Larger discrepancies between HULLAC and NIST transition energies appear mainly for transitions between different electron configurations, while relatively good agreement is obtained for forbidden fine-structure transitions within the same configuration. These discrepancies may reflect possible uncertainties in the relative energies of the configurations in the atomic structure calculations, however, they do not directly imply comparable uncertainties in the computed collision strengths.

For transitions that are allowed in both magnetic dipole and electric quadrupole radiation, both \(A_{M1}\) and \(A_{E2}\) are calculated, and the total transition probability in our model is taken as
\begin{equation}
g\bar{A} = gA_{M1} + gA_{E2}.
\end{equation}

Table \ref{tab:hullac} also shows the energy-averaged collision strength of the key elements. The collisional cross sections are computed with HULLAC for a sample of input projectile energies, from which we compute the energy-specific collision strength for each input energy, based on the relation
\begin{equation}
	\sigma_{12}(v) = \frac{\pi \hbar^2}{m_{\rm e}^2 v^2}\frac{\Omega_{12}}{g_1}.\label{eq:sigma}
\end{equation}
The energy-average collisional strength is then computed using Equation \ref{eq:upsilon} assuming \(T=2000\,\text{K}\).

\begin{deluxetable*}{cccccccc}
\tablecaption{Radiative transition probabilities and energy-average collision strengths calculated with HULLAC for featured species\label{tab:hullac}}
\tablewidth{0pt}
\tablehead{
\colhead{Ion} & \colhead{\(\lambda(\mu{\rm m})\)} &  \multicolumn{2}{c}{\(\Delta E\) (eV)\(^a\)} & \colhead{\(gA_{E1}\) (\({\rm s}^{-1}\))} & \colhead{\(gA_{M1}\) (\({\rm s}^{-1}\))} & \colhead{\(gA_{E2}\) (\({\rm s}^{-1}\))}  & \colhead{\(\Upsilon\)\(^b\)} \\
 & & \colhead{NIST} & \colhead{HULLAC} & & & & 
}
\startdata
Ge I & 1.4034 &  0.883 & 0.746 & - & - & \(3.42\times10^{-4}\) & 0.23 \\
Kr II & 1.8622 & 0.666 & 0.845 & - & 23.8 & \(0.18 \) & 1.24\\
Kr III & 2.1986 & 0.564 & 0.621 & - & 11.1 & \(9.65\times10^{-3}\) & 0.56\\
	& 1.8822 & 0.659 & 0.701 & - & - & \(7.67\times10^{-3}\) & 0.79 \\
Rb IV  & 1.5973 & 0.776 & 0.871 & - & 32.2 & \(3.92\times10^{-2}\) & 0.54 \\
Sn II & 2.3521 & 0.527 & 0.554 & - & 3.79 & \(2.66\times10^{-2}\) & 3.44 \\
Sb III & 1.5207 & 0.815 & 0.793 & - & 8.83 & \(4.81\times10^{-2}\) & 2.16\\
Te I & 2.1049 & 0.589 & 0.566 & - & 5.52 & \(1.42\times10^{-2}\) & 1.57\\
Te III & 2.1048 & 0.589 & 0.579 & - & 4.98 & - & 1.65\\
I II  & 1.4111 & 0.879 & 0.885 & - & 23.7 & \(7.93\times10^{-2}\) & 1.16 \\
La III & 1.3898 &  0.892 & 2.78 & \(6.20\times10^8\) & - & - & 3.02 \\
         & 1.4100 & 0.879 & 2.75 & \(9.41\times10^8\) & - & - & 3.11 \\
Ce III & 1.5852 & 0.782 &  1.26 & \(4.81\times 10^{7}\) & - & - & 4.40 \\
	& 1.5961 & 0.777 & 1.21 & \(3.32\times10^{7}\) & - & - & 4.39 \\
	& 1.6133 & 0.769 & 1.28 & \(6.80\times 10^{7}\) & - & - & 4.69 \\
	& 1.9146 & 0.648 & 0.718 & \(1.29\times10^{6}\) & - & - & 1.34\\
	& 1.9504 & 0.636 & 0.651 & \(2.96\times10^{5}\) & - & - & 0.75 \\
	& 1.9976 & 0.621 & 0.944 & - & 1.14 & \(2.07\times10^{-7}\) & \(7.15\times10^{-2}\)\\
	& 2.0691 & 0.599 & 0.638 & \(6.75\times10^{5}\) & - & - & 1.24\\
	& 2.0987 & 0.591 & 0.717 & - & \(2.21\times10^{-2}\) & \(9.34\times10^{-4}\) & 0.39 \\
Dy I & 1.4183 & 0.874 & 0.814 & - & - & \(9.38\times10^{-8}\) &  0.41\\
\enddata
\tablecomments{\(^a\) Transition energy from NIST atomic database and the computed transition energy with HULLAC \\ \(^b\) Energy-average collision strength (Equation \ref{eq:upsilon} with \(T=2000\,\)K) }
\end{deluxetable*}

\bibliography{kn_late}{}

@Misc{NIST_ASD,
	author = {A.~Kramida and {Yu.~Ralchenko} and J.~Reader and {and NIST ASD Team}},
	HOWPUBLISHED = {{NIST Atomic Spectra Database(ver. 5.12), [Online]. Available: {\tt{https://physics.nist.gov/asd}} [2025, December 31]. National Institute of Standards and Technology, Gaithersburg, MD.}},
	year = {2024},
}

@ARTICLE{domoto2022,
       author = {{Domoto}, Nanae and {Tanaka}, Masaomi and {Kato}, Daiji and {Kawaguchi}, Kyohei and {Hotokezaka}, Kenta and {Wanajo}, Shinya},
        title = "{Lanthanide Features in Near-infrared Spectra of Kilonovae}",
      journal = {\apj},
         year = 2022,
        month = nov,
       volume = {939},
       number = {1},
          eid = {8},
        pages = {8},
          doi = {10.3847/1538-4357/ac8c36},
archivePrefix = {arXiv},
       eprint = {2206.04232},
 primaryClass = {astro-ph.HE},
       adsurl = {https://ui.adsabs.harvard.edu/abs/2022ApJ...939....8D}
}

@Misc{scasa,
    author = {{Blaise}, Jean and {Wyart}, Jean-Fran{\c c}ois},
    HOWPUBLISHED = {Selected Constants Energy Levels and Atomic Spectra of Actinides, [Online]. Available: {\tt{http://www.lac.universite-paris-saclay.fr/Data/\\Database/}}},
    year = {1994},
}

@article{ding2024,
  title={Spectrum and energy levels of the low-lying configurations of Nd III},
  author={Ding, Milan and Ryabtsev, AN and Kononov, EY and Ryabchikova, T and Clear, CP and Concepcion, F and Pickering, JC},
  journal={Astronomy \& Astrophysics},
  volume={684},
  pages={A149},
  year={2024},
  publisher={EDP Sciences}
}

@ARTICLE{pognan2022opacity,
       author = {{Pognan}, Quentin and {Jerkstrand}, Anders and {Grumer}, Jon},
        title = "{NLTE effects on kilonova expansion opacities}",
      journal = {\mnras},
         year = 2022,
        month = jul,
       volume = {513},
       number = {4},
        pages = {5174-5197},
          doi = {10.1093/mnras/stac1253},
archivePrefix = {arXiv},
       eprint = {2202.09245},
 primaryClass = {astro-ph.HE},
       adsurl = {https://ui.adsabs.harvard.edu/abs/2022MNRAS.513.5174P}
}

@ARTICLE{hotokezaka2021,
       author = {{Hotokezaka}, Kenta and {Tanaka}, Masaomi and {Kato}, Daiji and {Gaigalas}, Gediminas},
        title = "{Nebular emission from lanthanide-rich ejecta of neutron star merger}",
      journal = {\mnras},
         year = 2021,
        month = oct,
       volume = {506},
       number = {4},
        pages = {5863-5877},
          doi = {10.1093/mnras/stab1975},
archivePrefix = {arXiv},
       eprint = {2102.07879},
 primaryClass = {astro-ph.HE},
       adsurl = {https://ui.adsabs.harvard.edu/abs/2021MNRAS.506.5863H}
}

@ARTICLE{pognan2023,
       author = {{Pognan}, Quentin and {Grumer}, Jon and {Jerkstrand}, Anders and {Wanajo}, Shinya},
        title = "{NLTE spectra of kilonovae}",
      journal = {\mnras},
         year = 2023,
        month = dec,
       volume = {526},
       number = {4},
        pages = {5220-5248},
          doi = {10.1093/mnras/stad3106},
archivePrefix = {arXiv},
       eprint = {2309.01134},
 primaryClass = {astro-ph.HE},
       adsurl = {https://ui.adsabs.harvard.edu/abs/2023MNRAS.526.5220P}
}

@ARTICLE{pognan2025,
       author = {{Pognan}, Quentin and {Kawaguchi}, Kyohei and {Wanajo}, Shinya and {Fujibayashi}, Sho and {Jerkstrand}, Anders},
        title = "{Lanthanide Impact on the Infra-Red Spectra of Nebular Phase Kilonovae}",
      journal = {arXiv e-prints},
         year = 2025,
        month = oct,
          eid = {arXiv:2510.12413},
        pages = {arXiv:2510.12413},
          doi = {10.48550/arXiv.2510.12413},
archivePrefix = {arXiv},
       eprint = {2510.12413},
 primaryClass = {astro-ph.HE},
       adsurl = {https://ui.adsabs.harvard.edu/abs/2025arXiv251012413P}
}

@ARTICLE{watson2019,
       author = {{Watson}, Darach and {Hansen}, Camilla J. and {Selsing}, Jonatan and {Koch}, Andreas and {Malesani}, Daniele B. and {Andersen}, Anja C. and {Fynbo}, Johan P.~U. and {Arcones}, Almudena and {Bauswein}, Andreas and {Covino}, Stefano and {Grado}, Aniello and {Heintz}, Kasper E. and {Hunt}, Leslie and {Kouveliotou}, Chryssa and {Leloudas}, Giorgos and {Levan}, Andrew J. and {Mazzali}, Paolo and {Pian}, Elena},
        title = "{Identification of strontium in the merger of two neutron stars}",
      journal = {\nat},
         year = 2019,
        month = oct,
       volume = {574},
       number = {7779},
        pages = {497-500},
          doi = {10.1038/s41586-019-1676-3},
archivePrefix = {arXiv},
       eprint = {1910.10510},
 primaryClass = {astro-ph.HE},
       adsurl = {https://ui.adsabs.harvard.edu/abs/2019Natur.574..497W}
}

@ARTICLE{sneppen2023,
       author = {{Sneppen}, Albert and {Watson}, Darach},
        title = "{Discovery of a 760 nm P Cygni line in AT2017gfo: Identification of yttrium in the kilonova photosphere}",
      journal = {\aap},
         year = 2023,
        month = jul,
       volume = {675},
          eid = {A194},
        pages = {A194},
          doi = {10.1051/0004-6361/202346421},
archivePrefix = {arXiv},
       eprint = {2306.14942},
 primaryClass = {astro-ph.HE},
       adsurl = {https://ui.adsabs.harvard.edu/abs/2023A&A...675A.194S}
}

@ARTICLE{domoto2025,
       author = {{Domoto}, Nanae and {Wanajo}, Shinya and {Tanaka}, Masaomi and {Kato}, Daiji and {Hotokezaka}, Kenta},
        title = "{Thorium in Kilonova Spectra: Exploring the Heaviest Detectable Element}",
      journal = {\apj},
         year = 2025,
        month = jan,
       volume = {978},
       number = {1},
          eid = {99},
        pages = {99},
          doi = {10.3847/1538-4357/ad96b3},
archivePrefix = {arXiv},
       eprint = {2411.16998},
 primaryClass = {astro-ph.HE},
       adsurl = {https://ui.adsabs.harvard.edu/abs/2025ApJ...978...99D}
}

@ARTICLE{gillanders2024,
       author = {{Gillanders}, J. H. and {Sim}, S. A. and {Smartt}, S. J. and {Goriely}, S. and {Bauswein}, A.},
        title = "{Modelling the spectra of the kilonova AT2017gfo - II. Beyond the photospheric epochs}",
      journal = {\mnras},
         year = 2024,
        month = apr,
       volume = {529},
       number = {3},
        pages = {2918-2945},
          doi = {10.1093/mnras/stad3688},
archivePrefix = {arXiv},
       eprint = {2306.15055},
 primaryClass = {astro-ph.HE},
       adsurl = {https://ui.adsabs.harvard.edu/abs/2024MNRAS.529.2918G}
}

@ARTICLE{flors2025,
       author = {{Fl{\"o}rs}, Andreas and {Ferreira da Silva}, Ricardo and {Marques}, Jos{\'e} P. and {Sampaio}, Jorge M. and {Mart{\'\i}nez-Pinedo}, Gabriel},
        title = "{Calibrated Lanthanide Atomic Data for Kilonova Radiative Transfer. I. Atomic Structure and Opacities}",
      journal = {arXiv e-prints},
         year = 2025,
        month = jul,
          eid = {arXiv:2507.07785},
        pages = {arXiv:2507.07785},
          doi = {10.48550/arXiv.2507.07785},
archivePrefix = {arXiv},
       eprint = {2507.07785},
 primaryClass = {astro-ph.HE},
       adsurl = {https://ui.adsabs.harvard.edu/abs/2025arXiv250707785F}
}

@ARTICLE{vald1,
       author = {{Piskunov}, N.~E. and {Kupka}, F. and {Ryabchikova}, T.~A. and {Weiss}, W.~W. and {Jeffery}, C.~S.},
        title = "{VALD: The Vienna Atomic Line Data Base.}",
      journal = {\aaps},
         year = 1995,
        month = sep,
       volume = {112},
        pages = {525},
       adsurl = {https://ui.adsabs.harvard.edu/abs/1995A&AS..112..525P}
}

@ARTICLE{vald2,
       author = {{Kupka}, F. and {Piskunov}, N. and {Ryabchikova}, T.~A. and {Stempels}, H.~C. and {Weiss}, W.~W.},
        title = "{VALD-2: Progress of the Vienna Atomic Line Data Base}",
      journal = {\aaps},
         year = 1999,
        month = jul,
       volume = {138},
        pages = {119-133},
          doi = {10.1051/aas:1999267},
       adsurl = {https://ui.adsabs.harvard.edu/abs/1999A&AS..138..119K}
}

@ARTICLE{vald3,
       author = {{Ryabchikova}, T. and {Piskunov}, N. and {Kurucz}, R.~L. and {Stempels}, H.~C. and {Heiter}, U. and {Pakhomov}, Yu and {Barklem}, P.~S.},
        title = "{A major upgrade of the VALD database}",
      journal = {\physscr},
         year = 2015,
        month = may,
       volume = {90},
       number = {5},
          eid = {054005},
        pages = {054005},
          doi = {10.1088/0031-8949/90/5/054005},
       adsurl = {https://ui.adsabs.harvard.edu/abs/2015PhyS...90e4005R}
}

@ARTICLE{tanaka2023,
       author = {{Tanaka}, Masaomi and {Domoto}, Nanae and {Aoki}, Wako and {Ishigaki}, Miho N. and {Wanajo}, Shinya and {Hotokezaka}, Kenta and {Kawaguchi}, Kyohei and {Kato}, Daiji and {Lee}, Jae-Joon and {Lee}, Ho-Gyu and {Hirano}, Teruyuki and {Kotani}, Takayuki and {Kuzuhara}, Masayuki and {Nishikawa}, Jun and {Omiya}, Masashi and {Tamura}, Motohide and {Ueda}, Akitoshi},
        title = "{Cerium Features in Kilonova Near-infrared Spectra: Implication from a Chemically Peculiar Star}",
      journal = {\apj},
         year = 2023,
        month = aug,
       volume = {953},
       number = {1},
          eid = {17},
        pages = {17},
          doi = {10.3847/1538-4357/acdc95},
archivePrefix = {arXiv},
       eprint = {2306.04697},
 primaryClass = {astro-ph.HE},
       adsurl = {https://ui.adsabs.harvard.edu/abs/2023ApJ...953...17T}
}

@ARTICLE{kawaguchi2018,
       author = {{Kawaguchi}, Kyohei and {Shibata}, Masaru and {Tanaka}, Masaomi},
        title = "{Radiative Transfer Simulation for the Optical and Near-infrared Electromagnetic Counterparts to GW170817}",
      journal = {\apjl},
         year = 2018,
        month = oct,
       volume = {865},
       number = {2},
          eid = {L21},
        pages = {L21},
          doi = {10.3847/2041-8213/aade02},
archivePrefix = {arXiv},
       eprint = {1806.04088},
 primaryClass = {astro-ph.HE},
       adsurl = {https://ui.adsabs.harvard.edu/abs/2018ApJ...865L..21K}
}

@ARTICLE{wanajo2014,
       author = {{Wanajo}, Shinya and {Sekiguchi}, Yuichiro and {Nishimura}, Nobuya and {Kiuchi}, Kenta and {Kyutoku}, Koutarou and {Shibata}, Masaru},
        title = "{Production of All the r-process Nuclides in the Dynamical Ejecta of Neutron Star Mergers}",
      journal = {\apjl},
         year = 2014,
        month = jul,
       volume = {789},
       number = {2},
          eid = {L39},
        pages = {L39},
          doi = {10.1088/2041-8205/789/2/L39},
archivePrefix = {arXiv},
       eprint = {1402.7317},
 primaryClass = {astro-ph.SR},
       adsurl = {https://ui.adsabs.harvard.edu/abs/2014ApJ...789L..39W}
}

@ARTICLE{tanvir2017,
       author = {{Tanvir}, N.~R. and {Levan}, A.~J. and {Gonz{\'a}lez-Fern{\'a}ndez}, C. and {Korobkin}, O. and {Mandel}, I. and {Rosswog}, S. and {Hjorth}, J. and {D'Avanzo}, P. and {Fruchter}, A.~S. and {Fryer}, C.~L. and {Kangas}, T. and {Milvang-Jensen}, B. and {Rosetti}, S. and {Steeghs}, D. and {Wollaeger}, R.~T. and {Cano}, Z. and {Copperwheat}, C.~M. and {Covino}, S. and {D'Elia}, V. and {de Ugarte Postigo}, A. and {Evans}, P.~A. and {Even}, W.~P. and {Fairhurst}, S. and {Figuera Jaimes}, R. and {Fontes}, C.~J. and {Fujii}, Y.~I. and {Fynbo}, J.~P.~U. and {Gompertz}, B.~P. and {Greiner}, J. and {Hodosan}, G. and {Irwin}, M.~J. and {Jakobsson}, P. and {J{\o}rgensen}, U.~G. and {Kann}, D.~A. and {Lyman}, J.~D. and {Malesani}, D. and {McMahon}, R.~G. and {Melandri}, A. and {O'Brien}, P.~T. and {Osborne}, J.~P. and {Palazzi}, E. and {Perley}, D.~A. and {Pian}, E. and {Piranomonte}, S. and {Rabus}, M. and {Rol}, E. and {Rowlinson}, A. and {Schulze}, S. and {Sutton}, P. and {Th{\"o}ne}, C.~C. and {Ulaczyk}, K. and {Watson}, D. and {Wiersema}, K. and {Wijers}, R.~A.~M.~J.},
        title = "{The Emergence of a Lanthanide-rich Kilonova Following the Merger of Two Neutron Stars}",
      journal = {\apjl},
         year = 2017,
        month = oct,
       volume = {848},
       number = {2},
          eid = {L27},
        pages = {L27},
          doi = {10.3847/2041-8213/aa90b6},
archivePrefix = {arXiv},
       eprint = {1710.05455},
 primaryClass = {astro-ph.HE},
       adsurl = {https://ui.adsabs.harvard.edu/abs/2017ApJ...848L..27T}
}

@ARTICLE{fujibayashi2023,
       author = {{Fujibayashi}, Sho and {Kiuchi}, Kenta and {Wanajo}, Shinya and {Kyutoku}, Koutarou and {Sekiguchi}, Yuichiro and {Shibata}, Masaru},
        title = "{Comprehensive Study of Mass Ejection and Nucleosynthesis in Binary Neutron Star Mergers Leaving Short-lived Massive Neutron Stars}",
      journal = {\apj},
         year = 2023,
        month = jan,
       volume = {942},
       number = {1},
          eid = {39},
        pages = {39},
          doi = {10.3847/1538-4357/ac9ce0},
archivePrefix = {arXiv},
       eprint = {2205.05557},
 primaryClass = {astro-ph.HE},
       adsurl = {https://ui.adsabs.harvard.edu/abs/2023ApJ...942...39F}
}

@ARTICLE{hotokezaka2023,
       author = {{Hotokezaka}, Kenta and {Tanaka}, Masaomi and {Kato}, Daiji and {Gaigalas}, Gediminas},
        title = "{Tellurium emission line in kilonova AT 2017gfo}",
      journal = {\mnras},
         year = 2023,
        month = nov,
       volume = {526},
       number = {1},
        pages = {L155-L159},
          doi = {10.1093/mnrasl/slad128},
archivePrefix = {arXiv},
       eprint = {2307.00988},
 primaryClass = {astro-ph.HE},
       adsurl = {https://ui.adsabs.harvard.edu/abs/2023MNRAS.526L.155H}
}

@ARTICLE{hotokezaka2022,
       author = {{Hotokezaka}, Kenta and {Tanaka}, Masaomi and {Kato}, Daiji and {Gaigalas}, Gediminas},
        title = "{Tungsten versus Selenium as a potential source of kilonova nebular emission observed by Spitzer}",
      journal = {\mnras},
         year = 2022,
        month = sep,
       volume = {515},
       number = {1},
        pages = {L89-L93},
          doi = {10.1093/mnrasl/slac071},
archivePrefix = {arXiv},
       eprint = {2204.00737},
 primaryClass = {astro-ph.HE},
       adsurl = {https://ui.adsabs.harvard.edu/abs/2022MNRAS.515L..89H}
}

@ARTICLE{prantzos2020,
       author = {{Prantzos}, N. and {Abia}, C. and {Cristallo}, S. and {Limongi}, M. and {Chieffi}, A.},
        title = "{Chemical evolution with rotating massive star yields II. A new assessment of the solar s- and r-process components}",
      journal = {\mnras},
         year = 2020,
        month = jan,
       volume = {491},
       number = {2},
        pages = {1832-1850},
          doi = {10.1093/mnras/stz3154},
archivePrefix = {arXiv},
       eprint = {1911.02545},
 primaryClass = {astro-ph.GA},
       adsurl = {https://ui.adsabs.harvard.edu/abs/2020MNRAS.491.1832P}
}

@ARTICLE{regemorter,
       author = {{van Regemorter}, Henri},
        title = "{Rate of Collisional Excitation in Stellar Atmospheres.}",
      journal = {\apj},
         year = 1962,
        month = nov,
       volume = {136},
        pages = {906},
          doi = {10.1086/147445},
       adsurl = {https://ui.adsabs.harvard.edu/abs/1962ApJ...136..906V}
}

@ARTICLE{mulholland2025,
       author = {{Mulholland}, Leo P. and {Bromley}, Steven J. and {Ballance}, Connor P. and {Sim}, Stuart A. and {Ramsbottom}, Catherine A.},
        title = "{On the use of the Axelrod formula for thermal electron collisions in astrophysical modelling}",
      journal = {\jqsrt},
         year = 2025,
        month = nov,
       volume = {345},
          eid = {109545},
        pages = {109545},
          doi = {10.1016/j.jqsrt.2025.109545},
archivePrefix = {arXiv},
       eprint = {2503.05489},
 primaryClass = {astro-ph.HE},
       adsurl = {https://ui.adsabs.harvard.edu/abs/2025JQSRT.34509545M}
}

@ARTICLE{mulholland2024,
       author = {{Mulholland}, L.~P. and {McElroy}, N.~E. and {McNeill}, F.~L. and {Sim}, S.~A. and {Ballance}, C.~P. and {Ramsbottom}, C.~A.},
        title = "{New radiative and collisional atomic data for Sr II and Y II with application to Kilonova modelling}",
      journal = {\mnras},
         year = 2024,
        month = aug,
       volume = {532},
       number = {2},
        pages = {2289-2308},
          doi = {10.1093/mnras/stae1615},
archivePrefix = {arXiv},
       eprint = {2407.01398},
 primaryClass = {astro-ph.HE},
       adsurl = {https://ui.adsabs.harvard.edu/abs/2024MNRAS.532.2289M}
}

@ARTICLE{mulholland2025te,
       author = {{Mulholland}, Leo P. and {Ramsbottom}, Catherine A. and {Ballance}, Connor P. and {Sneppen}, Albert and {Sim}, Stuart A.},
        title = "{Electron impact excitation of Te IV and V and Level Resolved R-matrix Photoionization of Te I - IV with application to modelling of AT2017gfo}",
      journal = {arXiv e-prints},
         year = 2025,
        month = oct,
          eid = {arXiv:2510.17357},
        pages = {arXiv:2510.17357},
          doi = {10.48550/arXiv.2510.17357},
archivePrefix = {arXiv},
       eprint = {2510.17357},
 primaryClass = {astro-ph.HE},
       adsurl = {https://ui.adsabs.harvard.edu/abs/2025arXiv251017357M}
}

@ARTICLE{pasternack1940,
       author = {{Pasternack}, Simon},
        title = "{Transition Probabilities of Forbidden Lines.}",
      journal = {\apj},
         year = 1940,
        month = sep,
       volume = {92},
        pages = {129},
          doi = {10.1086/144208},
       adsurl = {https://ui.adsabs.harvard.edu/abs/1940ApJ....92..129P}
}

@article{shortley1940,
  title = {The Computation of Quadrupole and Magnetic-Dipole Transition Probabilities},
  author = {Shortley, George H.},
  journal = {Phys. Rev.},
  volume = {57},
  issue = {3},
  pages = {225--234},
  numpages = {0},
  year = {1940},
  month = {Feb},
  publisher = {American Physical Society},
  doi = {10.1103/PhysRev.57.225},
  url = {https://link.aps.org/doi/10.1103/PhysRev.57.225}
}

@ARTICLE{bahcall1968,
       author = {{Bahcall}, John N. and {Wolf}, Richard A.},
        title = "{Fine-Structure Transitions}",
      journal = {\apj},
         year = 1968,
        month = jun,
       volume = {152},
        pages = {701},
          doi = {10.1086/149589},
       adsurl = {https://ui.adsabs.harvard.edu/abs/1968ApJ...152..701B}
}

@ARTICLE{pognan2022steady,
       author = {{Pognan}, Quentin and {Jerkstrand}, Anders and {Grumer}, Jon},
        title = "{On the validity of steady-state for nebular phase kilonovae}",
      journal = {\mnras},
         year = 2022,
        month = mar,
       volume = {510},
       number = {3},
        pages = {3806-3837},
          doi = {10.1093/mnras/stab3674},
archivePrefix = {arXiv},
       eprint = {2112.07484},
 primaryClass = {astro-ph.HE},
       adsurl = {https://ui.adsabs.harvard.edu/abs/2022MNRAS.510.3806P}
}

@ARTICLE{roederer2018,
       author = {{Roederer}, Ian U. and {Sakari}, Charli M. and {Placco}, Vinicius M. and {Beers}, Timothy C. and {Ezzeddine}, Rana and {Frebel}, Anna and {Hansen}, Terese T.},
        title = "{The R-Process Alliance: A Comprehensive Abundance Analysis of HD 222925, a Metal-poor Star with an Extreme R-process Enhancement of [Eu/H] = -0.14}",
      journal = {\apj},
         year = 2018,
        month = oct,
       volume = {865},
       number = {2},
          eid = {129},
        pages = {129},
          doi = {10.3847/1538-4357/aadd92},
archivePrefix = {arXiv},
       eprint = {1808.09469},
 primaryClass = {astro-ph.SR},
       adsurl = {https://ui.adsabs.harvard.edu/abs/2018ApJ...865..129R}
}

@article{hullac,
title = {HULLAC, an integrated computer package for atomic processes in plasmas},
journal = {Journal of Quantitative Spectroscopy and Radiative Transfer},
volume = {71},
number = {2},
pages = {169-188},
year = {2001},
note = {Radiative Properties of Hot Dense Matter},
issn = {0022-4073},
doi = {https://doi.org/10.1016/S0022-4073(01)00066-8},
url = {https://www.sciencedirect.com/science/article/pii/S0022407301000668},
author = {A. Bar-Shalom and M. Klapisch and J. Oreg}
}

@ARTICLE{mooley2018,
       author = {{Mooley}, K.~P. and {Frail}, D.~A. and {Dobie}, D. and {Lenc}, E. and {Corsi}, A. and {De}, K. and {Nayana}, A.~J. and {Makhathini}, S. and {Heywood}, I. and {Murphy}, T. and {Kaplan}, D.~L. and {Chandra}, P. and {Smirnov}, O. and {Nakar}, E. and {Hallinan}, G. and {Camilo}, F. and {Fender}, R. and {Goedhart}, S. and {Groot}, P. and {Kasliwal}, M.~M. and {Kulkarni}, S.~R. and {Woudt}, P.~A.},
        title = "{A Strong Jet Signature in the Late-time Light Curve of GW170817}",
      journal = {\apjl},
         year = 2018,
        month = nov,
       volume = {868},
       number = {1},
          eid = {L11},
        pages = {L11},
          doi = {10.3847/2041-8213/aaeda7},
archivePrefix = {arXiv},
       eprint = {1810.12927},
 primaryClass = {astro-ph.HE},
       adsurl = {https://ui.adsabs.harvard.edu/abs/2018ApJ...868L..11M}
}

@ARTICLE{ghirlanda2019,
       author = {{Ghirlanda}, G. and {Salafia}, O.~S. and {Paragi}, Z. and {Giroletti}, M. and {Yang}, J. and {Marcote}, B. and {Blanchard}, J. and {Agudo}, I. and {An}, T. and {Bernardini}, M.~G. and {Beswick}, R. and {Branchesi}, M. and {Campana}, S. and {Casadio}, C. and {Chassande-Mottin}, E. and {Colpi}, M. and {Covino}, S. and {D'Avanzo}, P. and {D'Elia}, V. and {Frey}, S. and {Gawronski}, M. and {Ghisellini}, G. and {Gurvits}, L.~I. and {Jonker}, P.~G. and {van Langevelde}, H.~J. and {Melandri}, A. and {Moldon}, J. and {Nava}, L. and {Perego}, A. and {Perez-Torres}, M.~A. and {Reynolds}, C. and {Salvaterra}, R. and {Tagliaferri}, G. and {Venturi}, T. and {Vergani}, S.~D. and {Zhang}, M.},
        title = "{Compact radio emission indicates a structured jet was produced by a binary neutron star merger}",
      journal = {Science},
         year = 2019,
        month = mar,
       volume = {363},
       number = {6430},
        pages = {968-971},
          doi = {10.1126/science.aau8815},
archivePrefix = {arXiv},
       eprint = {1808.00469},
 primaryClass = {astro-ph.HE},
       adsurl = {https://ui.adsabs.harvard.edu/abs/2019Sci...363..968G}
}

@ARTICLE{lattimer1974,
       author = {{Lattimer}, J.~M. and {Schramm}, D.~N.},
        title = "{Black-Hole-Neutron-Star Collisions}",
      journal = {\apjl},
         year = 1974,
        month = sep,
       volume = {192},
        pages = {L145},
          doi = {10.1086/181612},
       adsurl = {https://ui.adsabs.harvard.edu/abs/1974ApJ...192L.145L}
}

@ARTICLE{symbalisty1982,
       author = {{Symbalisty}, E. and {Schramm}, D.~N.},
        title = "{Neutron Star Collisions and the r-Process}",
      journal = {aplett},
         year = 1982,
        month = jan,
       volume = {22},
        pages = {143},
       adsurl = {https://ui.adsabs.harvard.edu/abs/1982ApL....22..143S}
}

@ARTICLE{eichler1989,
       author = {{Eichler}, David and {Livio}, Mario and {Piran}, Tsvi and {Schramm}, David N.},
        title = "{Nucleosynthesis, neutrino bursts and {\ensuremath{\gamma}}-rays from coalescing neutron stars}",
      journal = {\nat},
         year = 1989,
        month = jul,
       volume = {340},
       number = {6229},
        pages = {126-128},
          doi = {10.1038/340126a0},
       adsurl = {https://ui.adsabs.harvard.edu/abs/1989Natur.340..126E}
}

@ARTICLE{freiburghaus1999,
       author = {{Freiburghaus}, C. and {Rosswog}, S. and {Thielemann}, F.-K.},
        title = "{R-Process in Neutron Star Mergers}",
      journal = {\apjl},
         year = 1999,
        month = nov,
       volume = {525},
       number = {2},
        pages = {L121-L124},
          doi = {10.1086/312343},
       adsurl = {https://ui.adsabs.harvard.edu/abs/1999ApJ...525L.121F}
}

@ARTICLE{goriely2011,
       author = {{Goriely}, Stephane and {Bauswein}, Andreas and {Janka}, Hans-Thomas},
        title = "{r-process Nucleosynthesis in Dynamically Ejected Matter of Neutron Star Mergers}",
      journal = {\apjl},
         year = 2011,
        month = sep,
       volume = {738},
       number = {2},
          eid = {L32},
        pages = {L32},
          doi = {10.1088/2041-8205/738/2/L32},
archivePrefix = {arXiv},
       eprint = {1107.0899},
 primaryClass = {astro-ph.SR},
       adsurl = {https://ui.adsabs.harvard.edu/abs/2011ApJ...738L..32G}
}

@ARTICLE{li1998,
       author = {{Li}, Li-Xin and {Paczy{\'n}ski}, Bohdan},
        title = "{Transient Events from Neutron Star Mergers}",
      journal = {\apjl},
         year = 1998,
        month = nov,
       volume = {507},
       number = {1},
        pages = {L59-L62},
          doi = {10.1086/311680},
archivePrefix = {arXiv},
       eprint = {astro-ph/9807272},
 primaryClass = {astro-ph},
       adsurl = {https://ui.adsabs.harvard.edu/abs/1998ApJ...507L..59L}
}

@ARTICLE{metzger2010,
       author = {{Metzger}, B.~D. and {Mart{\'\i}nez-Pinedo}, G. and {Darbha}, S. and {Quataert}, E. and {Arcones}, A. and {Kasen}, D. and {Thomas}, R. and {Nugent}, P. and {Panov}, I.~V. and {Zinner}, N.~T.},
        title = "{Electromagnetic counterparts of compact object mergers powered by the radioactive decay of r-process nuclei}",
      journal = {\mnras},
         year = 2010,
        month = aug,
       volume = {406},
       number = {4},
        pages = {2650-2662},
          doi = {10.1111/j.1365-2966.2010.16864.x},
archivePrefix = {arXiv},
       eprint = {1001.5029},
 primaryClass = {astro-ph.HE},
       adsurl = {https://ui.adsabs.harvard.edu/abs/2010MNRAS.406.2650M}
}

@ARTICLE{roberts2011,
       author = {{Roberts}, L.~F. and {Kasen}, D. and {Lee}, W.~H. and {Ramirez-Ruiz}, E.},
        title = "{Electromagnetic Transients Powered by Nuclear Decay in the Tidal Tails of Coalescing Compact Binaries}",
      journal = {\apjl},
         year = 2011,
        month = jul,
       volume = {736},
       number = {1},
          eid = {L21},
        pages = {L21},
          doi = {10.1088/2041-8205/736/1/L21},
archivePrefix = {arXiv},
       eprint = {1104.5504},
 primaryClass = {astro-ph.HE},
       adsurl = {https://ui.adsabs.harvard.edu/abs/2011ApJ...736L..21R}
}

@ARTICLE{abbott2017a,
       author = {{Abbott}, B.~P. and {Abbott}, R. and {Abbott}, T.~D. and {Acernese}, F. and {Ackley}, K. and {Adams}, C. and {Adams}, T. and {Addesso}, P. and {Adhikari}, R.~X. and {Adya}, V.~B. and {Affeldt}, C. and {Afrough}, M. and {Agarwal}, B. and {Agathos}, M. and {Agatsuma}, K. and {Aggarwal}, N. and {Aguiar}, O.~D. and {Aiello}, L. and {Ain}, A. and {Ajith}, P. and {Allen}, B. and {Allen}, G. and {Allocca}, A. and {Altin}, P.~A. and {Amato}, A. and {Ananyeva}, A. and {Anderson}, S.~B. and {Anderson}, W.~G. and {Angelova}, S.~V. and {Antier}, S. and {Appert}, S. and {Arai}, K. and {Araya}, M.~C. and {Areeda}, J.~S. and {Arnaud}, N. and {Arun}, K.~G. and {Ascenzi}, S. and {Ashton}, G. and {Ast}, M. and {Aston}, S.~M. and {Astone}, P. and {Atallah}, D.~V. and {Aufmuth}, P. and {Aulbert}, C. and {AultONeal}, K. and {Austin}, C. and {Avila-Alvarez}, A. and {Babak}, S. and {Bacon}, P. and {Bader}, M.~K.~M. and {Bae}, S. and {Bailes}, M. and {Baker}, P.~T. and {Baldaccini}, F. and {Ballardin}, G. and {Ballmer}, S.~W. and {Banagiri}, S. and {Barayoga}, J.~C. and {Barclay}, S.~E. and {Barish}, B.~C. and {Barker}, D. and {Barkett}, K. and {Barone}, F. and {Barr}, B. and {Barsotti}, L. and {Barsuglia}, M. and {Barta}, D. and {Barthelmy}, S.~D. and {Bartlett}, J. and {Bartos}, I. and {Bassiri}, R. and {Basti}, A. and {Batch}, J.~C. and {Bawaj}, M. and {Bayley}, J.~C. and {Bazzan}, M. and {B{\'e}csy}, B. and {Beer}, C. and {Bejger}, M. and {Belahcene}, I. and {Bell}, A.~S. and {Berger}, B.~K. and {Bergmann}, G. and {Bernuzzi}, S. and {Bero}, J.~J. and {Berry}, C.~P.~L. and {Bersanetti}, D. and {Bertolini}, A. and {Betzwieser}, J. and {Bhagwat}, S. and {Bhandare}, R. and {Bilenko}, I.~A. and {Billingsley}, G. and {Billman}, C.~R. and {Birch}, J. and {Birney}, R. and {Birnholtz}, O. and {Biscans}, S. and {Biscoveanu}, S. and {Bisht}, A. and {Bitossi}, M. and {Biwer}, C. and {Bizouard}, M.~A. and {Blackburn}, J.~K. and {Blackman}, J. and {Blair}, C.~D. and {Blair}, D.~G. and {Blair}, R.~M. and {Bloemen}, S. and {Bock}, O. and {Bode}, N. and {Boer}, M. and {Bogaert}, G. and {Bohe}, A. and {Bondu}, F. and {Bonilla}, E. and {Bonnand}, R. and {Boom}, B.~A. and {Bork}, R. and {Boschi}, V. and {Bose}, S. and {Bossie}, K. and {Bouffanais}, Y. and {Bozzi}, A. and {Bradaschia}, C. and {Brady}, P.~R. and {Branchesi}, M. and {Brau}, J.~E. and {Briant}, T. and {Brillet}, A. and {Brinkmann}, M. and {Brisson}, V. and {Brockill}, P. and {Broida}, J.~E. and {Brooks}, A.~F. and {Brown}, D.~A. and {Brown}, D.~D. and {Brunett}, S. and {Buchanan}, C.~C. and {Buikema}, A. and {Bulik}, T. and {Bulten}, H.~J. and {Buonanno}, A. and {Buskulic}, D. and {Buy}, C. and {Byer}, R.~L. and {Cabero}, M. and {Cadonati}, L. and {Cagnoli}, G. and {Cahillane}, C. and {Calder{\'o}n Bustillo}, J. and {Callister}, T.~A. and {Calloni}, E. and {Camp}, J.~B. and {Canepa}, M. and {Canizares}, P. and {Cannon}, K.~C. and {Cao}, H. and {Cao}, J. and {Capano}, C.~D. and {Capocasa}, E. and {Carbognani}, F. and {Caride}, S. and {Carney}, M.~F. and {Carullo}, G. and {Casanueva Diaz}, J. and {Casentini}, C. and {Caudill}, S. and {Cavagli{\`a}}, M. and {Cavalier}, F. and {Cavalieri}, R. and {Cella}, G. and {Cepeda}, C.~B. and {Cerd{\'a}-Dur{\'a}n}, P. and {Cerretani}, G. and {Cesarini}, E. and {Chamberlin}, S.~J. and {Chan}, M. and {Chao}, S. and {Charlton}, P. and {Chase}, E. and {Chassande-Mottin}, E. and {Chatterjee}, D. and {Chatziioannou}, K. and {Cheeseboro}, B.~D. and {Chen}, H.~Y. and {Chen}, X. and {Chen}, Y. and {Cheng}, H.-P. and {Chia}, H. and {Chincarini}, A. and {Chiummo}, A. and {Chmiel}, T. and {Cho}, H.~S. and {Cho}, M. and {Chow}, J.~H. and {Christensen}, N. and {Chu}, Q. and {Chua}, A.~J.~K. and {Chua}, S.},
        title = "{GW170817: Observation of Gravitational Waves from a Binary Neutron Star Inspiral}",
      journal = {\prl},
         year = 2017,
        month = oct,
       volume = {119},
       number = {16},
          eid = {161101},
        pages = {161101},
          doi = {10.1103/PhysRevLett.119.161101},
archivePrefix = {arXiv},
       eprint = {1710.05832},
 primaryClass = {gr-qc},
       adsurl = {https://ui.adsabs.harvard.edu/abs/2017PhRvL.119p1101A}
}

@ARTICLE{abbott2017b,
       author = {{Abbott}, B.~P. and {Abbott}, R. and {Abbott}, T.~D. and {Acernese}, F. and {Ackley}, K. and {Adams}, C. and {Adams}, T. and {Addesso}, P. and {Adhikari}, R.~X. and {Adya}, V.~B. and {Affeldt}, C. and {Afrough}, M. and {Agarwal}, B. and {Agathos}, M. and {Agatsuma}, K. and {Aggarwal}, N. and {Aguiar}, O.~D. and {Aiello}, L. and {Ain}, A. and {Ajith}, P. and {Allen}, B. and {Allen}, G. and {Allocca}, A. and {Altin}, P.~A. and {Amato}, A. and {Ananyeva}, A. and {Anderson}, S.~B. and {Anderson}, W.~G. and {Angelova}, S.~V. and {Antier}, S. and {Appert}, S. and {Arai}, K. and {Araya}, M.~C. and {Areeda}, J.~S. and {Arnaud}, N. and {Arun}, K.~G. and {Ascenzi}, S. and {Ashton}, G. and {Ast}, M. and {Aston}, S.~M. and {Astone}, P. and {Atallah}, D.~V. and {Aufmuth}, P. and {Aulbert}, C. and {AultONeal}, K. and {Austin}, C. and {Avila-Alvarez}, A. and {Babak}, S. and {Bacon}, P. and {Bader}, M.~K.~M. and {Bae}, S. and {Baker}, P.~T. and {Baldaccini}, F. and {Ballardin}, G. and {Ballmer}, S.~W. and {Banagiri}, S. and {Barayoga}, J.~C. and {Barclay}, S.~E. and {Barish}, B.~C. and {Barker}, D. and {Barkett}, K. and {Barone}, F. and {Barr}, B. and {Barsotti}, L. and {Barsuglia}, M. and {Barta}, D. and {Barthelmy}, S.~D. and {Bartlett}, J. and {Bartos}, I. and {Bassiri}, R. and {Basti}, A. and {Batch}, J.~C. and {Bawaj}, M. and {Bayley}, J.~C. and {Bazzan}, M. and {B{\'e}csy}, B. and {Beer}, C. and {Bejger}, M. and {Belahcene}, I. and {Bell}, A.~S. and {Berger}, B.~K. and {Bergmann}, G. and {Bero}, J.~J. and {Berry}, C.~P.~L. and {Bersanetti}, D. and {Bertolini}, A. and {Betzwieser}, J. and {Bhagwat}, S. and {Bhandare}, R. and {Bilenko}, I.~A. and {Billingsley}, G. and {Billman}, C.~R. and {Birch}, J. and {Birney}, R. and {Birnholtz}, O. and {Biscans}, S. and {Biscoveanu}, S. and {Bisht}, A. and {Bitossi}, M. and {Biwer}, C. and {Bizouard}, M.~A. and {Blackburn}, J.~K. and {Blackman}, J. and {Blair}, C.~D. and {Blair}, D.~G. and {Blair}, R.~M. and {Bloemen}, S. and {Bock}, O. and {Bode}, N. and {Boer}, M. and {Bogaert}, G. and {Bohe}, A. and {Bondu}, F. and {Bonilla}, E. and {Bonnand}, R. and {Boom}, B.~A. and {Bork}, R. and {Boschi}, V. and {Bose}, S. and {Bossie}, K. and {Bouffanais}, Y. and {Bozzi}, A. and {Bradaschia}, C. and {Brady}, P.~R. and {Branchesi}, M. and {Brau}, J.~E. and {Briant}, T. and {Brillet}, A. and {Brinkmann}, M. and {Brisson}, V. and {Brockill}, P. and {Broida}, J.~E. and {Brooks}, A.~F. and {Brown}, D.~A. and {Brown}, D.~D. and {Brunett}, S. and {Buchanan}, C.~C. and {Buikema}, A. and {Bulik}, T. and {Bulten}, H.~J. and {Buonanno}, A. and {Buskulic}, D. and {Buy}, C. and {Byer}, R.~L. and {Cabero}, M. and {Cadonati}, L. and {Cagnoli}, G. and {Cahillane}, C. and {Calder{\'o}n Bustillo}, J. and {Callister}, T.~A. and {Calloni}, E. and {Camp}, J.~B. and {Canepa}, M. and {Canizares}, P. and {Cannon}, K.~C. and {Cao}, H. and {Cao}, J. and {Capano}, C.~D. and {Capocasa}, E. and {Carbognani}, F. and {Caride}, S. and {Carney}, M.~F. and {Casanueva Diaz}, J. and {Casentini}, C. and {Caudill}, S. and {Cavagli{\`a}}, M. and {Cavalier}, F. and {Cavalieri}, R. and {Cella}, G. and {Cepeda}, C.~B. and {Cerd{\'a}-Dur{\'a}n}, P. and {Cerretani}, G. and {Cesarini}, E. and {Chamberlin}, S.~J. and {Chan}, M. and {Chao}, S. and {Charlton}, P. and {Chase}, E. and {Chassande-Mottin}, E. and {Chatterjee}, D. and {Chatziioannou}, K. and {Cheeseboro}, B.~D. and {Chen}, H.~Y. and {Chen}, X. and {Chen}, Y. and {Cheng}, H.-P. and {Chia}, H. and {Chincarini}, A. and {Chiummo}, A. and {Chmiel}, T. and {Cho}, H.~S. and {Cho}, M. and {Chow}, J.~H. and {Christensen}, N. and {Chu}, Q. and {Chua}, A.~J.~K. and {Chua}, S. and {Chung}, A.~K.~W. and {Chung}, S. and {Ciani}, G.},
        title = "{Multi-messenger Observations of a Binary Neutron Star Merger}",
      journal = {\apjl},
         year = 2017,
        month = oct,
       volume = {848},
       number = {2},
          eid = {L12},
        pages = {L12},
          doi = {10.3847/2041-8213/aa91c9},
archivePrefix = {arXiv},
       eprint = {1710.05833},
 primaryClass = {astro-ph.HE},
       adsurl = {https://ui.adsabs.harvard.edu/abs/2017ApJ...848L..12A}
}

@ARTICLE{arcavi2017,
       author = {{Arcavi}, Iair and {Hosseinzadeh}, Griffin and {Howell}, D. Andrew and {McCully}, Curtis and {Poznanski}, Dovi and {Kasen}, Daniel and {Barnes}, Jennifer and {Zaltzman}, Michael and {Vasylyev}, Sergiy and {Maoz}, Dan and {Valenti}, Stefano},
        title = "{Optical emission from a kilonova following a gravitational-wave-detected neutron-star merger}",
      journal = {\nat},
         year = 2017,
        month = nov,
       volume = {551},
       number = {7678},
        pages = {64-66},
          doi = {10.1038/nature24291},
archivePrefix = {arXiv},
       eprint = {1710.05843},
 primaryClass = {astro-ph.HE},
       adsurl = {https://ui.adsabs.harvard.edu/abs/2017Natur.551...64A}
}

@ARTICLE{pian2017,
       author = {{Pian}, E. and {D'Avanzo}, P. and {Benetti}, S. and {Branchesi}, M. and {Brocato}, E. and {Campana}, S. and {Cappellaro}, E. and {Covino}, S. and {D'Elia}, V. and {Fynbo}, J.~P.~U. and {Getman}, F. and {Ghirlanda}, G. and {Ghisellini}, G. and {Grado}, A. and {Greco}, G. and {Hjorth}, J. and {Kouveliotou}, C. and {Levan}, A. and {Limatola}, L. and {Malesani}, D. and {Mazzali}, P.~A. and {Melandri}, A. and {M{\o}ller}, P. and {Nicastro}, L. and {Palazzi}, E. and {Piranomonte}, S. and {Rossi}, A. and {Salafia}, O.~S. and {Selsing}, J. and {Stratta}, G. and {Tanaka}, M. and {Tanvir}, N.~R. and {Tomasella}, L. and {Watson}, D. and {Yang}, S. and {Amati}, L. and {Antonelli}, L.~A. and {Ascenzi}, S. and {Bernardini}, M.~G. and {Bo{\"e}r}, M. and {Bufano}, F. and {Bulgarelli}, A. and {Capaccioli}, M. and {Casella}, P. and {Castro-Tirado}, A.~J. and {Chassande-Mottin}, E. and {Ciolfi}, R. and {Copperwheat}, C.~M. and {Dadina}, M. and {De Cesare}, G. and {di Paola}, A. and {Fan}, Y.~Z. and {Gendre}, B. and {Giuffrida}, G. and {Giunta}, A. and {Hunt}, L.~K. and {Israel}, G.~L. and {Jin}, Z.-P. and {Kasliwal}, M.~M. and {Klose}, S. and {Lisi}, M. and {Longo}, F. and {Maiorano}, E. and {Mapelli}, M. and {Masetti}, N. and {Nava}, L. and {Patricelli}, B. and {Perley}, D. and {Pescalli}, A. and {Piran}, T. and {Possenti}, A. and {Pulone}, L. and {Razzano}, M. and {Salvaterra}, R. and {Schipani}, P. and {Spera}, M. and {Stamerra}, A. and {Stella}, L. and {Tagliaferri}, G. and {Testa}, V. and {Troja}, E. and {Turatto}, M. and {Vergani}, S.~D. and {Vergani}, D.},
        title = "{Spectroscopic identification of r-process nucleosynthesis in a double neutron-star merger}",
      journal = {\nat},
         year = 2017,
        month = nov,
       volume = {551},
       number = {7678},
        pages = {67-70},
          doi = {10.1038/nature24298},
archivePrefix = {arXiv},
       eprint = {1710.05858},
 primaryClass = {astro-ph.HE},
       adsurl = {https://ui.adsabs.harvard.edu/abs/2017Natur.551...67P}
}

@ARTICLE{smartt2017,
       author = {{Smartt}, S. J. and {Chen}, T.-W. and {Jerkstrand}, A. and {Coughlin}, M. and {Kankare}, E. and {Sim}, S.~A. and {Fraser}, M. and {Inserra}, C. and {Maguire}, K. and {Chambers}, K.~C. and {Huber}, M.~E. and {Kr{\"u}hler}, T. and {Leloudas}, G. and {Magee}, M. and {Shingles}, L.~J. and {Smith}, K.~W. and {Young}, D.~R. and {Tonry}, J. and {Kotak}, R. and {Gal-Yam}, A. and {Lyman}, J.~D. and {Homan}, D.~S. and {Agliozzo}, C. and {Anderson}, J.~P. and {Angus}, C.~R. and {Ashall}, C. and {Barbarino}, C. and {Bauer}, F.~E. and {Berton}, M. and {Botticella}, M.~T. and {Bulla}, M. and {Bulger}, J. and {Cannizzaro}, G. and {Cano}, Z. and {Cartier}, R. and {Cikota}, A. and {Clark}, P. and {De Cia}, A. and {Della Valle}, M. and {Denneau}, L. and {Dennefeld}, M. and {Dessart}, L. and {Dimitriadis}, G. and {Elias-Rosa}, N. and {Firth}, R.~E. and {Flewelling}, H. and {Fl{\"o}rs}, A. and {Franckowiak}, A. and {Frohmaier}, C. and {Galbany}, L. and {Gonz{\'a}lez-Gait{\'a}n}, S. and {Greiner}, J. and {Gromadzki}, M. and {Guelbenzu}, A. Nicuesa and {Guti{\'e}rrez}, C.~P. and {Hamanowicz}, A. and {Hanlon}, L. and {Harmanen}, J. and {Heintz}, K.~E. and {Heinze}, A. and {Hernandez}, M.-S. and {Hodgkin}, S.~T. and {Hook}, I.~M. and {Izzo}, L. and {James}, P.~A. and {Jonker}, P.~G. and {Kerzendorf}, W.~E. and {Klose}, S. and {Kostrzewa-Rutkowska}, Z. and {Kowalski}, M. and {Kromer}, M. and {Kuncarayakti}, H. and {Lawrence}, A. and {Lowe}, T.~B. and {Magnier}, E.~A. and {Manulis}, I. and {Martin-Carrillo}, A. and {Mattila}, S. and {McBrien}, O. and {M{\"u}ller}, A. and {Nordin}, J. and {O'Neill}, D. and {Onori}, F. and {Palmerio}, J.~T. and {Pastorello}, A. and {Patat}, F. and {Pignata}, G. and {Podsiadlowski}, Ph. and {Pumo}, M.~L. and {Prentice}, S.~J. and {Rau}, A. and {Razza}, A. and {Rest}, A. and {Reynolds}, T. and {Roy}, R. and {Ruiter}, A.~J. and {Rybicki}, K.~A. and {Salmon}, L. and {Schady}, P. and {Schultz}, A.~S.~B. and {Schweyer}, T. and {Seitenzahl}, I.~R. and {Smith}, M. and {Sollerman}, J. and {Stalder}, B. and {Stubbs}, C.~W. and {Sullivan}, M. and {Szegedi}, H. and {Taddia}, F. and {Taubenberger}, S. and {Terreran}, G. and {van Soelen}, B. and {Vos}, J. and {Wainscoat}, R.~J. and {Walton}, N.~A. and {Waters}, C. and {Weiland}, H. and {Willman}, M. and {Wiseman}, P. and {Wright}, D.~E. and {Wyrzykowski}, {\L}. and {Yaron}, O.},
        title = "{A kilonova as the electromagnetic counterpart to a gravitational-wave source}",
      journal = {\nat},
         year = 2017,
        month = nov,
       volume = {551},
       number = {7678},
        pages = {75-79},
          doi = {10.1038/nature24303},
archivePrefix = {arXiv},
       eprint = {1710.05841},
 primaryClass = {astro-ph.HE},
       adsurl = {https://ui.adsabs.harvard.edu/abs/2017Natur.551...75S}
}

@ARTICLE{utsumi2017,
       author = {{Utsumi}, Yousuke and {Tanaka}, Masaomi and {Tominaga}, Nozomu and {Yoshida}, Michitoshi and {Barway}, Sudhanshu and {Nagayama}, Takahiro and {Zenko}, Tetsuya and {Aoki}, Kentaro and {Fujiyoshi}, Takuya and {Furusawa}, Hisanori and {Kawabata}, Koji S. and {Koshida}, Shintaro and {Lee}, Chien-Hsiu and {Morokuma}, Tomoki and {Motohara}, Kentaro and {Nakata}, Fumiaki and {Ohsawa}, Ryou and {Ohta}, Kouji and {Okita}, Hirofumi and {Tajitsu}, Akito and {Tanaka}, Ichi and {Terai}, Tsuyoshi and {Yasuda}, Naoki and {Abe}, Fumio and {Asakura}, Yuichiro and {Bond}, Ian A. and {Miyazaki}, Shota and {Sumi}, Takahiro and {Tristram}, Paul J. and {Honda}, Satoshi and {Itoh}, Ryosuke and {Itoh}, Yoichi and {Kawabata}, Miho and {Morihana}, Kumiko and {Nagashima}, Hiroki and {Nakaoka}, Tatsuya and {Ohshima}, Tomohito and {Takahashi}, Jun and {Takayama}, Masaki and {Aoki}, Wako and {Baar}, Stefan and {Doi}, Mamoru and {Finet}, Francois and {Kanda}, Nobuyuki and {Kawai}, Nobuyuki and {Kim}, Ji Hoon and {Kuroda}, Daisuke and {Liu}, Wei and {Matsubayashi}, Kazuya and {Murata}, Katsuhiro L. and {Nagai}, Hiroshi and {Saito}, Tomoki and {Saito}, Yoshihiko and {Sako}, Shigeyuki and {Sekiguchi}, Yuichiro and {Tamura}, Yoichi and {Tanaka}, Masayuki and {Uemura}, Makoto and {Yamaguchi}, Masaki S.},
        title = "{J-GEM observations of an electromagnetic counterpart to the neutron star merger GW170817}",
      journal = {\pasj},
         year = 2017,
        month = dec,
       volume = {69},
       number = {6},
          eid = {101},
        pages = {101},
          doi = {10.1093/pasj/psx118},
archivePrefix = {arXiv},
       eprint = {1710.05848},
 primaryClass = {astro-ph.HE},
       adsurl = {https://ui.adsabs.harvard.edu/abs/2017PASJ...69..101U}
}

@ARTICLE{kasen2017,
       author = {{Kasen}, Daniel and {Metzger}, Brian and {Barnes}, Jennifer and {Quataert}, Eliot and {Ramirez-Ruiz}, Enrico},
        title = "{Origin of the heavy elements in binary neutron-star mergers from a gravitational-wave event}",
      journal = {\nat},
         year = 2017,
        month = nov,
       volume = {551},
       number = {7678},
        pages = {80-84},
          doi = {10.1038/nature24453},
archivePrefix = {arXiv},
       eprint = {1710.05463},
 primaryClass = {astro-ph.HE},
       adsurl = {https://ui.adsabs.harvard.edu/abs/2017Natur.551...80K}
}

@ARTICLE{shibata2017,
       author = {{Shibata}, Masaru and {Fujibayashi}, Sho and {Hotokezaka}, Kenta and {Kiuchi}, Kenta and {Kyutoku}, Koutarou and {Sekiguchi}, Yuichiro and {Tanaka}, Masaomi},
        title = "{Modeling GW170817 based on numerical relativity and its implications}",
      journal = {\prd},
         year = 2017,
        month = dec,
       volume = {96},
       number = {12},
          eid = {123012},
        pages = {123012},
          doi = {10.1103/PhysRevD.96.123012},
archivePrefix = {arXiv},
       eprint = {1710.07579},
 primaryClass = {astro-ph.HE},
       adsurl = {https://ui.adsabs.harvard.edu/abs/2017PhRvD..96l3012S}
}

@ARTICLE{tanaka2017,
       author = {{Tanaka}, Masaomi and {Utsumi}, Yousuke and {Mazzali}, Paolo A. and {Tominaga}, Nozomu and {Yoshida}, Michitoshi and {Sekiguchi}, Yuichiro and {Morokuma}, Tomoki and {Motohara}, Kentaro and {Ohta}, Kouji and {Kawabata}, Koji S. and {Abe}, Fumio and {Aoki}, Kentaro and {Asakura}, Yuichiro and {Baar}, Stefan and {Barway}, Sudhanshu and {Bond}, Ian A. and {Doi}, Mamoru and {Fujiyoshi}, Takuya and {Furusawa}, Hisanori and {Honda}, Satoshi and {Itoh}, Yoichi and {Kawabata}, Miho and {Kawai}, Nobuyuki and {Kim}, Ji Hoon and {Lee}, Chien-Hsiu and {Miyazaki}, Shota and {Morihana}, Kumiko and {Nagashima}, Hiroki and {Nagayama}, Takahiro and {Nakaoka}, Tatsuya and {Nakata}, Fumiaki and {Ohsawa}, Ryou and {Ohshima}, Tomohito and {Okita}, Hirofumi and {Saito}, Tomoki and {Sumi}, Takahiro and {Tajitsu}, Akito and {Takahashi}, Jun and {Takayama}, Masaki and {Tamura}, Yoichi and {Tanaka}, Ichi and {Terai}, Tsuyoshi and {Tristram}, Paul J. and {Yasuda}, Naoki and {Zenko}, Tetsuya},
        title = "{Kilonova from post-merger ejecta as an optical and near-Infrared counterpart of GW170817}",
      journal = {\pasj},
         year = 2017,
        month = dec,
       volume = {69},
       number = {6},
          eid = {102},
        pages = {102},
          doi = {10.1093/pasj/psx121},
archivePrefix = {arXiv},
       eprint = {1710.05850},
 primaryClass = {astro-ph.HE},
       adsurl = {https://ui.adsabs.harvard.edu/abs/2017PASJ...69..102T}
}

@ARTICLE{gillanders2021,
       author = {{Gillanders}, J.~H. and {Smartt}, S.~J. and {Sim}, S.~A. and {Bauswein}, A. and {Goriely}, S.},
        title = "{Modelling the spectra of the kilonova AT2017gfo - I. The photospheric epochs}",
      journal = {\mnras},
         year = 2022,
        month = sep,
       volume = {515},
       number = {1},
        pages = {631-651},
          doi = {10.1093/mnras/stac1258},
archivePrefix = {arXiv},
       eprint = {2202.01786},
 primaryClass = {astro-ph.HE},
       adsurl = {https://ui.adsabs.harvard.edu/abs/2022MNRAS.515..631G}
}

@ARTICLE{tarumi2023,
       author = {{Tarumi}, Yuta and {Hotokezaka}, Kenta and {Domoto}, Nanae and {Tanaka}, Masaomi},
        title = "{Non-LTE analysis for Helium and Strontium lines in the kilonova AT2017gfo}",
      journal = {arXiv e-prints},
         year = 2023,
        month = feb,
          eid = {arXiv:2302.13061},
        pages = {arXiv:2302.13061},
          doi = {10.48550/arXiv.2302.13061},
archivePrefix = {arXiv},
       eprint = {2302.13061},
 primaryClass = {astro-ph.HE},
       adsurl = {https://ui.adsabs.harvard.edu/abs/2023arXiv230213061T}
}

@ARTICLE{sneppen2024he,
       author = {{Sneppen}, Albert and {Damgaard}, Rasmus and {Watson}, Darach and {Collins}, Christine E. and {Shingles}, Luke and {Sim}, Stuart A.},
        title = "{Helium features are inconsistent with the spectral evolution of the kilonova AT2017gfo}",
      journal = {\aap},
         year = 2024,
        month = dec,
       volume = {692},
          eid = {A134},
        pages = {A134},
          doi = {10.1051/0004-6361/202451450},
archivePrefix = {arXiv},
       eprint = {2407.12907},
 primaryClass = {astro-ph.HE},
       adsurl = {https://ui.adsabs.harvard.edu/abs/2024A&A...692A.134S}
}

@ARTICLE{perego2022,
       author = {{Perego}, Albino and {Vescovi}, Diego and {Fiore}, Achille and {Chiesa}, Leonardo and {Vogl}, Christian and {Benetti}, Stefano and {Bernuzzi}, Sebastiano and {Branchesi}, Marica and {Cappellaro}, Enrico and {Cristallo}, Sergio and {Fl{\"o}rs}, Andreas and {Kerzendorf}, Wolfgang E. and {Radice}, David},
        title = "{Production of Very Light Elements and Strontium in the Early Ejecta of Neutron Star Mergers}",
      journal = {\apj},
         year = 2022,
        month = jan,
       volume = {925},
       number = {1},
          eid = {22},
        pages = {22},
          doi = {10.3847/1538-4357/ac3751},
archivePrefix = {arXiv},
       eprint = {2009.08988},
 primaryClass = {astro-ph.HE},
       adsurl = {https://ui.adsabs.harvard.edu/abs/2022ApJ...925...22P}
}

@ARTICLE{jerkstrand2025,
       author = {{Jerkstrand}, Anders and {Pognan}, Quentin and {Banerjee}, Smaranika and {Sterling}, Nicholas and {Grumer}, Jon and {Ferguson}, Niamh and {Butler}, Keith and {Gillanders}, James and {Smartt}, Stephen and {Kawaguchi}, Kyohei and {Vilagos}, Blanka},
        title = "{Infrared spectral signatures of light r-process elements in kilonovae}",
      journal = {arXiv e-prints},
         year = 2025,
        month = oct,
          eid = {arXiv:2510.12410},
        pages = {arXiv:2510.12410},
          doi = {10.48550/arXiv.2510.12410},
archivePrefix = {arXiv},
       eprint = {2510.12410},
 primaryClass = {astro-ph.SR},
       adsurl = {https://ui.adsabs.harvard.edu/abs/2025arXiv251012410J}
}

@ARTICLE{domoto2021,
       author = {{Domoto}, Nanae and {Tanaka}, Masaomi and {Wanajo}, Shinya and {Kawaguchi}, Kyohei},
        title = "{Signatures of r-process Elements in Kilonova Spectra}",
      journal = {\apj},
         year = 2021,
        month = may,
       volume = {913},
       number = {1},
          eid = {26},
        pages = {26},
          doi = {10.3847/1538-4357/abf358},
archivePrefix = {arXiv},
       eprint = {2103.15284},
 primaryClass = {astro-ph.HE},
       adsurl = {https://ui.adsabs.harvard.edu/abs/2021ApJ...913...26D}
}

@ARTICLE{rahmouni2025,
       author = {{Rahmouni}, Salma and {Tanaka}, Masaomi and {Domoto}, Nanae and {Kato}, Daiji and {Hotokezaka}, Kenta and {Aoki}, Wako and {Hirano}, Teruyuki and {Kotani}, Takayuki and {Kuzuhara}, Masayuki and {Tamura}, Motohide},
        title = "{Revisiting Near-infrared Features of Kilonovae: The Importance of Gadolinium}",
      journal = {\apj},
         year = 2025,
        month = feb,
       volume = {980},
       number = {1},
          eid = {43},
        pages = {43},
          doi = {10.3847/1538-4357/ada251},
archivePrefix = {arXiv},
       eprint = {2412.14597},
 primaryClass = {astro-ph.HE},
       adsurl = {https://ui.adsabs.harvard.edu/abs/2025ApJ...980...43R}
}

@ARTICLE{mulholland2024te,
       author = {{Mulholland}, L.~P. and {McNeill}, F. and {Sim}, S.~A. and {Ballance}, C.~P. and {Ramsbottom}, C.~A.},
        title = "{Collisional and radiative data for tellurium ions in kilonovae modelling and laboratory benchmarks}",
      journal = {\mnras},
         year = 2024,
        month = nov,
       volume = {534},
       number = {4},
        pages = {3423-3438},
          doi = {10.1093/mnras/stae2331},
archivePrefix = {arXiv},
       eprint = {2410.05958},
 primaryClass = {astro-ph.SR},
       adsurl = {https://ui.adsabs.harvard.edu/abs/2024MNRAS.534.3423M}
}

@ARTICLE{kato2024,
       author = {{Kato}, Daiji and {Tanaka}, Masaomi and {Gaigalas}, Gediminas and {Kitovien{\.{e}}}, Laima and {Rynkun}, Pavel},
        title = "{Systematic opacity calculations for kilonovae - II. Improved atomic data for singly ionized lanthanides}",
      journal = {\mnras},
         year = 2024,
        month = dec,
       volume = {535},
       number = {3},
        pages = {2670-2686},
          doi = {10.1093/mnras/stae2504},
archivePrefix = {arXiv},
       eprint = {2501.13286},
 primaryClass = {astro-ph.HE},
       adsurl = {https://ui.adsabs.harvard.edu/abs/2024MNRAS.535.2670K}
}

@ARTICLE{chiba2026,
       author = {{Chiba}, Koya and {Tanaka}, Masaomi and {Wanajo}, Shinya and {Fujibayashi}, Sho and {Kawaguchi}, Kyohei and {Hotokezaka}, Kenta},
        title = "{Non-LTE Ionization Modeling for Helium and Strontium in Neutron Star Merger Ejecta}",
      journal = {arXiv e-prints},
         year = 2026,
        month = apr,
          eid = {arXiv:2604.05703},
        pages = {arXiv:2604.05703},
          doi = {10.48550/arXiv.2604.05703},
archivePrefix = {arXiv},
       eprint = {2604.05703},
 primaryClass = {astro-ph.HE},
       adsurl = {https://ui.adsabs.harvard.edu/abs/2026arXiv260405703C}
}

@ARTICLE{sneden2003,
       author = {{Sneden}, Christopher and {Cowan}, John J. and {Lawler}, James E. and {Ivans}, Inese I. and {Burles}, Scott and {Beers}, Timothy C. and {Primas}, Francesca and {Hill}, Vanessa and {Truran}, James W. and {Fuller}, George M. and {Pfeiffer}, Bernd and {Kratz}, Karl-Ludwig},
        title = "{The Extremely Metal-poor, Neutron Capture-rich Star CS 22892-052: A Comprehensive Abundance Analysis}",
      journal = {\apj},
         year = 2003,
        month = jul,
       volume = {591},
       number = {2},
        pages = {936-953},
          doi = {10.1086/375491},
archivePrefix = {arXiv},
       eprint = {astro-ph/0303542},
 primaryClass = {astro-ph},
       adsurl = {https://ui.adsabs.harvard.edu/abs/2003ApJ...591..936S}
}

@ARTICLE{roederer2022,
       author = {{Roederer}, Ian U. and {Lawler}, James E. and {Den Hartog}, Elizabeth A. and {Placco}, Vinicius M. and {Surman}, Rebecca and {Beers}, Timothy C. and {Ezzeddine}, Rana and {Frebel}, Anna and {Hansen}, Terese T. and {Hattori}, Kohei and {Holmbeck}, Erika M. and {Sakari}, Charli M.},
        title = "{The R-process Alliance: A Nearly Complete R-process Abundance Template Derived from Ultraviolet Spectroscopy of the R-process-enhanced Metal-poor Star HD 222925}",
      journal = {\apjs},
         year = 2022,
        month = jun,
       volume = {260},
       number = {2},
          eid = {27},
        pages = {27},
          doi = {10.3847/1538-4365/ac5cbc},
archivePrefix = {arXiv},
       eprint = {2205.03426},
 primaryClass = {astro-ph.SR},
       adsurl = {https://ui.adsabs.harvard.edu/abs/2022ApJS..260...27R}
}

@ARTICLE{ji2019,
       author = {{Ji}, Alexander P. and {Drout}, Maria R. and {Hansen}, Terese T.},
        title = "{The Lanthanide Fraction Distribution in Metal-poor Stars: A Test of Neutron Star Mergers as the Dominant r-process Site}",
      journal = {\apj},
         year = 2019,
        month = sep,
       volume = {882},
       number = {1},
          eid = {40},
        pages = {40},
          doi = {10.3847/1538-4357/ab3291},
archivePrefix = {arXiv},
       eprint = {1905.01814},
 primaryClass = {astro-ph.HE},
       adsurl = {https://ui.adsabs.harvard.edu/abs/2019ApJ...882...40J}
}

@ARTICLE{kilpatrick2017,
       author = {{Kilpatrick}, C.~D. and {Foley}, R.~J. and {Kasen}, D. and {Murguia-Berthier}, A. and {Ramirez-Ruiz}, E. and {Coulter}, D.~A. and {Drout}, M.~R. and {Piro}, A.~L. and {Shappee}, B.~J. and {Boutsia}, K. and {Contreras}, C. and {Di Mille}, F. and {Madore}, B.~F. and {Morrell}, N. and {Pan}, Y.-C. and {Prochaska}, J.~X. and {Rest}, A. and {Rojas-Bravo}, C. and {Siebert}, M.~R. and {Simon}, J.~D. and {Ulloa}, N.},
        title = "{Electromagnetic evidence that SSS17a is the result of a binary neutron star merger}",
      journal = {Science},
         year = 2017,
        month = dec,
       volume = {358},
       number = {6370},
        pages = {1583-1587},
          doi = {10.1126/science.aaq0073},
archivePrefix = {arXiv},
       eprint = {1710.05434},
 primaryClass = {astro-ph.HE},
       adsurl = {https://ui.adsabs.harvard.edu/abs/2017Sci...358.1583K}
}

@ARTICLE{chornock2017,
       author = {{Chornock}, R. and {Berger}, E. and {Kasen}, D. and {Cowperthwaite}, P.~S. and {Nicholl}, M. and {Villar}, V.~A. and {Alexander}, K.~D. and {Blanchard}, P.~K. and {Eftekhari}, T. and {Fong}, W. and {Margutti}, R. and {Williams}, P.~K.~G. and {Annis}, J. and {Brout}, D. and {Brown}, D.~A. and {Chen}, H.-Y. and {Drout}, M.~R. and {Farr}, B. and {Foley}, R.~J. and {Frieman}, J.~A. and {Fryer}, C.~L. and {Herner}, K. and {Holz}, D.~E. and {Kessler}, R. and {Matheson}, T. and {Metzger}, B.~D. and {Quataert}, E. and {Rest}, A. and {Sako}, M. and {Scolnic}, D.~M. and {Smith}, N. and {Soares-Santos}, M.},
        title = "{The Electromagnetic Counterpart of the Binary Neutron Star Merger LIGO/Virgo GW170817. IV. Detection of Near-infrared Signatures of r-process Nucleosynthesis with Gemini-South}",
      journal = {\apjl},
         year = 2017,
        month = oct,
       volume = {848},
       number = {2},
          eid = {L19},
        pages = {L19},
          doi = {10.3847/2041-8213/aa905c},
archivePrefix = {arXiv},
       eprint = {1710.05454},
 primaryClass = {astro-ph.HE},
       adsurl = {https://ui.adsabs.harvard.edu/abs/2017ApJ...848L..19C}
}

@ARTICLE{kitamura2025,
       author = {{Kitamura}, Ayari and {Kawaguchi}, Kyohei and {Tanaka}, Masaomi and {Fujibayashi}, Sho},
        title = "{Linking Analytic Light-curve Models to Physical Properties of Kilonovae}",
      journal = {\apj},
         year = 2025,
        month = apr,
       volume = {982},
       number = {2},
          eid = {97},
        pages = {97},
          doi = {10.3847/1538-4357/adb62c},
archivePrefix = {arXiv},
       eprint = {2502.10021},
 primaryClass = {astro-ph.HE},
       adsurl = {https://ui.adsabs.harvard.edu/abs/2025ApJ...982...97K}
}

@ARTICLE{gillanders2025,
       author = {{Gillanders}, J.~H. and {Flors}, A. and {Ferreira da Silva}, R.},
        title = "{Improved lanthanide constraints for the kilonova AT 2017gfo}",
      journal = {arXiv e-prints},
         year = 2025,
        month = dec,
          eid = {arXiv:2512.24257},
        pages = {arXiv:2512.24257},
          doi = {10.48550/arXiv.2512.24257},
archivePrefix = {arXiv},
       eprint = {2512.24257},
 primaryClass = {astro-ph.HE},
       adsurl = {https://ui.adsabs.harvard.edu/abs/2025arXiv251224257G}
}

@ARTICLE{fishbach2026,
       author = {{Fishbach}, Maya and {Ji}, Alexander P. and {Fong}, Wen-fai and {Wu}, Tom Y. and {Rastinejad}, Jillian C. and {Vijaykumar}, Aditya and {Chen}, Hsin-Yu},
        title = "{Implications of low neutron star merger rates for gamma-ray bursts, r-process production and Galactic double neutron stars}",
      journal = {arXiv e-prints},
         year = 2026,
        month = apr,
          eid = {arXiv:2604.05059},
        pages = {arXiv:2604.05059},
          doi = {10.48550/arXiv.2604.05059},
archivePrefix = {arXiv},
       eprint = {2604.05059},
 primaryClass = {astro-ph.HE},
       adsurl = {https://ui.adsabs.harvard.edu/abs/2026arXiv260405059F}
}

@ARTICLE{rosswog2018,
       author = {{Rosswog}, S. and {Sollerman}, J. and {Feindt}, U. and {Goobar}, A. and {Korobkin}, O. and {Wollaeger}, R. and {Fremling}, C. and {Kasliwal}, M.~M.},
        title = "{The first direct double neutron star merger detection: Implications for cosmic nucleosynthesis}",
      journal = {\aap},
         year = 2018,
        month = jul,
       volume = {615},
          eid = {A132},
        pages = {A132},
          doi = {10.1051/0004-6361/201732117},
archivePrefix = {arXiv},
       eprint = {1710.05445},
 primaryClass = {astro-ph.HE},
       adsurl = {https://ui.adsabs.harvard.edu/abs/2018A&A...615A.132R}
}

@ARTICLE{perego2019,
       author = {{Perego}, Albino and {Bernuzzi}, Sebastiano and {Radice}, David},
        title = "{Thermodynamics conditions of matter in neutron star mergers}",
      journal = {European Physical Journal A},
         year = 2019,
        month = aug,
       volume = {55},
       number = {8},
          eid = {124},
        pages = {124},
          doi = {10.1140/epja/i2019-12810-7},
archivePrefix = {arXiv},
       eprint = {1903.07898},
 primaryClass = {gr-qc},
       adsurl = {https://ui.adsabs.harvard.edu/abs/2019EPJA...55..124P}
}

@ARTICLE{hotokezaka2018,
       author = {{Hotokezaka}, Kenta and {Beniamini}, Paz and {Piran}, Tsvi},
        title = "{Neutron star mergers as sites of r-process nucleosynthesis and short gamma-ray bursts}",
      journal = {International Journal of Modern Physics D},
         year = 2018,
        month = jan,
       volume = {27},
       number = {13},
          eid = {1842005},
        pages = {1842005},
          doi = {10.1142/S0218271818420051},
archivePrefix = {arXiv},
       eprint = {1801.01141},
 primaryClass = {astro-ph.HE},
       adsurl = {https://ui.adsabs.harvard.edu/abs/2018IJMPD..2742005H}
}

@ARTICLE{ligo2025,
       author = {{The LIGO Scientific Collaboration} and {the Virgo Collaboration} and {the KAGRA Collaboration} and {Abac}, A.~G. and {Abouelfettouh}, I. and {Acernese}, F. and {Ackley}, K. and {Adamcewicz}, C. and {Adhicary}, S. and {Adhikari}, D. and {Adhikari}, N. and {Adhikari}, R.~X. and {Adkins}, V.~K. and {Afroz}, S. and {Agarwal}, D. and {Agathos}, M. and {Aghaei Abchouyeh}, M. and {Aguiar}, O.~D. and {Ahmadzadeh}, S. and {Aiello}, L. and {Ain}, A. and {Ajith}, P. and {Akutsu}, T. and {Albanesi}, S. and {Alfaidi}, R.~A. and {Al-Jodah}, A. and {All{\'e}n{\'e}}, C. and {Allocca}, A. and {Al-Shammari}, S. and {Altin}, P.~A. and {Alvarez-Lopez}, S. and {Amarasinghe}, O. and {Amato}, A. and {Amra}, C. and {Ananyeva}, A. and {Anderson}, S.~B. and {Anderson}, W.~G. and {Andia}, M. and {Ando}, M. and {Andrade}, T. and {Andr{\'e}s-Carcasona}, M. and {Andri{\'c}}, T. and {Anglin}, J. and {Ansoldi}, S. and {Antelis}, J.~M. and {Antier}, S. and {Aoumi}, M. and {Appavuravther}, E.~Z. and {Appert}, S. and {Apple}, S.~K. and {Arai}, K. and {Araya}, A. and {Araya}, M.~C. and {Arca Sedda}, M. and {Areeda}, J.~S. and {Argianas}, L. and {Aritomi}, N. and {Armato}, F. and {Armstrong}, S. and {Arnaud}, N. and {Arogeti}, M. and {Aronson}, S.~M. and {Arun}, K.~G. and {Ashton}, G. and {Aso}, Y. and {Assiduo}, M. and {Assis de Souza Melo}, S. and {Aston}, S.~M. and {Astone}, P. and {Attadio}, F. and {Aubin}, F. and {AultONeal}, K. and {Avallone}, G. and {Babak}, S. and {Badaracco}, F. and {Badger}, C. and {Bae}, S. and {Bagnasco}, S. and {Bagui}, E. and {Baiotti}, L. and {Bajpai}, R. and {Baka}, T. and {Baker}, T. and {Ball}, M. and {Ballardin}, G. and {Ballmer}, S.~W. and {Banagiri}, S. and {Banerjee}, B. and {Bankar}, D. and {Baptiste}, T.~M. and {Baral}, P. and {Barayoga}, J.~C. and {Barish}, B.~C. and {Barker}, D. and {Barman}, N. and {Barneo}, P. and {Barone}, F. and {Barr}, B. and {Barsotti}, L. and {Barsuglia}, M. and {Barta}, D. and {Bartoletti}, A.~M. and {Barton}, M.~A. and {Bartos}, I. and {Basak}, S. and {Basalaev}, A. and {Bassiri}, R. and {Basti}, A. and {Bates}, D.~E. and {Bawaj}, M. and {Baxi}, P. and {Bayley}, J.~C. and {Baylor}, A.~C. and {Baynard}, II, P.~A. and {Bazzan}, M. and {Bedakihale}, V.~M. and {Beirnaert}, F. and {Bejger}, M. and {Belardinelli}, D. and {Bell}, A.~S. and {Bellie}, D.~S. and {Bellizzi}, L. and {Beltran-Martinez}, D. and {Benoit}, W. and {Bentara}, I. and {Bentley}, J.~D. and {Ben Yaala}, M. and {Bera}, S. and {Bergamin}, F. and {Berger}, B.~K. and {Bernuzzi}, S. and {Beroiz}, M. and {Berry}, C.~P.~L. and {Bersanetti}, D. and {Bertolini}, A. and {Betzwieser}, J. and {Beveridge}, D. and {Bevilacqua}, G. and {Bevins}, N. and {Bhandare}, R. and {Bhatt}, R. and {Bhattacharjee}, D. and {Bhaumik}, S. and {Bhowmick}, S. and {Biancalana}, V. and {Bianchi}, A. and {Bilenko}, I.~A. and {Billingsley}, G. and {Binetti}, A. and {Bini}, S. and {Binu}, C. and {Birnholtz}, O. and {Biscoveanu}, S. and {Bisht}, A. and {Bitossi}, M. and {Bizouard}, M.-A. and {Blaber}, S. and {Blackburn}, J.~K. and {Blagg}, L.~A. and {Blair}, C.~D. and {Blair}, D.~G. and {Bobba}, F. and {Bode}, N. and {Boileau}, G. and {Boldrini}, M. and {Bolingbroke}, G.~N. and {Bolliand}, A. and {Bonavena}, L.~D. and {Bondarescu}, R. and {Bondu}, F. and {Bonilla}, E. and {Bonilla}, M.~S. and {Bonino}, A. and {Bonnand}, R. and {Booker}, P. and {Borchers}, A. and {Borhanian}, S. and {Boschi}, V. and {Bose}, S. and {Bossilkov}, V. and {Boudon}, A. and {Bozzi}, A. and {Bradaschia}, C. and {Brady}, P.~R. and {Branch}, A. and {Branchesi}, M. and {Braun}, I. and {Briant}, T. and {Brillet}, A. and {Brinkmann}, M. and {Brockill}, P. and {Brockmueller}, E. and {Brooks}, A.~F. and {Brown}, B.~C. and {Brown}, D.~D. and {Brozzetti}, M.~L. and {Brunett}, S. and {Bruno}, G. and {Bruntz}, R. and {Bryant}, J.},
        title = "{GWTC-4.0: Population Properties of Merging Compact Binaries}",
      journal = {arXiv e-prints},
         year = 2025,
        month = aug,
          eid = {arXiv:2508.18083},
        pages = {arXiv:2508.18083},
          doi = {10.48550/arXiv.2508.18083},
archivePrefix = {arXiv},
       eprint = {2508.18083},
 primaryClass = {astro-ph.HE},
       adsurl = {https://ui.adsabs.harvard.edu/abs/2025arXiv250818083T}
}

@ARTICLE{arya2026,
       author = {{Arya}, Aayush and {Damgaard}, Rasmus and {Sneppen}, Albert and {Dougan}, David J. and {Sim}, Stuart A. and {Ballance}, Connor P. and {Watson}, Darach},
        title = "{Strontium and helium in the kilonova AT2017gfo: Origin of the 1{\ensuremath{\mu}}m feature constrained via NLTE calculations}",
      journal = {arXiv e-prints},
         year = 2026,
        month = apr,
          eid = {arXiv:2604.05812},
        pages = {arXiv:2604.05812},
          doi = {10.48550/arXiv.2604.05812},
archivePrefix = {arXiv},
       eprint = {2604.05812},
 primaryClass = {astro-ph.HE},
       adsurl = {https://ui.adsabs.harvard.edu/abs/2026arXiv260405812A}
}

@ARTICLE{domoto2023,
       author = {{Domoto}, Nanae and {Lee}, Jae-Joon and {Tanaka}, Masaomi and {Lee}, Ho-Gyu and {Aoki}, Wako and {Ishigaki}, Miho N. and {Wanajo}, Shinya and {Kato}, Daiji and {Hotokezaka}, Kenta},
        title = "{Transition Probabilities of Near-infrared Ce III Lines from Stellar Spectra: Applications to Kilonovae}",
      journal = {\apj},
         year = 2023,
        month = oct,
       volume = {956},
       number = {2},
          eid = {113},
        pages = {113},
          doi = {10.3847/1538-4357/acf65a},
archivePrefix = {arXiv},
       eprint = {2309.01198},
 primaryClass = {astro-ph.HE},
       adsurl = {https://ui.adsabs.harvard.edu/abs/2023ApJ...956..113D}
}

@ARTICLE{dougan2025,
       author = {{Dougan}, David J. and {McElroy}, Niall E. and {Ballance}, Connor P. and {Ramsbottom}, Catherine A.},
        title = "{Strontium I, III, IV, and V: electron impact excitation data for Kilonovae and white dwarf diagnostic applications}",
      journal = {\mnras},
         year = 2025,
        month = jul,
       volume = {541},
       number = {1},
        pages = {367-383},
          doi = {10.1093/mnras/staf1013},
archivePrefix = {arXiv},
       eprint = {2505.09788},
 primaryClass = {astro-ph.SR},
       adsurl = {https://ui.adsabs.harvard.edu/abs/2025MNRAS.541..367D}
}

@ARTICLE{bromley2026,
       author = {{Bromley}, S. and {Garbe}, E. and {McElroy}, N. and {Ballance}, C. and {Fogle}, M. and {Stancil}, P. and {Loch}, S.},
        title = "{Atomic Data for Non-Equilibrium Modeling of Kilonovae: The Ionization Properties of Te I - III}",
      journal = {arXiv e-prints},
         year = 2026,
        month = feb,
          eid = {arXiv:2603.00346},
        pages = {arXiv:2603.00346},
          doi = {10.48550/arXiv.2603.00346},
archivePrefix = {arXiv},
       eprint = {2603.00346},
 primaryClass = {physics.atom-ph},
       adsurl = {https://ui.adsabs.harvard.edu/abs/2026arXiv260300346B}
}

@ARTICLE{levan2023,
       author = {{Levan}, Andrew J. and {Gompertz}, Benjamin P. and {Salafia}, Om Sharan and {Bulla}, Mattia and {Burns}, Eric and {Hotokezaka}, Kenta and {Izzo}, Luca and {Lamb}, Gavin P. and {Malesani}, Daniele B. and {Oates}, Samantha R. and {Ravasio}, Maria Edvige and {Rouco Escorial}, Alicia and {Schneider}, Benjamin and {Sarin}, Nikhil and {Schulze}, Steve and {Tanvir}, Nial R. and {Ackley}, Kendall and {Anderson}, Gemma and {Brammer}, Gabriel B. and {Christensen}, Lise and {Dhillon}, Vikram S. and {Evans}, Phil A. and {Fausnaugh}, Michael and {Fong}, Wen-fai and {Fruchter}, Andrew S. and {Fryer}, Chris and {Fynbo}, Johan P.~U. and {Gaspari}, Nicola and {Heintz}, Kasper E. and {Hjorth}, Jens and {Kennea}, Jamie A. and {Kennedy}, Mark R. and {Laskar}, Tanmoy and {Leloudas}, Giorgos and {Mandel}, Ilya and {Martin-Carrillo}, Antonio and {Metzger}, Brian D. and {Nicholl}, Matt and {Nugent}, Anya and {Palmerio}, Jesse T. and {Pugliese}, Giovanna and {Rastinejad}, Jillian and {Rhodes}, Lauren and {Rossi}, Andrea and {Saccardi}, Andrea and {Smartt}, Stephen J. and {Stevance}, Heloise F. and {Tohuvavohu}, Aaron and {van der Horst}, Alexander and {Vergani}, Susanna D. and {Watson}, Darach and {Barclay}, Thomas and {Bhirombhakdi}, Kornpob and {Breedt}, Elm{\'e} and {Breeveld}, Alice A. and {Brown}, Alexander J. and {Campana}, Sergio and {Chrimes}, Ashley A. and {D'Avanzo}, Paolo and {D'Elia}, Valerio and {De Pasquale}, Massimiliano and {Dyer}, Martin J. and {Galloway}, Duncan K. and {Garbutt}, James A. and {Green}, Matthew J. and {Hartmann}, Dieter H. and {Jakobsson}, P{\'a}ll and {Kerry}, Paul and {Kouveliotou}, Chryssa and {Langeroodi}, Danial and {Le Floc'h}, Emeric and {Leung}, James K. and {Littlefair}, Stuart P. and {Munday}, James and {O'Brien}, Paul and {Parsons}, Steven G. and {Pelisoli}, Ingrid and {Sahman}, David I. and {Salvaterra}, Ruben and {Sbarufatti}, Boris and {Steeghs}, Danny and {Tagliaferri}, Gianpiero and {Th{\"o}ne}, Christina C. and {de Ugarte Postigo}, Antonio and {Kann}, David Alexander},
        title = "{Heavy-element production in a compact object merger observed by JWST}",
      journal = {\nat},
         year = 2024,
        month = feb,
       volume = {626},
       number = {8000},
        pages = {737-741},
          doi = {10.1038/s41586-023-06759-1},
archivePrefix = {arXiv},
       eprint = {2307.02098},
 primaryClass = {astro-ph.HE},
       adsurl = {https://ui.adsabs.harvard.edu/abs/2024Natur.626..737L}
}

@ARTICLE{gillanders2024grb,
       author = {{Gillanders}, J.~H. and {Smartt}, S.~J.},
        title = "{Analysis of the JWST spectra of the kilonova AT 2023vfi accompanying GRB 230307A}",
      journal = {\mnras},
         year = 2025,
        month = apr,
       volume = {538},
       number = {3},
        pages = {1663-1689},
          doi = {10.1093/mnras/staf287},
archivePrefix = {arXiv},
       eprint = {2408.11093},
 primaryClass = {astro-ph.HE},
       adsurl = {https://ui.adsabs.harvard.edu/abs/2025MNRAS.538.1663G}
}

@ARTICLE{banerjee2025,
       author = {{Banerjee}, Smaranika and {Jerkstrand}, Anders and {Badnell}, Nigel and {Pognan}, Quentin and {Ferguson}, Niamh and {Grumer}, Jon},
        title = "{Nebular Spectra of Kilonovae with Detailed Recombination Rates. I. Light r-process Composition}",
      journal = {\apj},
         year = 2025,
        month = oct,
       volume = {992},
       number = {1},
          eid = {19},
        pages = {19},
          doi = {10.3847/1538-4357/adf6ba},
archivePrefix = {arXiv},
       eprint = {2501.18345},
 primaryClass = {astro-ph.HE},
       adsurl = {https://ui.adsabs.harvard.edu/abs/2025ApJ...992...19B}
}

@ARTICLE{gillanders2023grb,
       author = {{Gillanders}, James H. and {Troja}, Eleonora and {Fryer}, Chris L. and {Ristic}, Marko and {O'Connor}, Brendan and {Fontes}, Christopher J. and {Yang}, Yu-Han and {Domoto}, Nanae and {Rahmouni}, Salma and {Tanaka}, Masaomi and {Fox}, Ori D. and {Dichiara}, Simone},
        title = "{Heavy element nucleosynthesis associated with a gamma-ray burst}",
      journal = {arXiv e-prints},
         year = 2023,
        month = aug,
          eid = {arXiv:2308.00633},
        pages = {arXiv:2308.00633},
          doi = {10.48550/arXiv.2308.00633},
archivePrefix = {arXiv},
       eprint = {2308.00633},
 primaryClass = {astro-ph.HE},
       adsurl = {https://ui.adsabs.harvard.edu/abs/2023arXiv230800633G}
}

@ARTICLE{hotokezaka2020,
       author = {{Hotokezaka}, Kenta and {Nakar}, Ehud},
        title = "{Radioactive Heating Rate of r-process Elements and Macronova Light Curve}",
      journal = {\apj},
         year = 2020,
        month = mar,
       volume = {891},
       number = {2},
          eid = {152},
        pages = {152},
          doi = {10.3847/1538-4357/ab6a98},
archivePrefix = {arXiv},
       eprint = {1909.02581},
 primaryClass = {astro-ph.HE},
       adsurl = {https://ui.adsabs.harvard.edu/abs/2020ApJ...891..152H}
}

@ARTICLE{schoning1997,
       author = {{Schoning}, T.},
        title = "{Effective collision strengths for transitions in the 4p\^k (k = 2-4) ground configurations of KR III, KR IV and KR V}",
      journal = {\aaps},
         year = 1997,
        month = apr,
       volume = {122},
        pages = {277-283},
          doi = {10.1051/aas:1997133},
       adsurl = {https://ui.adsabs.harvard.edu/abs/1997A&AS..122..277S}
}

@ARTICLE{domoto2026,
       author = {{Domoto}, Nanae and {Hotokezaka}, Kenta and {Kasen}, Daniel},
        title = "{Heavy element dust explains the late-time spectra of kilonovae}",
      journal = {arXiv e-prints},
         year = 2026,
        month = jul,
          eid = {arXiv:2607.00433},
        pages = {arXiv:2607.00433},
          doi = {10.48550/arXiv.2607.00433},
archivePrefix = {arXiv},
       eprint = {2607.00433},
 primaryClass = {astro-ph.HE},
       adsurl = {https://ui.adsabs.harvard.edu/abs/2026arXiv260700433D}
}

@ARTICLE{mcmillan2011,
       author = {{McMillan}, Paul J.},
        title = "{Mass models of the Milky Way}",
      journal = {\mnras},
         year = 2011,
        month = jul,
       volume = {414},
       number = {3},
        pages = {2446-2457},
          doi = {10.1111/j.1365-2966.2011.18564.x},
archivePrefix = {arXiv},
       eprint = {1102.4340},
 primaryClass = {astro-ph.GA},
       adsurl = {https://ui.adsabs.harvard.edu/abs/2011MNRAS.414.2446M}
}

@ARTICLE{ricigliano2025,
       author = {{Ricigliano}, Giacomo and {Hotokezaka}, Kenta and {Arcones}, Almudena},
        title = "{Modelling the emission lines from r-process elements in supernova nebulae}",
      journal = {\mnras},
         year = 2025,
        month = nov,
       volume = {543},
       number = {3},
        pages = {2534-2552},
          doi = {10.1093/mnras/staf1577},
archivePrefix = {arXiv},
       eprint = {2502.15896},
 primaryClass = {astro-ph.HE},
       adsurl = {https://ui.adsabs.harvard.edu/abs/2025MNRAS.543.2534R}
}

@article{gaigalas2026jj2lsj,
  title={The jj2lsj transformation for Hullac},
  author={Gaigalas, G and Kato, D and Tanaka, M},
  journal={Journal of Quantitative Spectroscopy and Radiative Transfer},
  pages={110058},
  year={2026},
  publisher={Elsevier}
}

@ARTICLE{mulholland2026ce,
       author = {{Mulholland}, Leo Patrick and {Ferguson}, Niamh and {Shingles}, Luke J. and {Ramsbottom}, Catherine A. and {Ballance}, Connor P. and {Sim}, Stuart A.},
        title = "{Ce II-IV emission in kilonovae with R-matrix collision strengths and distorted wave recombination rates}",
      journal = {arXiv e-prints},
         year = 2026,
        month = sep,
          eid = {arXiv:2609.15900},
        pages = {arXiv:2609.15900},
archivePrefix = {arXiv},
       eprint = {2609.15900},
 primaryClass = {astro-ph.HE},
       adsurl = {https://ui.adsabs.harvard.edu/abs/2026arXiv260915900M}
}
\bibliographystyle{aasjournalv7}

\end{document}